\documentclass[11pt,onecolumn,draftclsnofoot]{IEEEtran}
\usepackage{amsthm}

\usepackage[T1]{fontenc}
\usepackage[utf8]{inputenc}
\usepackage{lmodern}
\usepackage{amsmath,amssymb,amsthm}
\usepackage{amsmath,amssymb,booktabs}
 \usepackage{tikz}
\usetikzlibrary{arrows.meta,positioning,fit,backgrounds,calc} \usetikzlibrary{arrows.meta,positioning,calc,shapes.multipart}

 \definecolor{serverfill}{RGB}{232,230,186}

\usepackage{amsmath,amssymb,amsfonts,mathtools,bm,amsthm}

\usepackage{enumitem}
\usepackage{booktabs}
\usepackage{graphicx}
\usepackage[caption=false,font=footnotesize]{subfig} 

\usepackage{tikz}
\usetikzlibrary{arrows.meta,positioning,decorations.pathreplacing,fit,calc}

\usepackage{microtype}
\usepackage{etoolbox}
\usepackage{perpage}
\MakePerPage{footnote}
\usepackage{color}
\usepackage[normalem]{ulem}
\usepackage{makeidx}
\usepackage{xspace}
\usepackage{wrapfig}
\usepackage{lipsum}
\usepackage{cuted}
\usepackage{cite}
\usepackage[ruled,vlined]{algorithm2e} 

\usepackage[hidelinks]{hyperref}

\theoremstyle{plain}
\newtheorem{theorem}{Theorem}
\newtheorem{lemma}{Lemma}

\newtheorem{proposition}{Proposition}

\theoremstyle{definition}

\theoremstyle{remark}
\newtheorem{remark}{Remark}

\definecolor{darkred}{rgb}{1, 0.1, 0.3}
\definecolor{darkblue}{rgb}{0.1, 0.1, 1}
\definecolor{darkgreen}{rgb}{0,0.6,0.5}
\def\BibTeX{{\rm B\kern-.05em{\sc i\kern-.025em b}\kern-.08em
    T\kern-.1667em\lower.7ex\hbox{E}\kern-.125emX}}

\DeclareMathOperator{\Argmin}{Argmin}

\DeclareMathOperator{\rank}{rank}

\newcommand{\R}{\mathbb{R}}

\newcommand{\E}{\mathbb{E}}
\newcommand{\MP}{\mathrm{MP}}

\def \w {\mathbf{w}}

\newcommand{\ind}{\mathbf{1}}
\newcommand{\mrecv}{\underline{m}} 
\allowdisplaybreaks

\begin{document}
\title{{General Multi-User Distributed Computing}\\[0.4em]
\large A Learning-Theoretic RKHS Framework for Generic Nonlinear Target Functions with Topology-Aware Risk Analysis}

\author{Ali Khalesi\\
\small Institut Polytechnique des Sciences Avanc\'ees (IPSA) and LINCS Lab, Paris, France\\
\small \texttt{ali.khalesi@ipsa.fr}
}

\maketitle
\vspace{-2em}
\begin{abstract}
This paper studies multi-user distributed computation over shared real-valued subfunctions under computation and communication constraints. We consider a \emph{General Multi-User Distributed Computing (GMUDC)} model in which different users request heterogeneous target functions represented in the reproducing-kernel Hilbert space of a shift-invariant kernel, thereby covering generic nonlinear target mappings beyond linearly separable tasks. Unlike tessellated distributed computing frameworks that rely on disjoint-support topologies in their native setting, the GMUDC model allows arbitrary task-assignment and connectivity topologies subject to per-server computation and communication budgets~$\Gamma$ and~$\Delta$. We analyze two complementary regimes. In the \emph{quenched} regime, the assignment and communication topology are fixed, and we derive upper and lower bounds on the resulting reconstruction risk that separate a spectral approximation term from a topology-dependent coverage term. In the \emph{annealed} regime, the assignment and links are drawn uniformly at random from a prescribed ensemble, and we characterize the corresponding average-risk scaling together with a topology-dependent coverage threshold. These results provide a topology-aware characterization of the computation--communication--accuracy trade-off for approximate multi-user distributed computation. They identify fundamental limits, up to constants and logarithmic factors, under the model assumptions adopted in the paper. In the shared linear/isotropic comparison regime, the framework also recovers the relevant tessellated distributed computing benchmark, while the broader GMUDC formulation applies to generic nonlinear target functions in kernel RKHSs and accommodates a wider range of task-assignment and communication topologies than disjoint-support constructions alone.
\end{abstract}
\section{Introduction}
The increasing complexity of distributed learning, sensing, and computation in modern networked systems demands architectures that can simultaneously balance computation, communication, and reliability under tight resource constraints such as power and data. As tasks scale across multiple nodes, users, and layers of processing, understanding the limits of distributed computation under explicit resource constraints becomes increasingly important. The well-known computation--communication tradeoff stands at the heart of distributed computing and appears across a broad range of settings~\cite{verbraeken2020survey,ulukus2022private,wang2018fundamental,li2017fundamental,yu2017polynomial,reisizadeh2021codedreduce,chen2021distributed}, where computation and communication act as two coupled bottlenecks governing overall system efficiency. The balance between these resources shapes the achievable performance of distributed processing and learning systems, including coded computing, federated learning, and edge inference paradigms.

A further key dimension concerns \emph{computational accuracy}---that is, the ability to recover the desired user tasks with small error or distortion under limited resources. Numerous recent works~\cite{khalesi2024tessellated,Malak,woodruff2014sketching,jahani2021codedsketch,RaviApproximated,CharalambidesApproximated,RamchandranApproximated,StarkApproximated,ZhuRamchandranApproximated,TayyebehBerrutMaddah-Ali,CadambeApproximated,NarayananKrishna,RashmiApproximated,ZhuJinggeApproximatedLearning,JahaniNezhadApproximated,MaddahAliApproximated,omidvar2024hybrid} have proposed approximate or randomized computation mechanisms to improve this accuracy.

While classical coded distributed computing frameworks were originally designed around \emph{finite-field algebraic codes}, such as Reed--Solomon and polynomial-based constructions, their applicability to modern large-scale learning and inference systems remains limited. These algebraic approaches rely on \emph{exact recovery} over discrete alphabets and are therefore not naturally matched to real-valued, noisy, and inherently approximate computations of the kind that arise in machine learning and signal processing workloads. Motivated by these limitations, a recent line of research has sought to extend coded computing beyond finite-field algebra to operate directly over the reals, embracing approximate computation as a natural property rather than a defect. This direction includes Learning-Theoretic Coded Computing~\cite{moradi2024learningtheoretic}, which formulates coded computation as a functional optimization problem, as well as follow-up works on probabilistic straggler and adversarial regimes~\cite{moradi2025probabilistic,moradi2025adversarial}. Collectively, these works point toward a broader learning-theoretic view of real-valued coded computation.

In essence, distributed computation in such settings operates at the intersection of three tightly interlinked quantities---accuracy, communication, and computation. In this work, we study this interplay in the context of \emph{multi-user distributed computing}, with the goal of characterizing how computation and communication budgets shape the achievable approximation risk across heterogeneous users and network topologies.

\begin{figure}
      \centering
\includegraphics[scale=0.6]{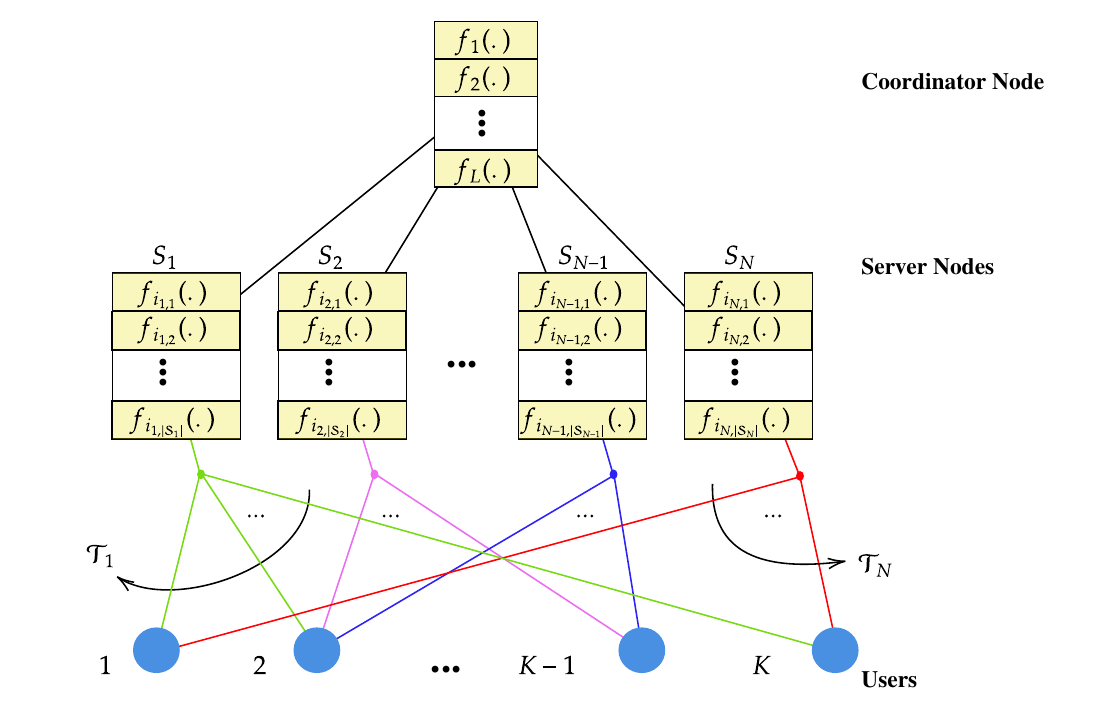}
\vspace{-29pt}
      \caption{The $K$-user, $N$-server, $T$-shot \emph{General Multi-User Distributed Computing} framework considers a setting where each server $n$ computes a subset of subfunctions $\mathcal{S}_n=\{f_{i_{n,1}}(.),f_{i_{n,2}}(.),\hdots , f_{i_{n,|\mathcal{S}_n|}}(.)\}$ and communicates the corresponding results to a subset of users $\mathcal{T}_{n}$. This operation is performed under the computational constraint $|\mathcal{S}_n|\leq \Gamma\leq L$ and the communication constraint $|\mathcal{T}_n|\leq \Delta \leq K$, which correspond to the normalized budgets $\gamma = \Gamma/L$ and $\delta = \Delta/K$. The system aims to minimize the population risk $\mathcal{R}\big(\mathcal{D}\mid \mathsf{A},\mathsf{L}\big)$, representing the average reconstruction error of the users with respect to their desired target functions. Here, $\mathcal{D}$ denotes the overall encoding and decoding strategy, $\mathsf{A}=(\mathcal{S}_1,\ldots,\mathcal{S}_N)$ specifies the computation assignments, and $\mathsf{L}=(\mathcal{T}_1,\ldots,\mathcal{T}_N)$ defines the communication topology between the servers and the users.  }
      \label{Fig: System Model}
  \end{figure}

In particular, our setting considers a coordinator node that manages $N$ server nodes which must contribute in a distributed manner to the computation of the target functions requested by $K$ different users, where the target functions can be obtained as general (not necessarily linear) transformations of at most $L$ output subfunctions (see Fig.~\ref{Fig: System Model}). The subfunctions are computed by the servers. More precisely, user $k \in \{1,2,\hdots, K\}$ demands a function
\[
F_{k}(x)=h_k(f_1(x),f_2(x), \hdots, f_{L}(x)), \qquad x \in \mathcal{X} \subseteq \mathbb{R}^{d},\; d \in \mathbb{N},
\]
where $h_k(\cdot)$ belongs to the \emph{Reproducing Kernel Hilbert Space} (RKHS) $\mathcal{H}_K$ induced by a bounded, shift-invariant kernel $K(\cdot,\cdot)$. This formulation captures a broad class of target functions, including linear, polynomial, and smooth nonlinear mappings. For example, under the linear kernel $K(u,v)=u^\top v$, the user functions reduce to linear combinations of the subfunction outputs, recovering the linearly separable regime studied in~\cite{wan2021distributed,Khalesi-perfect,khalesi2,khalesi2024tessellated}; under a polynomial kernel $K(u,v)=(1+u^\top v)^q$, the user functions encompass all multivariate polynomials of degree up to $q$; and under Gaussian or Laplacian kernels, the targets can represent smooth nonlinear transformations. Thus the framework covers both classical linearly decomposable models and broader nonlinear task families arising in distributed learning and inference. Moreover, even when the true user function $F_k$ does not belong exactly to $\mathcal{H}_K$, the analysis can be extended by introducing an additional misspecification term.

The above target-function formulation serves as an abstraction of distributed matrix--vector and matrix--matrix multiplication~\cite{yu2017polynomial,ramamoorthy2019universally}, distributed gradient coding~\cite{ye2018communication,raviv2020gradient}, distributed linear transforms~\cite{wang2018fundamental}, and multivariate polynomial computation~\cite{yu2020straggler}. Beyond these classical coded-computing problems, it also connects naturally to distributed learning settings such as federated learning~\cite{kairouz2021advances}, distributed kernel regression and Gaussian-process inference~\cite{park2021communication}, and large-scale ridge regression under communication constraints~\cite{lee2021distributed}. In addition, the same abstraction can serve as a lightweight modeling template for certain distributed estimation and control problems in networked sensing systems, where local nodes compute subfunction-like summaries of measurements and users or coordinators aggregate them under communication constraints~\cite{olfati2007consensus,cattivelli2010diffusion,beard2002coordination,sun2019distributed,li2020decentralized}. We use such examples here as motivating abstractions rather than as claims of a fully realistic physical-system model.

In this work, similarly to~\cite{khalesi2024tessellated}, we assume that the system operates in three sequential phases. In the \emph{assignment phase}, the subfunctions are allocated to the servers, represented by
$\mathsf{A}\triangleq(\mathcal{S}_1,\ldots,\mathcal{S}_N)$, where each $\mathcal{S}_n\subseteq\{1,2,\ldots,L\}$ denotes the set of subfunctions assigned to server $n$. The assignment must satisfy a computation budget $\Gamma\in\{1,2,\ldots,L\}$, which specifies the maximum number of subfunctions that can be computed by any single server. In the subsequent \emph{computation/encoding phase}, each server computes its assigned subfunctions and applies an encoder to generate symbols that are transmitted over error-free links in $T$ shots to the connected users. Finally, in the \emph{communication/decoding phase}, the encoded symbols are delivered to users according to the network's link structure, which must satisfy the communication budget $\Delta$, representing the maximum number of user links per server. Formally, each server $n$ can communicate with a subset of users $\mathcal{T}_n\subseteq\{1,2,\ldots,K\}$, and we denote the overall link configuration by $\mathsf{L}=(\mathcal{T}_1,\ldots,\mathcal{T}_N)$. After receiving the encoded symbols, each user applies a decoder to reconstruct an approximation $\widehat{F}_k$ of its desired target function $F_k$.

Throughout the paper, we work under the nominal system assumptions stated in Section~\ref{System-Model}, in particular the synchronous no-straggler setting and the equal-significance abstraction for subfunction computation cost. These assumptions are adopted to isolate the topology--risk trade-off studied in this paper and to keep the analytical dependence on $(\Gamma,\Delta,T)$ transparent. In particular, the present results do not model asynchronous server behavior, random straggling, or heterogeneous per-server/per-subfunction computational costs. Such effects are practically important in distributed-learning and edge-computing systems, but incorporating them would require a modified formulation in which the effective communicated-feature budget and the relevant coverage events become random and workload-dependent. We therefore treat these effects as extensions beyond the nominal GMUDC model considered here, while reporting supplementary robustness-sensitivity diagnostics outside the nominal assumptions in Appendix~\ref{subsec:sim_robustness_monotone}.These assumptions allow us to isolate the topology--risk tradeoff studied here.

We measure the performance of each design $\mathcal{D}$---which consists solely of the encoding and decoding functions of the servers and users---under two complementary regimes. In the first regime, we evaluate the performance for a fixed realization of the computation assignments and communication links $(\mathsf{A},\mathsf{L})$ through the \emph{quenched population risk}, defined as
\begin{equation*}
\mathcal{R}\big(\mathcal{D}\mid \mathsf{A},\mathsf{L}\big)
=\mathbb{E}_{x\sim\mathbb{P}_X}\!\left[
\frac{1}{K}\sum_{k=1}^K
\big\|\widehat{F}_k(x) - F_k(x)\big\|_2^2
\right],
\end{equation*}
where $\mathbb{P}_X$ denotes the input distribution over $\mathcal{X}$. This quantity captures the reconstruction error of all users for a specific system realization.

The second regime, referred to as the \emph{annealed population risk}, averages over the randomness of the uniformly distributed computation assignments and communication links, each constrained by the budgets $\Gamma$ and $\Delta$, respectively:
\begin{equation*}
\overline{\mathcal{R}}(\mathcal{D})
\triangleq
\mathbb{E}_{\mathsf{A},\mathsf{L}}\!\left[\mathcal{R}\big(\mathcal{D}\mid \mathsf{A},\mathsf{L}\big)\right].
\end{equation*}
This quantity represents the expected performance of the system over typical random realizations of the network topology. Together, the quenched and annealed population risks provide two complementary characterizations of performance in GMUDC: the former captures \emph{instance-wise} behavior under a fixed network realization, while the latter describes the \emph{average-case} behavior across random topologies. In what follows, we formally define the two optimization problems associated with these performance metrics. The first corresponds to the \emph{quenched design}, where the assignment and link topology are fixed, while the second corresponds to the \emph{annealed design}, which averages performance over random network realizations.

\paragraph*{P1 (Quenched design)}
Given a fixed realization $(\mathsf{A},\mathsf{L})$, the objective is to find encoder and decoder mappings that minimize the quenched population risk:
\begin{equation*}
\begin{aligned}
\inf_{\mathcal{D}}\quad &
\mathcal{R}\big(\mathcal{D}\mid \mathsf{A},\mathsf{L}\big).
\end{aligned}
\end{equation*}

\paragraph*{P2 (Annealed design)}
In this case, the goal is to minimize the expected (annealed) population risk averaged over all possible random assignments and link configurations:
\begin{equation*}
\begin{aligned}
\inf_{\mathcal{D}}\quad &
\overline{\mathcal{R}}(\mathcal{D}).
\end{aligned}
\end{equation*}
Here, the design is not conditioned on a specific realization of $(\mathsf{A},\mathsf{L})$, but rather aims to minimize the expected population risk over their random draws. The encoders and decoders may depend on model parameters such as $(\Gamma,\Delta,T)$ and on the data distribution $\mathbb{P}_X$, but not on any particular instance of the topology.

\paragraph{Methods and novelties}
Our approach combines an achievability construction and a converse argument tailored to multi-user distributed computing. On the achievability side, we instantiate \emph{masked random Fourier feature (RFF)} encoders that respect the per-server compute mask of size $\Gamma$ across $T$ shots, paired with user-wise ridge-regression decoders. This construction yields a Monte Carlo approximation of the kernel on subfunction outputs while explicitly quantifying the effect of masking and limited fan-out on the resulting variance. On the converse side, we develop a \emph{spectral--coverage} argument: \emph{(i)} a spectral lower bound driven by the eigen-tail of the kernel integral operator beyond the communicated-feature budget, and \emph{(ii)} a coverage lower bound that enforces a nonzero floor whenever essential subfunction coordinates for a user are not observed by the realized assignment/link topology. In the averaged regime, the coverage penalty decays exponentially in $\gamma N\delta$ (with $\gamma=\Gamma/L$ and $\delta=\Delta/K$).

\paragraph{Meaning of tightness}
The main results are \emph{tight} in the sense relevant to the present model: the achievability and converse bounds are governed by the same communicated-feature scale and the same spectral approximation mechanism, with any residual mismatch limited to constant and logarithmic factors together with the explicit coverage term. We do not claim exact equality of the upper and lower bounds for all parameter choices. Rather, the results identify the fundamental tradeoff structure under the GMUDC model assumptions. In particular, for Theorem~\ref{thm:P1} the two sides match at the same per-user feature scale once the realized coverage floor vanishes, while for Theorem~\ref{thm:P2} they match up to constants and logarithmic factors once $\gamma N\delta$ exceeds the logarithmic coverage threshold and the exponential coverage penalty becomes negligible.

\paragraph{Quenched and annealed bounds}
For the quenched regime, Theorem~\ref{thm:P1} shows that, with high probability,
\[
\mathcal{R}(\mathcal{D}\mid \mathsf{A},\mathsf{L})
=
\tilde{\mathcal{O}}\!\left(
\frac{B^2}{\gamma\,m_{\mathrm{harm}}}
+
\frac{\sigma^2 d_\lambda}{M}
+
B^2\lambda
\right),
\]
while every design obeys the lower bound
\[
\mathcal{R}(\mathcal{D}\mid \mathsf{A},\mathsf{L})
=
\Omega\!\Bigg(
\frac{1}{K}\sum_{k=1}^{K}
\Big[
B^2\!\sum_{j>\mrecv_k(\mathcal{T})}\lambda_j
\;\vee\;
\varepsilon_{\mathrm{cov},k}(\mathsf{A},\mathsf{L})
\Big]
\Bigg).
\]
Thus, for a fixed realized topology, the risk is controlled by the communicated-feature scale together with the spectral tail of the target class, unless a topology-induced coverage floor is present.

For the annealed regime, Theorem~\ref{thm:P2} gives
\[
B^2\!\sum_{j>TN\delta}\!\lambda_j
\ \vee\
\Omega(1)\cdot\mathbf{1}\!\{\gamma N\delta \lesssim \log r_{\mathrm{avg}}\}
\ \le\
\inf_{\mathcal{D}}\overline{\mathcal{R}}(\mathcal{D})
\ \le\
\tilde{\mathcal{O}}\!\left(
\frac{B^2}{\gamma\,TN\delta}
+\frac{\sigma^2 d_\lambda}{M}
+B^2\lambda
+r_{\mathrm{avg}}e^{-\gamma N\delta}
\right),
\]
where $r_{\mathrm{avg}}=\tfrac{1}{K}\sum_{k=1}^K r_k$ denotes the average essential-coordinate size across users. Once $\gamma N\delta \gtrsim \log r_{\mathrm{avg}}$, the coverage probability becomes overwhelming, the coverage penalty becomes negligible, and the achievable risk matches the spectral lower bound up to constants and logarithmic factors.
\paragraph{Design law and significance}
Combining both regimes yields the unified scaling
\[
\text{Risk}\ \asymp\
\underbrace{\frac{1}{\gamma\,N\,\delta\,T}}_{\text{resource-driven rate}}
\;+\;
\underbrace{\text{(kernel spectral tail)}}_{\text{task complexity}}
\;+\;
\underbrace{e^{-\gamma N\delta}}_{\text{coverage reliability}},
\]
showing that our achievable construction attains the fundamental computation–communication–accuracy tradeoff up to constants and logarithms. 
Practically: increase $T$, $N$, or $\delta$ to multiply the received features ($TN\delta$); increase $\gamma$ to mitigate masking loss; ensure $\gamma N\delta \gtrsim \log r_{\mathrm{avg}}$ to eliminate coverage floors; and tune $(M,\lambda)$ to control variance/bias via $d_\lambda$.

\paragraph{Comparison with previous real-valued multi-user distributed computing}
To compare the present framework with prior work---most notably \emph{Tessellated Distributed Computing (TDC)}~\cite{khalesi2024tessellated}---we adopt a kernel-theoretic and ridge-regression perspective that extends naturally beyond the linearly separable setting of TDC. Our formulation accommodates generic nonlinear, smooth, and power-limited targets in kernel RKHSs, while preserving linear user-side decoding. At the same time, unlike TDC's disjoint-support or cyclic-assignment structural assumptions, the GMUDC model allows arbitrary assignment and link topologies provided they obey the computation and communication budgets $(\Gamma,\Delta)$, thereby covering heterogeneous, overlapping, and unbalanced networks.

To obtain a direct comparison with TDC, Section~\ref{sec:comparison} specializes to the shared linear/isotropic regime used in~\cite[Thm.\,2]{khalesi2024tessellated}. In that regime, our scheme's quenched distortion is exactly the lower-tail truncated first moment of the user Gram spectrum, which enables a sharp comparison to the Marchenko--Pastur (MP) benchmark. Under disjoint-and-balanced supports, the achievable quenched distortion matches the MP envelope asymptotically. For general $(\Gamma,\Delta)$-regular topologies, Theorem~\ref{thm:gap-bounds} defines the \emph{quenched MP--gap}, gives an exact integral representation via lower-tail quantiles, proves nonnegativity, and quantifies how overlap or imbalance changes the lower spectral tail relative to the MP benchmark.

\paragraph*{Main Contributions}
The main contributions of the paper are as follows:
\begin{itemize}[leftmargin=1.3em]

\item \textbf{A learning-theoretic GMUDC model.}
We formulate multi-user distributed computation as an inference problem in RKHSs, thereby connecting distributed computation, kernel approximation, and statistical learning within a single real-valued framework.

\item \textbf{Masked random-feature achievability with ridge decoding.}
We develop an achievable scheme based on masked random Fourier feature encoders and user-wise ridge-regression decoders, leading to explicit accuracy--resource tradeoffs in terms of the masking rate $\gamma=\Gamma/L$ and the fan-out $\delta=\Delta/K$.

\item \textbf{A spectral--coverage converse mechanism.}
We derive lower bounds that separate a spectral obstruction, through the eigen-tail of the kernel integral operator beyond the communicated-feature budget, from a topology-induced coverage obstruction when essential coordinates remain unobserved.

\item \textbf{Quenched and annealed risk characterizations.}
We establish complementary bounds for fixed realized topologies and for random topology ensembles, revealing an explicit computation--communication--accuracy law with an exponential coverage phase transition.

\item \textbf{A comparison tool in the linear/isotropic regime.}
In the regime shared with TDC, we introduce the quenched MP--gap as a spectral measure of the loss relative to the Marchenko--Pastur benchmark, and use it to compare disjoint-support and overlapping topologies.

\item \textbf{A broader function and topology class.}
The framework accommodates generic nonlinear target functions represented in kernel RKHSs and remains valid for arbitrary assignment and link topologies obeying the budgets $(\Gamma,\Delta)$.

\item \textbf{Worked examples and numerical validation.}
We provide concrete quenched and annealed worked examples that fully instantiate the abstract model, together with theorem-driven synthetic experiments validating the main mechanisms of Theorems~\ref{thm:P1},~\ref{thm:P2}, and~\ref{thm:gap-bounds} at finite dimensions.

\end{itemize}

\subsection{Paper Organization}
Section~\ref{System-Model} introduces the GMUDC system model, formalizes the quenched and annealed risks, and states the optimization problems \textbf{P1} and \textbf{P2}. Section~\ref{Results} presents the main results: \textbf{Theorem~\ref{thm:P1}} for the quenched regime and \textbf{Theorem~\ref{thm:P2}} for the annealed regime, together with proof sketches and two worked examples that fully instantiate the abstract model in fixed-topology and random-topology settings. Section~\ref{sec:comparison} specializes the framework to the shared linear/isotropic regime used for comparison with Tessellated Distributed Computing~\cite{khalesi2024tessellated} and introduces \textbf{Theorem~\ref{thm:gap-bounds}} on the quenched MP--gap. Section~\ref{sec:conclusion} concludes the paper. The appendices contain the technical proofs, the annealed counterpart of the worked example, and a numerical-validation section reporting theorem-driven synthetic experiments and supplementary diagnostics.

\paragraph*{Notation}
Scalars are plain (e.g., $a,b,\lambda$); vectors are bold (e.g., $\mathbf{x},\mathbf{w}$); matrices use uppercase roman (e.g., $G$); sets are calligraphic (e.g., $\mathcal{S},\mathcal{T},\mathcal{X}$). Indices: users $k\in[K]\triangleq\{1,\dots,K\}$, servers $n\in[N]$, coordinates $\ell\in[L]$, shots $t\in[T]$, samples $i\in[M]$. For $x\in \mathcal{X} \subseteq \mathbb{R}^d$, the subfunction outputs are $w_{\ell}=f_{\ell}(x)$ and $\mathbf{w}(x)=(w_1,\hdots,w_L)\in\mathbb{R}^{L}$; all such vectors are column vectors, and $\mathbf{w}(x)_{-\ell}$ denotes the vector $\mathbf{w}(x)$ with its $\ell$-th coordinate removed. $K(\cdot,\cdot)$ is a positive-definite kernel with RKHS $\mathcal{H}_K$ and norm $\|\cdot\|_{\mathcal{H}_K}$; the associated integral operator $T_K$ on $L^2(\mu_{\mathbf{w}})$ has nonincreasing eigenvalues $\{\lambda_j\}_{j\ge1}$, where $\mu_{\mathbf{w}}$ denotes the law (pushforward measure) of $\mathbf{w}(X)$ induced by $\mathbb{P}_X$. Probability and asymptotics: $\mathbb{P}(\cdot)$ is probability, $\mathbb{E}[\cdot]$ is expectation, $\ind\{\cdot\}$ is the indicator, $g\lesssim h$ means $g\le C h$ for an absolute constant $C$, and $g\asymp h$ denotes two-sided $\lesssim$. Norms: $\|\cdot\|_2$ is Euclidean, $\|\cdot\|_{\mathrm{op}}$ is the operator norm. All logarithms are natural unless stated otherwise. For scalars $a,b\in\mathbb{R}$ we use $a\vee b \triangleq \max\{a,b\}$ and $a\wedge b \triangleq \min\{a,b\}$. For $f\in\mathcal{H}_K$, the RKHS norm $\|f\|_{\mathcal{H}_K}$ is the Hilbert norm induced by $\langle\cdot,\cdot\rangle_{\mathcal{H}_K}$ and is characterized by the reproducing property
$f(x)=\langle f, K(\cdot,x)\rangle_{\mathcal{H}_K}$ for all $x\in\mathcal{X}$. $\mathbf{1}_{\mathcal{R}}\in\{0,1\}^{K}$, $\mathcal{R}\subseteq[K]$, is the coordinate indicator of $\mathcal{R}$, with $(\mathbf{1}_{\mathcal{R}})_k=1$ if $k\in\mathcal{R}$ and $0$ otherwise. Operator $\odot$ denotes Hadamard product. $\|S\|_{\psi_2}$ denotes the sub-Gaussian Orlicz norm of a random variable $S$, defined as $\|S\|_{\psi_2}=\inf\{c>0:\E[\exp(S^2/c^2)]\le 2\}$, equivalently $\|S\|_{\psi_2}\asymp \sup_{p\ge 1} (\E|S|^p)^{1/p}/\sqrt{p}$. For a linear map $A:\mathbb{R}^p\!\to\!\mathbb{R}^q$, the \emph{range} is $\mathrm{Range}(A)=\{Ax:x\in\mathbb{R}^p\}\subseteq\mathbb{R}^q$. The space $L_2(\mathbb{P}_X)$ is the Hilbert space of real-valued, square-integrable functions of the random variable $X$. More generally, for a measure $\rho$, $L_2(\rho)$ denotes the Hilbert space of square-integrable functions with respect to $\rho$. For $f,g\in L_2(\rho)$, we also write $f\otimes g$ for the kernel
$(f\otimes g)(u,v)\triangleq f(u)g(v)$,
which corresponds to the operator $h\mapsto \langle h,g\rangle_{L_2(\rho)}f$. $A \succeq B$ means $A-B$ is positive semidefinite (PSD), and $A \succ B$ means $A-B$ is positive definite (PD).

\section{System Model}
\label{System-Model}
We consider a multi-user distributed computing system with $K$ users, $N$ servers, and $L$ subfunctions. 

For each input sample $x\in\mathcal{X}\subseteq\mathbb{R}^d$ (with $d\in\mathbb{N}$), define $w_\ell \triangleq f_\ell(x), \ell\in[L],$ and collect $\mathbf{w}(x)=(w_1,\dots,w_L)\in\mathbb{R}^L$. Note that $f_\ell(x)$ is a subfunction to be computed by the servers. 
Each user $k\in[K]$ requests a target function in the RKHS of a bounded, shift-invariant kernel $K$, interpreted coordinate-wise:
\[
F_k(x)=h_k(\mathbf{w}(x)),\qquad h_k=(h_{k,1},\ldots,h_{k,m_k}),\ \ h_{k,j}\in\mathcal{H}_K.
\]
We assume there are no stragglers and all $N$ servers are active in all shots. This abstraction cleanly separates the {computational layer}, where subfunctions $f_\ell$ are evaluated by the servers, from the {task layer}, where users request target functions $F_k$, thereby accommodating general, possibly nonlinear transformations $h_k$. The system operates in three sequential phases once the coordinator node receives the users’ function requests. In the {assignment phase}, subfunctions are allocated to servers under a computation budget by the coordinator node; in the {computation/encoding phase}, each server computes its assigned subfunctions and applies an encoder to produce symbols to be transmitted via the error-free links in  $T$ shots to the connected users; and in the {communication/decoding phase}, encoded symbols are delivered to users according to the link structure, after which each user decodes to approximate $F_k$. During the assignment phase, each server $n\in[N]$ is assigned a set $\mathcal{S}_n\subseteq[L]$ of subfunctions to compute, subject to
\begin{equation}
|\mathcal{S}_n|\le \Gamma. \label{eq:comp-budget}
\end{equation}
We assume each subfunction has an equally significant computational cost. The above assumptions define the nominal analytical model used throughout the paper. They should be interpreted as a deliberate abstraction that separates topology, feature budget, and approximation effects from additional scheduling and workload heterogeneity issues. In particular, if servers were allowed to straggle or if subfunctions had nonuniform costs, then both the received-feature counts and the coordinate-coverage events would become random objects coupled to the execution dynamics. The present theorems do not characterize that more general setting. Rather, they identify the topology-aware risk trade-off under the synchronous homogeneous-cost model, which serves as the baseline regime for the subsequent achievability, converse, and comparison results. In this sense, we define the normalized computation cost to be
\[
\gamma \triangleq \frac{\Gamma}{L},
\]
so $\gamma$ represents the average replication level (fraction of subfunctions per server), governing computation–communication–risk trade-offs. We assume $1\le \Gamma\le L$, hence $0<\gamma\le 1$.

In each of $T$ shots, server $n$ applies an encoder to its computed subfunctions:
\begin{equation}
z_{n,t}(x) = \phi_{n,t}\big(\{w_\ell(x):\ell\in\mathcal{S}_n\}\big)\in\mathbb{R},\qquad t\in[T],\qquad n \in [N]. \label{eq:encoder}
\end{equation}
The encoded symbols of server $n$ are $\mathbf{z}_n(x)=(z_{n,1}(x),\dots,z_{n,T}(x))$. Encoders $\phi_{n,t}^{(\mathcal S_n)}$ may use internal randomness independent of $(\mathsf A,\mathsf L, \mathsf X)$.

In the communication phase, each symbol $z_{n,t}(x)$ is made available to a subset of users through error-free links $\mathcal{T}_{n,t}\subseteq[K]$.
We assume shot-agnostic links for clarity, i.e., $\mathcal{T}_{n,t}=\mathcal{T}_n$ for all $t\in[T]$. The aggregate user set served by server $n$ obeys
\begin{equation}
\mathcal{T}_n \subseteq [K],\qquad |\mathcal{T}_n| \le \Delta.
\label{eq:comm-budget-}
\end{equation}
with normalized communication cost
\[
\delta \triangleq \frac{\Delta}{K},
\]
the per-server fan-out fraction (fraction of users a server can reach). We assume $1\le \Delta\le K$, hence $0<\delta\le 1$. User $k$ receives the symbols
\begin{equation}
\mathcal{R}_k(x) = \{\,z_{n,t}(x):\ k\in\mathcal{T}_{n}\text{ and } t\in[T]\,\}, \label{eq:recv}
\end{equation}
and applies a (possibly nonlinear) decoder to get an estimate of its desired function.
\begin{equation}
\widehat{F}_k(x) = \psi_k\big(\mathcal{R}_k(x)\big)\in\mathbb{R}^{m_k}. \label{eq:decoder}
\end{equation}

\subsection{Problem Formulation}
In our annealed analysis, to expose typical-case trade-offs cleanly—without committing to a specific topology—we model assignment and links as uniformly random:
\begin{enumerate}
\item \emph{Random computation assignment:}
\[
\mathcal{S}_n \sim \mathrm{Unif}\!\big(\{S\subseteq[L]:\, |S|=\Gamma\}\big)\qquad\text{i.i.d.\ over }n.
\]
\item \emph{Random user links:}
\[
\mathcal{T}_n \sim \mathrm{Unif}\!\big(\{U\subseteq[K]:\, |U|=\Delta\}\big)\qquad\text{i.d.d.\ over }n.
\]
\end{enumerate}
Throughout, $(\mathsf{A},\mathsf{L})$ are drawn independently of the data $X\sim\mathbb{P}_X$ and of any encoder randomness.
This random graph model serves as an average-case benchmark, isolating the impact of $(\Gamma,\Delta,T)$ on performance. Note that under shot-agnostic links,
\begin{align}
\mrecv_k(\mathcal{T}) \triangleq |\mathcal{R}_k| \;=\; T\,\big|\{\,n:\ k\in\mathcal{T}_n\,\}\big|, \label{def:m_k(T)}
\end{align}
where $\mathbb{E}[\mrecv_k(\mathcal{T})]=T\,\frac{N\Delta}{K}=T\,N\delta$, which makes $N\delta$ the expected number of servers linked to any given user\footnote{Standard concentration (e.g., Hoeffding/Chernoff) implies $\mrecv_k(\mathcal{T})$ concentrates around $T\,N\delta$ for each user $k$ when $N$ is moderate-to-large.}.
Under the above random assignment and links (not design variables), a design only specifies the encoders and decoders as follows:
\begin{itemize}[leftmargin=1.2em]
\item \emph{Server encoders:} for each $n$ and $t\in[T]$ choose a measurable
\[
\phi_{n,t}^{(\mathcal{S}_n)}:\ \mathbb{R}^{\Gamma}\to\mathbb{R},\qquad
z_{n,t}(x)=\phi_{n,t}^{(\mathcal{S}_n)}\!\big(\mathbf{w}_{\mathcal{S}_n}(x)\big),
\]
and collect $\mathbf{z}_n(x)=(z_{n,1}(x),\ldots,z_{n,T}(x))\in\mathbb{R}^{T}$.\footnote{Equivalently, view a single map $\phi_{n,t}:\mathbb{R}^L\times 2^{[L]}\to\mathbb{R}$ that depends only on coordinates in $\mathcal{S}_n$.}
\item \emph{User decoders:} each user $k$ applies
\[
\psi_k:\ \mathbb{R}^{\mrecv_k(\mathcal{T})}\to\mathbb{R}^{m_k},\qquad
\widehat{F}_k(x)=\psi_k\!\big(\mathcal{R}_k(x)\big).
\]
\end{itemize}
We can see that here the encoding and decoding procedure formulation differs from \cite{khalesi2024tessellated} as it is not restricted to the set of linear functions. Now we formulate the risk (distortion) criteria.

\subsection{Risk Criteria (Quenched vs.\ Annealed)}
Let $\mathbb{P}_X$ be the input distribution on $\mathcal{X}$, and let
$\mathsf{A}=(\mathcal{S}_1,\ldots,\mathcal{S}_N)$,
$\mathsf{L}=(\mathcal{T}_1,\ldots,\mathcal{T}_N)$
denote the realized assignment and links.

\paragraph{Quenched (conditional) population risk}
For fixed $(\mathsf{A},\mathsf{L})$ and a design
$\mathcal{D}=\big(\{\phi_{n,t}^{(\cdot)}\}_{n=1,t=1}^{N,T},\,\{\psi_k\}_{k=1}^K\big)$, define
\begin{equation}
\label{eq:quenched-risk}
\mathcal{R}\big(\mathcal{D}\mid \mathsf{A},\mathsf{L}\big)
=\mathbb{E}_{x\sim\mathbb{P}_X}\!\left[
\frac{1}{K}\sum_{k=1}^K
\big\|\psi_k(\mathcal{R}_k(x)) - h_k(\mathbf{w}(x))\big\|_2^2
\right].
\end{equation}
This evaluates performance for the realized assignment/link graph.

\paragraph{Annealed (average-over-graph) population risk}
Averaging over the assignment and link randomness,
\begin{equation}
\label{eq:annealed-risk3}
\overline{\mathcal{R}}(\mathcal{D})
\triangleq
\mathbb{E}_{\mathsf{A},\mathsf{L}}\!\left[\mathcal{R}\big(\mathcal{D}\mid \mathsf{A},\mathsf{L}\big)\right],
\end{equation}
which captures expected performance over typical random realizations. In what follows, we precisely formulate the optimization problems that interest us:

\subsection{Optimization Problems}

\paragraph*{P1 (Quenched design)}
Given $(\mathsf{A},\mathsf{L})$, find encoders/decoders that minimize \eqref{eq:quenched-risk}:
\begin{equation}
\label{eq:P1}
\begin{aligned}
\inf_{\mathcal{D}}\quad &
\mathcal{R}\big(\mathcal{D}\mid \mathsf{A},\mathsf{L}\big)\\
\text{s.t.}\quad &
\phi_{n,t}^{(\mathcal{S}_n)}:\mathbb{R}^{\Gamma}\!\to\mathbb{R}\ \text{measurable},\quad
\psi_k:\mathbb{R}^{\mrecv_k(\mathcal{T})}\!\to\mathbb{R}^{m_k}\ \text{measurable}.\\
\end{aligned}
\end{equation}

\paragraph*{P2 (Annealed design)}
The goal is to minimize the average quenched risk:
\begin{equation}
\label{eq:P1b}
\begin{aligned}
\inf_{\mathcal{D}}\quad &
\overline{\mathcal{R}}(\mathcal{D})\\
\text{s.t.}\quad
& \phi_{n,t}^{(\mathcal{S}_n)}:\mathbb{R}^{\Gamma}\!\to\mathbb{R}\ \text{measurable},\quad
\psi_k:\mathbb{R}^{\mrecv_k(\mathcal{T})}\!\to\mathbb{R}^{m_k}\ \text{measurable},
\end{aligned}
\end{equation}
In other words, P2 selects encoder/decoder families {without conditioning on a specific realization} of $(\mathsf{A},\mathsf{L})$, aiming to minimize the expected (annealed) population risk over the randomness of assignments and links. Designs may depend on model parameters (e.g., $\Gamma,\Delta,T$) and on $\mathbb{P}_X$, but not on the realized $(\mathsf{A},\mathsf{L})$.

\section{Results}\label{Results}
In this section, we first present the encoding and decoding setups in Subsection~\ref{setup}, and then state and explain the main theorems in Subsection~\ref{main-results}. Finally, in Subsection~\ref{proof-sketch}, we provide proof sketches of the main results, including the key achievability and converse arguments as well as the resulting design takeaways. In addition, we present a worked quenched example illustrating Theorem~\ref{thm:P1}; the corresponding annealed example for Theorem~\ref{thm:P2} is given in Appendix~\ref{subsec:worked_example_thm2}.

\subsection{Setup and Assumptions}\label{setup}

We instantiate the encoders in \eqref{eq:encoder} by masked random Fourier features (RFF) and the user decoders by ridge regression, and we benchmark against converse bounds combining an RKHS eigen-tail lower bound with a coverage floor. Let $K(\cdot,\cdot)$ be a shift-invariant kernel on $\R^{L}$ with spectral measure $\mu$ (e.g., Gaussian RBF). Let $\mathcal{H}_K$ denote the RKHS induced by $K(\cdot,\cdot)$, and assume each target $F_k$ belongs to $\mathcal{H}_K$ with $\|F_k\|_{\mathcal{H}_K}\le B$ \cite{ScholkopfSmola2002}. Training samples $\{x^{(i)}\}_{i=1}^M, M \in \mathbb{N}$ are i.i.d.\ from $\mathbb{P}_X$; labels may be noisy with sub-Gaussian noise of proxy variance $\sigma^2$. We write $\{\lambda_j\}_{j\ge 1}$ for the eigenvalues of the kernel integral operator (with respect to the pushforward of $\mathbb{P}_X$ through $\mathbf{w}$), ordered nonincreasingly \cite{BerlinetThomasAgnan2004}. For each server $n$ with assignment $\mathcal{S}_n$ (of size $\Gamma$) and each shot $t\in[T]$, draw independently
\[
\omega_{n,t}\sim \mu,\qquad b_{n,t}\sim \mathrm{Unif}[0,2\pi),
\]
form the masked frequency $\widetilde{\omega}_{n,t}=\omega_{n,t}\odot \mathbf{1}_{\mathcal{S}_n}, \mathbf{1}_{\mathcal{S}_n} \in \{0,1\}^L$, and emit
\begin{equation}
z_{n,t}(x)
=\sqrt{\tfrac{2}{\gamma}}\,
\cos\!\big(\widetilde{\omega}_{n,t}^{\top}\mathbf{w}(x)+b_{n,t}\big),\qquad
\gamma\triangleq \Gamma/L.
\label{eq:masked-rff}
\end{equation}
This scaling ensures (under standard masked-Bochner sampling) that $\E[z_{n,t}(x)z_{n,t}(x')]=K(\mathbf{w}(x),\mathbf{w}(x'))$ in expectation over $(\omega,b)$; more generally, it yields unbiasedness up to a masking constant absorbed by the variance proxy (cf. Lemma~\ref{lem:mask-variance-fact})\footnote{
Formally, each encoded feature can be written as 
$z_{n,t}(x)=\alpha\,M_{n,t}\,\varphi_{\omega,b}(\mathbf{w}(x))$, 
where $M_{n,t}\!\in\!\{0,1\}^{m_k}$ is a binary mask selecting the subset of features 
computed by server~$n$ at shot~$t$. 
The mask satisfies $\|M_{n,t}\|_0 = \Gamma m_k$, 
so that only a fraction $\gamma=\Gamma$ of features is evaluated per server. 
Under {masked Bochner sampling} for a shift-invariant kernel 
$K(u,u')=k(u-u')$ with 
$\varphi_{\omega,b}(u)=\sqrt{2}\cos(\omega^\top u+b)$, 
$b\!\sim\!\mathrm{Unif}[0,2\pi]$, and $\omega$ drawn from the spectral measure of~$k$, 
we have $\E_{\omega,b}[\,\varphi_{\omega,b}(u)\varphi_{\omega,b}(u')\,]=K(u,u')$. 
The additional scaling $\sqrt{2/\gamma}$ compensates for the masking density, 
preserving the kernel expectation up to a constant factor 
$C_{\text{mask}}\!\ge\!1$ that controls the induced variance.
}.

Each user $k$ concatenates the symbols it receives, yielding $\mrecv_k(\mathcal{T})$ features as in \eqref{eq:recv}. Given the received feature vector $\mathbf{z}_k(x)\in\R^{\mrecv_k(\mathcal{T})}$, user $k$ implements
\begin{equation}
\widehat{F}_k(x)=\bm{\beta}_k^\top \mathbf{z}_k(x),\qquad
\bm{\beta}_k=\arg\min_{\beta\in\R^{\mrecv_k(\mathcal{T})}}\ \frac{1}{M}\sum_{i=1}^M\big(\beta^\top\mathbf{z}_k(x^{(i)})-F_k(x^{(i)})\big)^2+\lambda\|\beta\|_2^2.
\label{eq:ridge}
\end{equation}
Let $d_{\lambda,k}$ denote the effective dimension of ridge at regularization $\lambda$ (nonincreasing in $\lambda$, bounded by $\mrecv_k(\mathcal{T})$).\footnote{Let $\Phi_k(x) = \big(z_{n,t}(x)\big)_{(n,t)\,:\,k\in\mathcal{T}_{n,t}}$ and $G_k=\mathbb{E}[\Phi_k(x)\Phi_k(x)^\top]\in\mathbb{R}^{\mrecv_k(\mathcal{T})\times \mrecv_k(\mathcal{T})}$ be the feature covariance for user $k$, with eigenvalues $\{\mu_{j,k}\}$. The effective dimension at ridge parameter $\lambda>0$ is
$d_{\lambda,k} \triangleq \mathrm{tr}\!\big(G_k(G_k+\lambda I)^{-1}\big)=\sum_{j}\frac{\mu_{j,k}}{\mu_{j;k}+\lambda}$.
Each summand decreases with $\lambda$, hence $d_{\lambda,k}$ is nonincreasing in $\lambda$ and satisfies $0\le d_{\lambda,k}\le \mathrm{rank}(G_k)\le \mrecv_k(\mathcal{T})$.}

For user $k\in[K]$, an essential coordinate set $\mathcal{S}_k^\star\subseteq[L]$ for the target $F_k$ is any set with the property that for every $\ell\in\mathcal{S}_k^\star$ there exist inputs $x,x'\in\mathcal{X}$ that agree on all subfunction outputs except possibly coordinate $\ell$ (i.e., $\mathbf{w}(x)_{-\ell}=\mathbf{w}(x')_{-\ell}$) and
$\|F_k(x)-F_k(x')\|_2>0$.
We define
\[
r_k \;\triangleq\; \min\big\{\,|\mathcal{S}|:\ \mathcal{S}\subseteq[L]\ \text{is essential for }F_k\,\big\},
\]
with the convention $r_k=+\infty$ if no finite essential set exists.
Intuitively, $r_k$ is the smallest number of subfunction coordinates on which $F_k$ essentially depends. For the identity task $F_k(x)=\mathbf{w}(x)$, $r_k=L$; for a sparse linear task
$F_k(x)=\sum_{\ell\in S} a_\ell\,w_\ell(x)$ with all $a_\ell\neq0$, $r_k=|S|$. Here, $m_{\mathrm{harm}}$ denotes the harmonic mean of the per-user feature dimensions $\{m_k\}_{k=1}^K$, i.e.
\[
m_{\mathrm{harm}} = \Big(\frac{1}{K}\sum_{k=1}^K \frac{1}{m_k}\Big)^{-1}.
\]
In the random assignment regime, $m_{\mathrm{harm}}$ is asymptotically equivalent to the arithmetic mean $m_{\mathrm{avg}}$. We assume also $\gamma\ge c_0>0$, i.e., each server computes at least a constant fraction $c_0$ of subfunctions. 
Otherwise the bound becomes vacuous as $\gamma\to0$.

\subsection{Main Results}\label{main-results}
Theorem \ref{thm:P1} describes the quenched achievability and converse for the problem P1 in \eqref{eq:P1}:
\begin{theorem}\label{thm:P1}
Fix a realization $(\mathsf{A},\mathsf{L})$ and assume 
\eqref{eq:masked-rff}--\eqref{eq:ridge}. 
There exist absolute constants $C_1,C_2,C_3>0$ such that, 
for any $\delta'\in(0,1)$ and suitable $\lambda=\lambda(M,\mrecv_{1}(\mathcal{T}),\ldots,\mrecv_{K}(\mathcal{T}))$, 
the following holds with probability at least $1-\delta'$ over the encoder 
randomness $\{(\omega_{n,t},b_{n,t})\}$:
\begin{align}
\mathcal{R}\big(\mathcal{D}\mid \mathsf{A},\mathsf{L}\big)
&\le \left(\frac{2}{\gamma}+C_1\right)\,
   \frac{B^2}{\gamma\,m_{\mathrm{harm}}}\,
   \log\!\frac{2}{\delta'}
   \;+\;
   C_2\,\frac{\sigma^2\,d_\lambda}{M}
   \;+\;
   C_3\,B^2\,\lambda,
\label{eq:P1-ach}
\end{align}
where 
$m_{\mathrm{harm}} \;\triangleq\; 
\Big(\tfrac{1}{K}\sum_{k=1}^K \tfrac{1}{\mrecv_k(\mathcal{T})}\Big)^{-1}$
is the harmonic mean of the received-feature budgets, and
$d_{\lambda} = 1/K \sum^{K}_{k=1}d_{\lambda,k}$, where $d_{\lambda,k} \;=\; \mathrm{tr}\!\big(G_k(G_k+\lambda I)^{-1}\big)$ is the effective dimension for user $k$. Moreover, for any encoder/decoder design that delivers $\mrecv_k(\mathcal{T})$ 
real scalars to user $k$, the following 
quenched converse bound holds for the average risk:
\begin{align}
\inf_{\mathcal{D}}
\mathcal{R}\big(\mathcal{D}\mid \mathsf{A},\mathsf{L}\big)
&\;\ge\;
\frac{B^2}{K}\sum_{k=1}^K \ \Big(\sum_{j> \mrecv_k(\mathcal{T})}\! \lambda_j
\ \vee\
\varepsilon_{\mathrm{cov},k}(\mathsf{A},\mathsf{L})\Big),
\label{eq:P1-conv}
\end{align}
where $(\lambda_j)_{j\ge1}$ are the eigenvalues of the kernel integral operator 
(in nonincreasing order), and $\varepsilon_{\mathrm{cov},k}(\mathsf{A},\mathsf{L})>0$ 
whenever, for the realized $(\mathsf{A},\mathsf{L})$, the set of server assignments 
fails to cover the essential coordinate set of user $k$’s target.
\end{theorem}

\begin{remark}
The normalized computation cost $\gamma=\Gamma/L$ captures the masking penalty (each feature “sees” a fraction $\gamma$ of coordinates of $\mathbf{w}(x)$). The achievability bound \eqref{eq:P1-ach} improves with more received features $\mrecv_k(\mathcal{T})=T|\{n:\,k\in\mathcal{T}_n\}|$, larger compute fraction $\gamma$, more samples $M$, and proper $\lambda$. The converse \eqref{eq:P1-conv} matches the best $\mrecv_k$-rank kernel approximation and enforces a nonzero floor if coverage fails in the realized graph. Note that in the quenched setting the normalized communication budget $\delta$ does not appear explicitly; its effect is entirely mediated through the realized feature count $\mrecv_k(\mathcal{T})$ and the coverage event. By contrast, in the annealed regime $\delta$ re-emerges in the expectation $\E[\mrecv_k(\mathcal{T})]=T N \delta$ and in the exponential coverage term.
\end{remark}

The next result gives annealed upper and lower bounds for P2 in \eqref{eq:P1b}.
\begin{theorem}
\label{thm:P2}
Under the assumptions of Theorem~\ref{thm:P1}, and after averaging over the random computation assignments $\mathsf{A}$ and link patterns $\mathsf{L}$, the optimal annealed population risk satisfies
\begin{align}
B^2 \sum_{j>m_k^{\mathrm{avg}}}\!\lambda_j
\;\vee\;
c'\,\mathbf{1}\!\left\{\gamma\,N\,\delta \,\lesssim\, \log(r_{\mathrm{avg}})\right\}
\ \le\
\inf_{\mathcal{D}}\overline{\mathcal{R}}(\mathcal{D})
\ \le\
\mathsf{U}_{\mathrm{ann}},
\end{align}
with
\begin{equation}
\label{eq:ann-upper-degree-corr}
\mathsf{U}_{\mathrm{ann}}
\ \le\
\left(\frac{2}{\gamma}+C_1\right)\frac{B^2}{\gamma\,T\,(1-\varepsilon)\,N\,\delta}
\;+\;
C_2\,\frac{\sigma^2\,d_\lambda}{M}
\;+\;
C_3\,B^2\,\lambda
\;+\;
r_{\mathrm{avg}}\,\mathrm{e}^{-\gamma\,N\,\delta}
\;+\;
\exp\!\big(-c\,\varepsilon^2 N\delta\big)\,C_\star,
\end{equation}
where $m_k^{\mathrm{avg}}\triangleq T N \delta$, and $r_{\mathrm{avg}}\triangleq \frac{1}{K}\sum_{k=1}^K r_k$, and $\epsilon \in (0,1)$ is a small positive deviation parameter. Here, $\gamma=\Gamma/L$ and $\delta=\Delta/K$ are the normalized computation and communication costs; $\{\lambda_j\}_{j\ge1}$ are the nonincreasing eigenvalues of the kernel integral operator associated with $K$ under the law of $\mathbf{w}(X)$; $r_k$ is the size of an essential coordinate set for $F_k$; and $C_1,C_2,C_3,c,c',C_\star>0$ are absolute constants independent of $(N,K,L,\Gamma,\Delta,T,M)$.
\end{theorem}

\begin{remark}
In the annealed regime, $\mrecv_k(\mathcal{T})$ concentrates around $m_k^{\mathrm{avg}}=T N\delta$, so the achievability bound behaves like $(2/\gamma +C_1)B^2/(\gamma\,m_k^{\mathrm{avg}})$ up to logarithmic factors, while the eigen-tail converse uses the same feature budget. The small additive term $r_{\mathrm{avg}}\,\mathrm{e}^{-c\,\gamma\,N\delta}$ accounts for the (averaged) exponentially small chance of a coverage miss. Hence, once $\gamma\,N\delta$ exceeds the (logarithmic) coverage threshold in $\log r_{\mathrm{avg}}$, the scheme is order-optimal up to constants.
\end{remark}
\begin{remark}
Theorem~\ref{thm:P2} is tight in the model-relevant sense: both the converse and achievability are governed by the same effective communicated-feature scale \(m_k^{\mathrm{avg}} \asymp TN\delta\) and by the same kernel spectral obstruction beyond that scale. More precisely, once \(\gamma N\delta \gtrsim \log r_{\mathrm{avg}}\), the coverage term becomes negligible, so the upper and lower bounds differ only by absolute constants and logarithmic factors, together with the standard ridge bias--variance terms. Hence Theorem~\ref{thm:P2} identifies the correct annealed computation--communication--accuracy law under the GMUDC assumptions.
\end{remark}
\subsection{Proof Sketches of Theorems~\ref{thm:P1} and~\ref{thm:P2}}\label{proof-sketch}

The proof of Theorem~\ref{thm:P1} (quenched bounds) and Theorem~\ref{thm:P2} (annealed bounds) follows from two complementary arguments: an achievability construction based on masked random-feature encoders and a converse argument based on spectral and coverage limitations. Both parts rely on the spectral structure of the user-dependent Gram matrices and the effective feature dimensions induced by the computation and communication budgets.

For Theorem~\ref{thm:P1}, the achievability argument fixes the topology $(\mathsf{A},\mathsf{L})$ and shows that the proposed masked random-feature encoder, combined with the ridge decoder, attains the upper bound in~\eqref{eq:P1-ach}. Each server uses a mask of size $\Gamma$ and emits features $z_{n,t}(x)=\sqrt{2/\gamma}\cos((\omega_{n,t}\odot 1_{S_n})^\top w(x)+b_{n,t})$ with $\gamma=\Gamma/L$. This yields an unbiased Monte Carlo approximation of the target kernel for each user once the $m_k(\mathcal{T})=T|\{n:k\in\mathcal{T}_n\}|$ received scalars are concatenated. A Bernstein-style concentration argument with masking variance control gives a kernel approximation error of order $(2/\gamma^2 m_k(\mathcal{T})+C'/(\gamma m_k(\mathcal{T})))\log(2/\delta')$. Substituting this into the standard kernel ridge decomposition yields the prediction risk bound $\mathcal{R}(\mathcal{D}\mid \mathsf{A},\mathsf{L})\lesssim (C_1+1/\gamma)B^2/(\gamma m_k(\mathcal{T}))+\sigma^2 d_\lambda/M+B^2\lambda$, where $d_{\lambda} = 1/K \sum^{K}_{k=1}d_{\lambda,k} ,d_{\lambda,k}=\mathrm{tr}(G_k(G_k+\lambda I)^{-1})$ is the ridge effective dimension. Averaging over users introduces the harmonic mean $m_{\mathrm{harm}}$ and gives the quenched achievability inequality~\eqref{eq:P1-ach}.

The converse of Theorem~\ref{thm:P1} combines a spectral tail limitation with a coverage floor. Any scheme that communicates at most $m_k(\mathcal{T})$ scalars per user confines the reconstruction to an $m_k(\mathcal{T})$-dimensional subspace of $L_2(\mathbb{P}_X)$. By the theory of Kolmogorov $m$-widths for RKHS balls, the minimum achievable risk equals the energy of the discarded spectral tail $B^2\sum_{j>m_k(T)}\lambda_j$. If an essential coordinate of user $k$ is never computed and transmitted in the realized topology, then inputs differing only on that coordinate are indistinguishable, producing a coverage floor $\varepsilon_{\mathrm{cov},k}(\mathsf{A},\mathsf{L})>0$. Combining both limitations yields the quenched converse bound~\eqref{eq:P1-conv}.

Theorem~\ref{thm:P2} concerns the annealed regime, obtained by averaging over the random assignment and link topologies. For the achievability part, the number of features received by each user satisfies $|\{n:k\in\mathcal{T}_n\}|\sim\mathrm{Binomial}(N,\delta)$, so $m_k(\mathcal{T})$ concentrates around $m_k^{\mathrm{avg}}=TN\delta$. Substituting $m_k^{\mathrm{avg}}$ into the quenched bound and accounting for the exponentially small deviation probability $\exp(-c\varepsilon^2 N\delta)$ gives the annealed upper bound $\mathsf{U}_{\mathrm{ann}}$ in~\eqref{eq:ann-upper-degree-corr}. The additional coverage term $r_{\mathrm{avg}}\exp(-\gamma N\delta)$ captures the small probability of missing an essential coordinate across users. Collecting all constants and averaging over $(\mathsf{A},\mathsf{L})$ yields the complete achievability part of Theorem~\ref{thm:P2}.

The converse in the annealed regime follows by averaging the spectral tail bound over $(\mathsf{A},\mathsf{L})$. Taking the expectation and using the convexity of $m\mapsto\sum_{j>m}\lambda_j$ gives $\inf_{\mathcal{D}}\overline{\mathcal{R}}(\mathcal{D})\ge B^2\sum_{j>m_k^{\mathrm{avg}}}\lambda_j$, with $m_k^{\mathrm{avg}}=TN\delta$. When $\gamma N\delta$ is below the logarithmic threshold $\log(r_{\mathrm{avg}})$, coverage failures occur with non-negligible probability and impose a constant risk floor, represented by the term $c'\mathbf{1}\{\gamma N\delta\lesssim\log(r_{\mathrm{avg}})\}$. Combining these elements establishes the annealed lower bound and completes the proof of Theorem~\ref{thm:P2}.

\subsection{Worked Example for Theorem~\ref{thm:P1}: A Fixed Quenched Gaussian-RKHS GMUDC Instance}
\label{subsec:worked_example_thm1}

We now present a concrete Gaussian-RKHS instance of the quenched design problem \textbf{P1} in \eqref{eq:P1}. The purpose of this example is to make the quantities in Theorem~\ref{thm:P1} explicit in a fully specified GMUDC setting. In particular, the input law, the subfunctions, the user target functions, the realized assignment/link topology $(\mathsf A,\mathsf L)$, and the encoder/decoder construction are all fixed explicitly. We consider a GMUDC system with $K=6$ users, $N=8$ servers, $L=12$ subfunctions, computation budget $\Gamma=4$, communication budget $\Delta=3$, and $T$ communication shots. Hence $\gamma=\Gamma/L=1/3$ and $\delta=\Delta/K=1/2$. Throughout this example, we take scalar-valued user outputs, i.e., $m_k=1$ for all $k\in[K]$, and we use the Gaussian kernel
\begin{equation}
K_{\rho}(u,v)=\exp\!\left(-\frac{\|u-v\|_2^2}{2\rho^2}\right),\qquad \rho=1.1.
\label{eq:worked_gaussian_kernel}
\end{equation}

\subsubsection{Input law, subfunctions, and user targets}

Let $X=(X_1,\dots,X_5)\sim \mathcal N(0,I_5)$. For each input $x\in\mathbb R^5$, define latent mixed coordinates $u_\ell(x)=\sum_{j=1}^{5} B_{\ell j}x_j$ for $\ell\in[L]$, where $B_{\ell j}=0.55\cos((\ell+2j)/3)+0.35\sin((2\ell-j)/4)$. The subfunctions are then
\begin{align}
f_\ell(x)
&=
0.45\tanh\!\bigl(u_\ell(x)\bigr)
+0.20\sin\!\bigl(1.7u_\ell(x)\bigr)
\notag\\
&\quad
+0.15\cos\!\bigl(u_{\ell+1}(x)\bigr)
+0.10\tanh\!\bigl(u_\ell(x)u_{\ell+2}(x)\bigr)
+0.10e^{-|u_{\ell+3}(x)|},
\qquad \ell\in[L],
\label{eq:worked_subfunctions}
\end{align}
with indices interpreted cyclically modulo $L$. As in Section~\ref{System-Model}, we write $w_\ell=f_\ell(x)$ and $\mathbf w(x)=(w_1,\dots,w_L)^\top\in\mathbb R^L$.

For each user $k\in[K]$, we define the target function by $F_k(x)=h_k(\mathbf w(x))$, where $h_k\in\mathcal H_{K_\rho}$ is chosen as the dense Gaussian-RKHS expansion
\begin{equation}
h_k(\mathbf w)=\sum_{q=1}^{6} a_{kq}\exp\!\left(-\frac{\|\mathbf w-c_{kq}\|_2^2}{2\rho^2}\right).
\label{eq:worked_dense_target}
\end{equation}
The coefficients are $a_{kq}=(-1)^{q+1}(0.40+0.08q+0.03k)$, and the centers $c_{kq}\in\mathbb R^{12}$ are defined componentwise by
\begin{equation}
(c_{kq})_\ell
=
0.60\sin\!\bigl(0.35(k+q)\ell\bigr)
+
0.25\cos\!\bigl(0.50(k+2q)\ell\bigr),
\qquad \ell\in[L].
\label{eq:worked_centers}
\end{equation}
Since each $h_k$ depends on all coordinates of $\mathbf w$, the essential coordinate set of every user is $\mathcal S_k^\star=[L]$, and hence $r_k=L=12$ for all $k\in[K]$.

\subsubsection{Fixed quenched realization of $(\mathsf A,\mathsf L)$}

We now fix a specific realization of the assignment and communication topology
$\mathsf A=(\mathcal S_1,\dots,\mathcal S_8)$ and
$\mathsf L=(\mathcal T_1,\dots,\mathcal T_8)$, with
$\mathcal S_1=\{1,2,3,4\}$, $\mathcal S_2=\{3,4,5,6\}$, $\mathcal S_3=\{5,6,7,8\}$, $\mathcal S_4=\{7,8,9,10\}$,
$\mathcal S_5=\{9,10,11,12\}$, $\mathcal S_6=\{11,12,1,2\}$, $\mathcal S_7=\{2,5,8,11\}$, $\mathcal S_8=\{1,4,7,10\}$,
and
$\mathcal T_1=\{1,2,3\}$, $\mathcal T_2=\{4,5,6\}$, $\mathcal T_3=\{1,2,3\}$, $\mathcal T_4=\{1,4,5\}$,
$\mathcal T_5=\{2,3,6\}$, $\mathcal T_6=\{1,4,5\}$, $\mathcal T_7=\{2,4,6\}$, and $\mathcal T_8=\{3,5,6\}$.
Thus each server computes exactly $\Gamma=4$ subfunctions and broadcasts to exactly $\Delta=3$ users. The corresponding fixed quenched topology is shown in Fig.~\ref{fig:worked_example_topology_clean}. The figure makes the realized objects $\mathcal S_n$ and $\mathcal T_n$ fully explicit and shows how the concrete communication pattern induces the received-feature budget of each user. Under this realized topology, every user is connected to exactly four servers, so by \eqref{def:m_k(T)},
\[
\mrecv_k(\mathcal T)=T\,|\{n:\,k\in\mathcal T_n\}|=4T,\qquad k\in[K].
\]
Therefore the harmonic mean in Theorem~\ref{thm:P1} is simply $m_{\mathrm{harm}}=4T$.

\begin{figure*}[t]
\centering
\resizebox{0.75\textwidth}{!}{%
\begin{tikzpicture}[x=0.75pt,y=0.75pt,yscale=-1,xscale=1]

\def\SrvW{60}
\def\SrvH{116}
\def\CoordW{70}
\def\CoordH{116}
\def\Marg{2}
\def\Gap{4}
\def\RowH{24}

\def\drawserver#1#2#3#4#5#6#7#8{%
    \draw [fill={rgb,255:red,255; green,255; blue,255}, fill opacity=1]
        ({#2},{#3}) rectangle ({#2+\SrvW},{#3+\SrvH});

    \draw [fill={rgb,255:red,234; green,229; blue,18}, fill opacity=0.27]
        ({#2+\Marg},{#3+\Marg}) rectangle ({#2+\SrvW-\Marg},{#3+\Marg+\RowH});
    \draw [fill={rgb,255:red,234; green,229; blue,18}, fill opacity=0.27]
        ({#2+\Marg},{#3+\Marg+\RowH+\Gap}) rectangle ({#2+\SrvW-\Marg},{#3+\Marg+2*\RowH+\Gap});
    \draw [fill={rgb,255:red,234; green,229; blue,18}, fill opacity=0.27]
        ({#2+\Marg},{#3+\Marg+2*\RowH+2*\Gap}) rectangle ({#2+\SrvW-\Marg},{#3+\Marg+3*\RowH+2*\Gap});
    \draw [fill={rgb,255:red,234; green,229; blue,18}, fill opacity=0.27]
        ({#2+\Marg},{#3+\Marg+3*\RowH+3*\Gap}) rectangle ({#2+\SrvW-\Marg},{#3+\Marg+4*\RowH+3*\Gap});

    \draw ({#2+0.5*\SrvW},{#3-18}) node {$#4$};

    \draw ({#2+0.5*\SrvW},{#3+\Marg+0.5*\RowH}) node {$#5$};
    \draw ({#2+0.5*\SrvW},{#3+\Marg+1.5*\RowH+\Gap}) node {$#6$};
    \draw ({#2+0.5*\SrvW},{#3+\Marg+2.5*\RowH+2*\Gap}) node {$#7$};
    \draw ({#2+0.5*\SrvW},{#3+\Marg+3.5*\RowH+3*\Gap}) node {$#8$};

    \coordinate (#1T) at ({#2+0.5*\SrvW},{#3});
    \coordinate (#1L) at ({#2+12},{#3+\SrvH});
    \coordinate (#1M) at ({#2+0.5*\SrvW},{#3+\SrvH});
    \coordinate (#1R) at ({#2+\SrvW-12},{#3+\SrvH});
}

\def\drawcoord#1#2{%
    \draw [fill={rgb,255:red,255; green,255; blue,255}, fill opacity=1]
        ({#1},{#2}) rectangle ({#1+\CoordW},{#2+\CoordH});

    \draw [fill={rgb,255:red,234; green,229; blue,18}, fill opacity=0.27]
        ({#1+\Marg},{#2+\Marg}) rectangle ({#1+\CoordW-\Marg},{#2+\Marg+\RowH});
    \draw [fill={rgb,255:red,234; green,229; blue,18}, fill opacity=0.27]
        ({#1+\Marg},{#2+\Marg+\RowH+\Gap}) rectangle ({#1+\CoordW-\Marg},{#2+\Marg+2*\RowH+\Gap});
    \draw [fill={rgb,255:red,234; green,229; blue,18}, fill opacity=0.27]
        ({#1+\Marg},{#2+\Marg+2*\RowH+2*\Gap}) rectangle ({#1+\CoordW-\Marg},{#2+\Marg+3*\RowH+2*\Gap});
    \draw [fill={rgb,255:red,234; green,229; blue,18}, fill opacity=0.27]
        ({#1+\Marg},{#2+\Marg+3*\RowH+3*\Gap}) rectangle ({#1+\CoordW-\Marg},{#2+\Marg+4*\RowH+3*\Gap});

    \draw ({#1+0.5*\CoordW},{#2+\Marg+0.5*\RowH}) node {$f_{1}(\cdot)$};
    \draw ({#1+0.5*\CoordW},{#2+\Marg+1.5*\RowH+\Gap}) node {$f_{2}(\cdot)$};
    \draw ({#1+0.5*\CoordW},{#2+\Marg+2.5*\RowH+2*\Gap}) node {{\LARGE \textbf{$\vdots$}}};
    \draw ({#1+0.5*\CoordW},{#2+\Marg+3.5*\RowH+3*\Gap}) node {$f_{12}(\cdot)$};

    \coordinate (c1) at ({#1+5},{#2+\CoordH});
    \coordinate (c2) at ({#1+13},{#2+\CoordH});
    \coordinate (c3) at ({#1+21},{#2+\CoordH});
    \coordinate (c4) at ({#1+29},{#2+\CoordH});
    \coordinate (c5) at ({#1+37},{#2+\CoordH});
    \coordinate (c6) at ({#1+45},{#2+\CoordH});
    \coordinate (c7) at ({#1+53},{#2+\CoordH});
    \coordinate (c8) at ({#1+61},{#2+\CoordH});
}

\def\drawuser#1#2#3{%
    \draw [color={rgb,255:red,74; green,144; blue,226}, draw opacity=1,
           fill={rgb,255:red,74; green,144; blue,226}, fill opacity=1]
           ({#2},{#3}) ellipse (14 and 14);
    \draw ({#2},{#3+1}) node [text=white] {$#1$};

    \coordinate (u#1a) at ({#2-10},{#3-13});
    \coordinate (u#1b) at ({#2-3},{#3-15});
    \coordinate (u#1c) at ({#2+3},{#3-15});
    \coordinate (u#1d) at ({#2+10},{#3-13});
}

\drawcoord{302}{20}

\drawserver{s1}{50}{190}{\mathcal S_{1}}{f_{1}(\cdot)}{f_{2}(\cdot)}{f_{3}(\cdot)}{f_{4}(\cdot)}
\drawserver{s2}{125}{190}{\mathcal S_{2}}{f_{3}(\cdot)}{f_{4}(\cdot)}{f_{5}(\cdot)}{f_{6}(\cdot)}
\drawserver{s3}{200}{190}{\mathcal S_{3}}{f_{5}(\cdot)}{f_{6}(\cdot)}{f_{7}(\cdot)}{f_{8}(\cdot)}
\drawserver{s4}{275}{190}{\mathcal S_{4}}{f_{7}(\cdot)}{f_{8}(\cdot)}{f_{9}(\cdot)}{f_{10}(\cdot)}
\drawserver{s5}{350}{190}{\mathcal S_{5}}{f_{9}(\cdot)}{f_{10}(\cdot)}{f_{11}(\cdot)}{f_{12}(\cdot)}
\drawserver{s6}{425}{190}{\mathcal S_{6}}{f_{11}(\cdot)}{f_{12}(\cdot)}{f_{1}(\cdot)}{f_{2}(\cdot)}
\drawserver{s7}{500}{190}{\mathcal S_{7}}{f_{2}(\cdot)}{f_{5}(\cdot)}{f_{8}(\cdot)}{f_{11}(\cdot)}
\drawserver{s8}{575}{190}{\mathcal S_{8}}{f_{1}(\cdot)}{f_{4}(\cdot)}{f_{7}(\cdot)}{f_{10}(\cdot)}

\drawuser{1}{95}{470}
\drawuser{2}{190}{470}
\drawuser{3}{285}{470}
\drawuser{4}{380}{470}
\drawuser{5}{475}{470}
\drawuser{6}{570}{470}

\draw (700,55)  node [anchor=west] {{\fontfamily{ptm}\selectfont \textbf{Coordinator Node}}};
\draw (700,245) node [anchor=west] {{\fontfamily{ptm}\selectfont \textbf{Server Nodes}}};
\draw (700,470) node [anchor=west] {{\fontfamily{ptm}\selectfont \textbf{Users}}};

\draw [color={rgb,255:red,0; green,0; blue,0}, draw opacity=1] (c1) -- (s1T);
\draw [color={rgb,255:red,0; green,0; blue,0}, draw opacity=1] (c2) -- (s2T);
\draw [color={rgb,255:red,0; green,0; blue,0}, draw opacity=1] (c3) -- (s3T);
\draw [color={rgb,255:red,0; green,0; blue,0}, draw opacity=1] (c4) -- (s4T);
\draw [color={rgb,255:red,0; green,0; blue,0}, draw opacity=1] (c5) -- (s5T);
\draw [color={rgb,255:red,0; green,0; blue,0}, draw opacity=1] (c6) -- (s6T);
\draw [color={rgb,255:red,0; green,0; blue,0}, draw opacity=1] (c7) -- (s7T);
\draw [color={rgb,255:red,0; green,0; blue,0}, draw opacity=1] (c8) -- (s8T);

\draw [color={rgb,255:red,120; green,217; blue,21}, draw opacity=1] (s1L) -- ++(0,42) -| (u1a);
\draw [color={rgb,255:red,120; green,217; blue,21}, draw opacity=1] (s1M) -- ++(0,42) -| (u2a);
\draw [color={rgb,255:red,120; green,217; blue,21}, draw opacity=1] (s1R) -- ++(0,42) -| (u3a);

\draw [color={rgb,255:red,0; green,180; blue,180}, draw opacity=1] (s2L) -- ++(0,54) -| (u4a);
\draw [color={rgb,255:red,0; green,180; blue,180}, draw opacity=1] (s2M) -- ++(0,54) -| (u5a);
\draw [color={rgb,255:red,0; green,180; blue,180}, draw opacity=1] (s2R) -- ++(0,54) -| (u6a);

\draw [color={rgb,255:red,235; green,110; blue,241}, draw opacity=1] (s3L) -- ++(0,66) -| (u1b);
\draw [color={rgb,255:red,235; green,110; blue,241}, draw opacity=1] (s3M) -- ++(0,66) -| (u2b);
\draw [color={rgb,255:red,235; green,110; blue,241}, draw opacity=1] (s3R) -- ++(0,66) -| (u3b);

\draw [color={rgb,255:red,255; green,0; blue,4}, draw opacity=1] (s4L) -- ++(0,78) -| (u1c);
\draw [color={rgb,255:red,255; green,0; blue,4}, draw opacity=1] (s4M) -- ++(0,78) -| (u4b);
\draw [color={rgb,255:red,255; green,0; blue,4}, draw opacity=1] (s4R) -- ++(0,78) -| (u5b);

\draw [color={rgb,255:red,45; green,35; blue,245}, draw opacity=1] (s5L) -- ++(0,90) -| (u2c);
\draw [color={rgb,255:red,45; green,35; blue,245}, draw opacity=1] (s5M) -- ++(0,90) -| (u3c);
\draw [color={rgb,255:red,45; green,35; blue,245}, draw opacity=1] (s5R) -- ++(0,90) -| (u6b);

\draw [color={rgb,255:red,245; green,166; blue,35}, draw opacity=1] (s6L) -- ++(0,102) -| (u1d);
\draw [color={rgb,255:red,245; green,166; blue,35}, draw opacity=1] (s6M) -- ++(0,102) -| (u4c);
\draw [color={rgb,255:red,245; green,166; blue,35}, draw opacity=1] (s6R) -- ++(0,102) -| (u5c);

\draw [color={rgb,255:red,180; green,60; blue,200}, draw opacity=1] (s7L) -- ++(0,114) -| (u2d);
\draw [color={rgb,255:red,180; green,60; blue,200}, draw opacity=1] (s7M) -- ++(0,114) -| (u4d);
\draw [color={rgb,255:red,180; green,60; blue,200}, draw opacity=1] (s7R) -- ++(0,114) -| (u6c);

\draw [color={rgb,255:red,120; green,90; blue,60}, draw opacity=1] (s8L) -- ++(0,126) -| (u3d);
\draw [color={rgb,255:red,120; green,90; blue,60}, draw opacity=1] (s8M) -- ++(0,126) -| (u5d);
\draw [color={rgb,255:red,120; green,90; blue,60}, draw opacity=1] (s8R) -- ++(0,126) -| (u6d);

\end{tikzpicture}%
}
\caption{Fixed quenched topology used in the worked example for Theorem~\ref{thm:P1}. The figure follows the same coordinator--server--user hierarchy as the system model and makes the assignment sets $\mathcal S_n$ explicit inside each server box. In each shot, all links emanating from the same server and drawn in the same color carry the same transmitted signal produced by that server, namely the common broadcast symbol sent to the users in its communication set.}
\label{fig:worked_example_topology_clean}
\end{figure*}

\subsubsection{Encoder, decoder, and training protocol}

We specialize the masked random Fourier feature encoder \eqref{eq:masked-rff} to the Gaussian kernel in \eqref{eq:worked_gaussian_kernel}. For each server $n\in[N]$ and shot $t\in[T]$, draw $\omega_{n,t}\sim\mathcal N(0,\rho^{-2}I_L)$ and $b_{n,t}\sim\mathrm{Unif}[0,2\pi]$ independently, define the masked frequency $\widetilde{\omega}_{n,t}=\omega_{n,t}\odot \mathbf 1_{\mathcal S_n}$, and transmit
\begin{equation}
z_{n,t}(x)=\sqrt{\frac{2}{\gamma}}\cos\!\bigl(\widetilde{\omega}_{n,t}^{\top}\mathbf w(x)+b_{n,t}\bigr).
\label{eq:worked_encoder}
\end{equation}
Since links are shot-agnostic, the same scalar $z_{n,t}(x)$ is sent to every user in $\mathcal T_n$ at shot $t$. Hence user $k$ receives the feature vector
\[
\Phi_k(x)=\bigl(z_{n,t}(x)\bigr)_{(n,t):\,k\in\mathcal T_n}\in\mathbb R^{\mrecv_k(\mathcal T)}=\mathbb R^{4T}.
\]

We train the decoder using $M=3000$ i.i.d. samples $\{x^{(i)}\}_{i=1}^{M}$ drawn from $\mathbb P_X$. The noisy label for user $k$ is $Y_k^{(i)}=F_k(x^{(i)})+\xi_k^{(i)}$, where $\xi_k^{(i)}\sim\mathcal N(0,\sigma^2)$, and the user-side decoder is the ridge rule of \eqref{eq:ridge}:
\begin{equation}
\widehat\beta_k
=
\arg\min_{\beta\in\mathbb R^{\mrecv_k(\mathcal T)}}
\frac{1}{M}\sum_{i=1}^{M}
\bigl(\beta^\top\Phi_k(x^{(i)})-Y_k^{(i)}\bigr)^2+\lambda\|\beta\|_2^2.
\label{eq:worked_ridge_decoder}
\end{equation}
The resulting estimate is $\widehat F_k(x)=\widehat\beta_k^\top\Phi_k(x)$. For population-risk evaluation we use an independent test set of size $M_{\mathrm{te}}=5000$.

This example is intentionally dense, since each user depends on all $L=12$ coordinates of $\mathbf w(x)$. However, the realized topology $(\mathsf A,\mathsf L)$ is chosen so that every user observes all essential coordinates, i.e., $\bigcup_{n:\,k\in\mathcal T_n}\mathcal S_n=[L]=\mathcal S_k^\star$ for all $k\in[K]$. Hence the coverage obstruction vanishes, so $\varepsilon_{\mathrm{cov},k}(\mathsf A,\mathsf L)=0$ for every user. As a result, this worked example isolates the feature-budget and spectral parts of Theorem~\ref{thm:P1} without any topology-induced coverage floor.

For the specific design $\mathcal D_{\mathrm{ex}}$, the quenched population risk is
\begin{equation}
\mathcal R\big(\mathcal D_{\mathrm{ex}}\mid \mathsf A,\mathsf L\big)
=
\mathbb E_{x\sim\mathbb P_X}\!\left[
\frac{1}{K}\sum_{k=1}^{K}\bigl|\widehat F_k(x)-F_k(x)\bigr|^2
\right].
\label{eq:worked_quenched_risk}
\end{equation}
Since $\mrecv_k(\mathcal T)=4T$ for all $k$ and therefore $m_{\mathrm{harm}}=4T$, the achievability bound in Theorem~\ref{thm:P1} becomes
\begin{equation}
\mathcal R\big(\mathcal D_{\mathrm{ex}}\mid \mathsf A,\mathsf L\big)
\le
\left(\frac{2}{\gamma}+C_1\right)\frac{B^2}{4\gamma T}\log\!\frac{2}{\delta'}
+
C_2\frac{\sigma^2 d_\lambda}{M}
+
C_3 B^2\lambda,
\label{eq:worked_achievability_specialized}
\end{equation}
while the converse reduces to the purely spectral lower bound
\begin{equation}
\inf_{\mathcal D}\mathcal R\big(\mathcal D\mid \mathsf A,\mathsf L\big)
\ge
\frac{B^2}{K}\sum_{k=1}^{K}\sum_{j>\mrecv_k(\mathcal T)}\lambda_j
=
B^2\sum_{j>4T}\lambda_j.
\label{eq:worked_converse_specialized}
\end{equation}
Thus, both sides are governed by the same communicated-feature scale $4T$.

It is also important to distinguish the kernel-operator spectrum $\{\lambda_j\}_{j\ge1}$ from the user-side feature covariance. For each user $k$, the latter is $G_k=\mathbb E[\Phi_k(X)\Phi_k(X)^\top]\in\mathbb R^{4T\times 4T}$, with effective ridge dimension $d_{\lambda,k}=\mathrm{tr}(G_k(G_k+\lambda I)^{-1})$. Hence $G_k$ controls the conditioning and statistical efficiency of the learned decoder, whereas the eigenvalues $\{\lambda_j\}$ determine the irreducible spectral approximation error beyond the communicated-feature budget.

\subsubsection{Numerical validation and interpretation}

\begin{figure}[t]
\centering
\includegraphics[width=0.45\columnwidth]{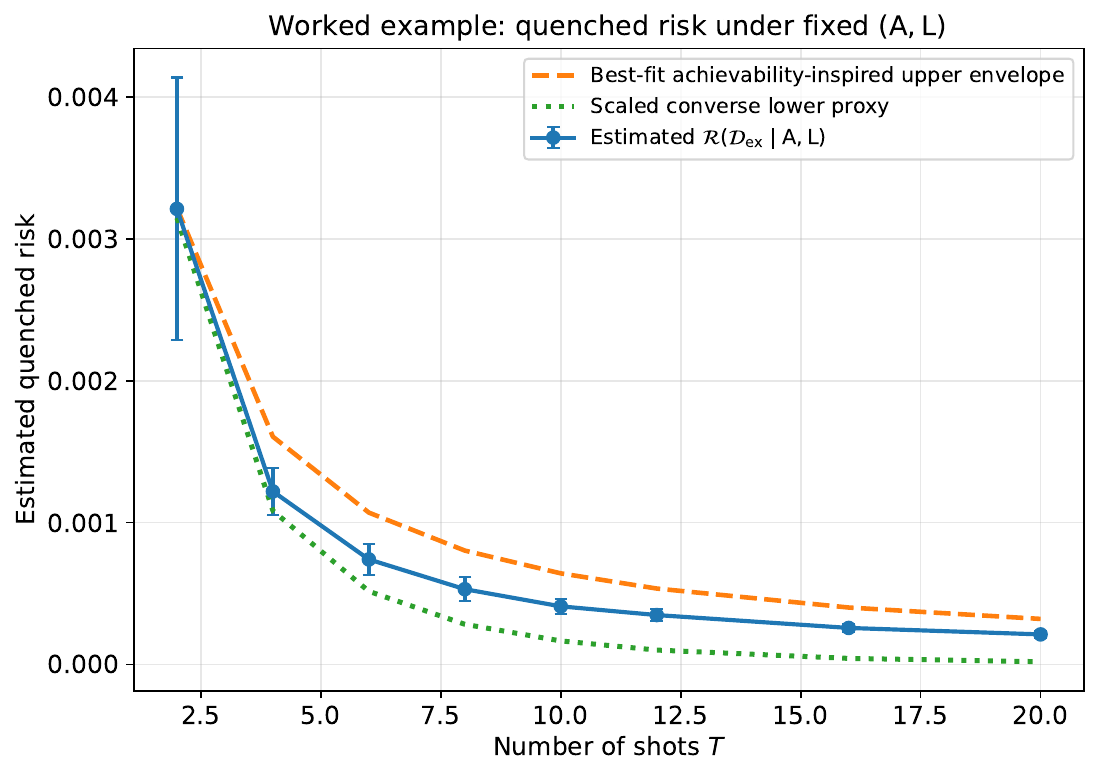}
\caption{Numerical validation of the worked example for Theorem~\ref{thm:P1} under the fixed quenched realization $(\mathsf A,\mathsf L)$ of Fig.~\ref{fig:worked_example_topology_clean}. The markers show the estimated quenched risk $\mathcal R(\mathcal D_{\mathrm{ex}}\mid \mathsf A,\mathsf L)$ as a function of the number of shots $T$. The dashed curve is a best-fit achievability-inspired upper envelope of the form $a\,3/(4T)+c$, while the dotted curve is a scaled converse lower proxy obtained from the empirical spectral tail of the Gaussian-kernel integral operator.}
\label{fig:worked_example_result}
\end{figure}

We validate the worked example by estimating the quenched risk \eqref{eq:worked_quenched_risk} as the number of shots \(T\) increases under the fixed realization \((\mathsf A,\mathsf L)\); the result is shown in Fig.~\ref{fig:worked_example_result}. The markers show the empirical values of \(\mathcal R(\mathcal D_{\mathrm{ex}}\mid \mathsf A,\mathsf L)\), the dashed curve is a best-fit achievability-inspired upper envelope of the form \(a\,3/(4T)+c\), and the dotted curve is a scaled converse proxy obtained from the empirical spectral tail of the Gaussian-kernel operator. Hence the simulation directly illustrates the two sides of Theorem~\ref{thm:P1} for this instantiated topology.

The figure shows three main facts. First, the quenched risk decreases monotonically with \(T\), consistent with the linear growth of the received-feature budget \(\mrecv_k(\mathcal T)=4T\). Second, no visible error floor appears over the tested range, in agreement with the fact that \(\varepsilon_{\mathrm{cov},k}(\mathsf A,\mathsf L)=0\). Third, the achievability-inspired envelope tracks the empirical curve closely, while the converse proxy remains below it, as predicted by the theorem. Thus, once coverage is guaranteed by the realized topology, the dominant improvement comes from the communicated-feature growth \(m_{\mathrm{harm}}=4T\), and the remaining obstruction is spectral rather than combinatorial.

In the simulation, the feature-side covariance is estimated by \(\widehat G_k=\frac{1}{M_{\mathrm{te}}}\sum_{i=1}^{M_{\mathrm{te}}}\Phi_k(x^{(i)})\Phi_k(x^{(i)})^\top\), whose eigenvalues provide a finite-sample proxy for the realized feature geometry. By contrast, the converse proxy is computed from the empirical eigenvalues of a Gaussian-kernel matrix built on independent samples of \(\mathbf w(X)\), and should therefore be interpreted only as a numerical approximation of the spectral tail in \eqref{eq:worked_converse_specialized}, rather than as the spectrum of \(\widehat G_k\) itself.

\paragraph*{Relation to the annealed regime}
The present example isolates the quenched mechanism of Theorem~\ref{thm:P1} by fixing a topology that guarantees complete coordinate coverage for every user. To complement this instance-wise picture, Appendix~\ref{subsec:worked_example_thm2} provides an annealed counterpart built from the same Gaussian-RKHS task model and the same system parameters $(K,N,L,\Gamma,\Delta)$, but with the topology drawn at random from the admissible ensemble. This appendix example shows how the performance picture changes once coverage is no longer guaranteed by construction and the average risk must reflect both the spectral approximation effect and the reliability of coordinate coverage, as predicted by Theorem~\ref{thm:P2}.
\section{Comparison}
\label{sec:comparison}
To compare the present framework with Theorem~2 of \cite{khalesi2024tessellated}, we specialize to the same linear/isotropic asymptotic regime considered there. In particular, we assume that the output subfunctions are linear and isotropic, and we then apply the achievable scheme of Theorem~\ref{thm:P1} using the general RKHS framework developed earlier, together with Lemma~\ref{lem:masked-rff} (kernel approximation error) and Lemma~\ref{lem:ridge-decomp} (ridge risk decomposition). This specialization isolates the spectral mechanism governing the error and enables a direct comparison with the Marchenko--Pastur (MP) benchmark in the regime shared with tessellated distributed computing. Throughout this comparison, we adopt the same limit order as in Theorem~2 of \cite{khalesi2024tessellated}:

\begin{itemize}[leftmargin=1.25em]
\item The input features are isotropic: $w(X)\in\R^L$ with $\E[w(X)]=0$ and $\E[w(X)w(X)^\top]=I_L$.
\item The ridge estimator is trained in the \emph{noiseless, infinite-sample} limit: training sample size $M\to\infty$ and ridge parameter $\lambda\downarrow 0$.
\item The system dimensions $(K,N,L)$ grow large with fixed ratios
\[
\gamma=\Gamma/L,\qquad \delta=\Delta/K,\qquad R=K/N.
\]
\end{itemize}

Now consider linearly separable targets for $K$ users
\[
F(x)=\mathbf{F}\,\mathbf{w}(x)\in\mathbb{R}^{K},\qquad \mathbf{F}\in\mathbb{R}^{K\times L},
\]
under the isotropic model described above. A linear scheme with computation/communication matrices $\mathbf{E}\in\mathbb{R}^{NT\times L}$ and $\mathbf{D}\in\mathbb{R}^{K\times NT}$ (supported by the bi-adjacency matrices of $(\mathsf{A},\mathsf{L})$) reconstructs
$\widehat{F}(x)=\mathbf{D}\mathbf{E}\mathbf{w}(x)$. Recall the \emph{quenched} population risk for a fixed topology $(\mathsf{A},\mathsf{L})$ (Def.~\eqref{eq:quenched-risk}):
\[
\mathcal{R}\big(\mathcal{D}\mid \mathsf{A},\mathsf{L}\big)
\;=\;
\mathbb{E}_{x}\!\left[\frac{1}{K}\sum_{k=1}^{K}\|\widehat{F}_k(x)-F_k(x)\|_2^2\right].
\]
In the same linear setting, TDC’s normalized distortion (cf.\ \cite{khalesi2024tessellated}) is
\begin{align}
D_{\mathrm{TDC}}
= \frac{1}{KL}\,\E_{\mathbf{F},\mathbf{w}}\!\left[\min_{\mathbf{D},\mathbf{E}}
\ \big\| \mathbf{D}\mathbf{E} - \mathbf{F}\big\|^2_{F}\right].\label{Definition_TDC}
\end{align}

When both sides are restricted to the same class of \emph{linear} designs supported by $(\mathsf{A},\mathsf{L})$, a direct comparison gives
\begin{align}
D_{\mathrm{TDC}} \ =\ \frac{1}{L}\,\mathcal{R}(\mathcal{D}\mid \mathsf{A},\mathsf{L}).\label{Normalized_risk}
\end{align}

We will show that in this limit, Lemma~\ref{lem:ridge-decomp} reduces the error to the spectral projection loss of the Gram operator $G_{\mathsf{A},\mathsf{L}}=\E[\Phi(X)\Phi(X)^\top]$, while Lemma~\ref{lem:masked-rff} ensures the Monte--Carlo kernel error vanishes. Thus the quenched distortion is determined entirely by the empirical spectrum of $G_{\mathsf{A},\mathsf{L}}$.

To be more precise, fix a user $k$. The $(\Gamma,\Delta)$--regular topology $(\mathsf A,\mathsf L)$\footnote{A pair \((\mathsf A,\mathsf L)\) is said to be \((\Gamma,\Delta)\)\text{--regular} if every server \(n \in [N]\) satisfies
$|\mathcal{S}_n| = \Gamma
\quad\text{and}\quad
|\mathcal{T}_n| = \Delta.$} determines
(i) the \emph{computation} supports of size $\Gamma$ per encoded row and
(ii) the \emph{communication} pattern that routes $T$ encoded shots per server to user $k$ with per-shot indegree $\Delta$ on average.
Stack the $m_k$ scalars received by user $k$ (over all servers and shots) into the feature map
\[
\Phi_k(x)\in\R^{m_k},\qquad \Phi_k(x)=E_k\,w(x),
\]
where $E_k\in\R^{m_k\times L}$ is supported by the bi-adjacency matrices of $(\mathsf A,\mathsf L)$ (rows have support size $\Gamma$). Note that, in this linear and isotropic specialization, the general Random Fourier Feature (RFF) construction used in the RKHS framework admits a simple linear limit.
Recall that for a shift-invariant kernel $K(x,x') = k(x-x')$, Bochner's theorem ensures the existence of a nonnegative spectral density $p(\omega)$ such that
\[
K(x,x') = \int_{\mathbb{R}^d} e^{j\,\omega^\top(x-x')} \, p(\omega)\, d\omega
= \mathbb{E}_{\omega\sim p(\omega)}[\cos(\omega^\top(x-x'))].
\]
Sampling frequencies $\omega_i \sim p(\omega)$ and random phases $b_i \sim \mathrm{Unif}[0,2\pi]$ yields the RFF embedding
\[
\Phi(x) = \frac{1}{\sqrt{m}}
\begin{bmatrix}
\sqrt{2}\cos(\omega_1^\top x + b_1) \\[-2pt]
\vdots \\[-2pt]
\sqrt{2}\cos(\omega_m^\top x + b_m)
\end{bmatrix},
\qquad
K(x,x') \approx \Phi(x)^\top \Phi(x').
\]
When the kernel bandwidth becomes large (for instance in the Gaussian kernel
$K(x,x')=\exp(-\|x-x'\|^2/2\sigma^2)$ as $\sigma\to\infty$),
the spectral measure $p(\omega)$ concentrates around $\omega=0$, and we may linearize $\cos(\omega_i^\top x+b_i)$ around the origin:
\[
\cos(\omega_i^\top x+b_i)
\;=\;
\cos(b_i) - \sin(b_i)\,\omega_i^\top x + O(\|\omega_i^\top x\|^2).
\]
After rescaling, the random feature vector becomes approximately linear in $x$, with covariance
\(\mathbb{E}[\Phi(x)\Phi(x')^\top] \approx x^\top x' I_m/m\).
Formally, taking the linear kernel limit $K(x,x') = x^\top x'$
corresponds to the degenerate spectral measure
$p(\omega) = \sum_{\ell=1}^d \delta(\omega-e_\ell)$,
so that the RFF map reduces exactly to the identity:
\[
\Phi(x) = [x_1, x_2, \ldots, x_d]^\top = x .
\]
In this limit, we model the random feature vector as an isotropic embedding
$w(x)\sim\mathcal{N}(0,I_L)$ satisfying $\mathbb{E}[w(x)w(x')^\top] = K(x,x') = x^\top x'$.
Hence, the relation $\Phi_k(x)=E_k w(x)$ simply expresses that user~$k$
has access to an encoded subset of the global RFF coordinates,
with the support pattern of $E_k$ determined by the computation and link topology
$(\mathsf{A},\mathsf{L})$; each nonzero entry of $E_k$ indicates a subfunction index
that is both computed and communicated to user~$k$. Even when the underlying kernel reduces to the linear case $K(x,x') = x^\top x'$ so that the associated feature map becomes the identity and the random feature vector satisfies $w(x)\sim\mathcal{N}(0,I_L)$, the encoding matrices $\{E_k\}$ remain nontrivial. Indeed, $E_k$ does not arise from the kernel itself but from the distributed system topology and resource constraints: it specifies which subset or linear combination of the $L$ orthogonal subfunction outputs is both computed and communicated to user~$k$. Formally,
\[
E_k(\ell,j) \neq 0
\quad\Longleftrightarrow\quad
\exists\, n\in[N]\ \text{s.t.}\ j\in\mathcal{S}_n,\ \ k\in\mathcal{T}_n,
\]
where $\mathcal{S}_n$ and $\mathcal{T}_n$ denote, respectively, the computation and link supports of server~$n$.
Hence, even though the global covariance $\mathbb{E}[w(x)w(x')^\top]=K(x,x')=x^\top x'$ is identity, each user observes only the projected feature vector $\Phi_k(x)=E_k w(x)$ whose covariance $G_k=E_kE_k^\top$ has rank limited by the normalized computation and communication budgets.
Only in the fully centralized regime $(\gamma=\delta=1)$, where every server computes and transmits all subfunctions to all users, does $E_k=I_L$ and the distortion vanish. Under isotropy, $\E[w]=0$ and $\E[ww^\top]=I_L$, the population Gram operator seen by the user is
\begin{equation}\label{eq:Gram-pre}
G_k\;\triangleq\;\E\big[\Phi_k(X)\Phi_k(X)^\top\big]\;=\;E_k\,E_k^\top\ \succeq\ 0.
\end{equation}
Let $\{\xi_{k,(i)}\}_{i=1}^{m_k}$ be the eigenvalues of $G_k$ in nonincreasing order, and let
\(
\mu_{k}\triangleq \tfrac{1}{m_k}\sum_{i=1}^{m_k}\delta_{\xi_{k,(i)}}
\)
be its empirical spectral distribution (ESD). For the linear/isotropic comparison below, we may drop the subscript $k$ when no confusion arises. The above is summarized in the following proposition:

\begin{proposition}
\label{prop:proj-limit}
Work under the setup and assumptions used in Theorem~\ref{thm:P1}:
(i)~fixed $(\mathsf A,\mathsf L)$ (quenched);
(ii)~linear/isotropic inputs $w(X)\in\R^{L}$ with $\E[w]=0$ and $\E[w w^\top]=I_L$;
(iii)~feature map for user $k$ given by $\Phi_k(x)=E_k\,w(x)\in\R^{m_k}$ with $E_k$ supported by $(\mathsf A,\mathsf L)$.
Consider the ridge predictor trained with $M$ i.i.d.\ samples and penalty $\lambda>0$:
\[
\widehat \beta_{M,\lambda}
\;\in\;
\arg\min_{\beta\in\R^{m_k}}
\frac1M\sum_{i=1}^M\big(\beta^\top \Phi_k(x_i)-F_k(x_i)\big)^2+\lambda\|\beta\|_2^2,
\quad
\widehat F_{M,\lambda}(x)=\widehat\beta_{M,\lambda}^{\!\top}\Phi_k(x).
\]
In the \emph{noiseless, infinite-sample} limit ($M\to\infty$ and $\lambda\downarrow 0$), the trained decoder converges in $L_2({\mathbb{P}}_X)$ to the orthogonal projector $P$ onto
$\mathrm{span}\{\Phi_k(x):x\in\mathcal X\}\subset L_2({\mathbb{P}}_X)$:
\[
\widehat F_{M,\lambda}\ \xrightarrow[M\to\infty,\ \lambda\downarrow 0]{L_2({\mathbb{P}}_X)}\ P F_k,
\qquad
P:\ L_2({\mathbb{P}}_X)\to \overline{\mathrm{span}\{\Phi_k(\cdot)\}}.
\]
Consequently, letting $G_k=\E[\Phi_k(X)\Phi_k(X)^\top]\succeq 0$ with ordered eigenvalues
$\xi_{(1)}\ge\cdots\ge \xi_{(m_k)}\ge 0$ and empirical spectral law
$\mu=\tfrac{1}{m_k}\sum_{i=1}^{m_k}\delta_{\xi_{(i)}}$,
the $1/(KL)$--normalized quenched distortion of our linear scheme equals the energy discarded by truncation at mass $\kappa=1-m_{\mathrm{eff}}$:
\begin{equation}\label{eq:Dq-eig-sum-prop}
D_{\mathrm{MUDC}}^{\mathrm q}(\mathsf A,\mathsf L)
\;=\;
\frac{1}{L}\sum_{i=m_k-\lceil \kappa m_k\rceil+1}^{m_k}\xi_{(i)}
\;=\;\int_{0}^{\kappa} Q_{\mu}^{\downarrow}(u)\,du,
\end{equation}
where $Q_{\mu}^{\downarrow}$ is the lower-tail quantile of $\mu$ and $m_{\mathrm{eff}}=\tfrac{T\gamma N}{K}$.
\end{proposition}

Proof of Proposition~\ref{prop:proj-limit} is provided in Appendix~\ref{proof:proj-limit}. In what follows, we argue that under disjoint and balanced supports (the structural assumption of TDC), the Gram spectrum converges to the Marchenko--Pastur law $\mathrm{MP}_{\lambda'}$ with aspect ratio $\lambda'=\delta K/(\gamma L)=\Delta/\Gamma$. The corresponding truncated first moment $\Phi_{\mathrm{MP},\lambda'}$ was shown in~\cite{khalesi2024tessellated} to serve as both an achievability result (for TDC) and a converse within that class. We then examine how deviations from the disjoint-support assumption---through overlapping subfunction assignments or nonuniform link topologies---affect performance, and how such structures modify the normalized distortion relative to the MP benchmark.

In particular, we compare $D_{\mathrm{MUDC}}^{\mathrm q}$ to the Marchenko--Pastur (MP) envelope at the same budgets\footnote{\emph{Notation:} we reserve $\lambda$ for ridge; we write $\lambda'$ for the MP aspect ratio.}:
\begin{equation}\label{eq:MP-params-pre}
\lambda'\;\triangleq\;\frac{\delta K}{\gamma L}\;=\;\frac{\Delta}{\Gamma},\qquad
r=(1-\sqrt{\lambda'})^2,\quad b=(1+\sqrt{\lambda'})^2.
\end{equation}
Let $f_{\mathrm{MP},\lambda'}$ and $F_{\mathrm{MP},\lambda'}$ be the MP density and CDF on $[r,b]$.
The MP truncation threshold $t$ is defined implicitly by
\begin{equation}\label{eq:MP-threshold-pre}
F_{\mathrm{MP},\lambda'}(t)\;=\;\kappa\;=\;1-\frac{T\gamma N}{K}.
\end{equation}
The corresponding truncated first moment (the MP benchmark) is
\begin{equation}\label{eq:PhiMP-pre}
\Phi_{\mathrm{MP},\lambda'}(t,r)
\;\triangleq\;\int_{r}^{t} x\,f_{\mathrm{MP},\lambda'}(x)\,dx
\;=\;\int_{0}^{\kappa} Q_{\mathrm{MP},\lambda'}^{\downarrow}(u)\,du.
\end{equation}

Given a fixed $(\Gamma,\Delta)$-regular topology $(\mathsf{A},\mathsf{L})$, we ask how the quenched distortion $D^{\mathrm{q}}_{\mathrm{MUDC}}(\mathsf{A},\mathsf{L})$, obtained from the achievable construction of this paper, compares with the MP envelope $\Phi_{\mathrm{MP},\lambda'}$. The difference
\[
G^{\mathrm{q}}(\mathsf{A},\mathsf{L})
=\Phi_{\mathrm{MP},\lambda'}-D^{\mathrm{q}}_{\mathrm{MUDC}}(\mathsf{A},\mathsf{L})
\]
is what we call the \emph{quenched MP--gap}. It measures the deviation of the realized topology from the MP benchmark in terms of lower-tail spectral mass. As we show below, this gap is nonnegative and vanishes for disjoint-and-balanced tessellations, while overlap or aliasing can make it strictly positive.

\medskip
With this in place, we proceed to the formal statement giving the exact gap representation and tight bounds.

\begin{theorem}
\label{thm:gap-bounds}
Work in the linear/isotropic setting and fix a $(\Gamma,\Delta)$--regular topology $(\mathsf A,\mathsf L)$.
Let $L$ be the ambient dimension for the user, and set
\[
m_{\mathrm{eff}}\;\triangleq\;\min\!\left\{1,\,\frac{T\gamma N}{K}\right\},
\qquad
\kappa\;\triangleq\;1-m_{\mathrm{eff}}.
\]
Let $G_{\mathsf A,\mathsf L}=\E[\Phi(X)\Phi(X)^\top]\in\R^{L\times L}$ be the user Gram operator and
$\mu_{\mathsf A,\mathsf L}\equiv \tfrac{1}{L}\sum_{i=1}^{L}\delta_{\xi_{(i)}}$ its empirical spectral law
with eigenvalues $\xi_{(1)}\!\ge\!\cdots\!\ge\!\xi_{(L)}\!\ge\!0$. Throughout, the eigenvalues are ordered in descending order, and the quantile
function $Q_{\mu}^{\downarrow}$ is defined accordingly as the \emph{nonincreasing}
(right-continuous) inverse of the CDF $F_{\mu}$, i.e.,
$Q_{\mu}^{\downarrow}(u)=\inf\{x:\,F_{\mu}(x)\ge u\}$.
Define the lower-tail quantile $Q_{\mu}^{\downarrow}(u)\triangleq\inf\{x:\,F_\mu(x)\ge u\}$ for $u\in[0,1]$. The $1/(KL)$-normalized quenched distortion of our linear scheme is
\[
D_{\mathrm{MUDC}}^{\mathrm q}(\mathsf A,\mathsf L)
\;=\;
\frac{1}{L}\sum_{i=L-\lceil \kappa L\rceil+1}^{L}\xi_{(i)}
\;=\;
\int_{0}^{\kappa} Q_{\mu_{\mathsf A,\mathsf L}}^{\downarrow}(u)\,du .
\]
Let $\lambda'=\delta K/(\gamma L)=\Delta/\Gamma$, let $\MP_{\lambda'}$ be the Marchenko--Pastur law with that aspect ratio, and define its lower-tail quantile $Q_{\mathrm{MP},\lambda'}^{\downarrow}$ and CDF $F_{\mathrm{MP},\lambda'}$.
Let $t$ solve $F_{\mathrm{MP},\lambda'}(t)=\kappa$ and define the MP benchmark
\[
\Phi_{\mathrm{MP},\lambda'}(\kappa)\;\triangleq\;\int_{0}^{\kappa} Q_{\mathrm{MP},\lambda'}^{\downarrow}(u)\,du
\;=\;\int x\,\mathbf 1_{\{x\le t\}}\,d\MP_{\lambda'}(x).
\]
Define the quenched MP--gap
\[
G^{\mathrm q}(\mathsf A,\mathsf L)\;\triangleq\;\Phi_{\mathrm{MP},\lambda'}(\kappa)-D_{\mathrm{MUDC}}^{\mathrm q}(\mathsf A,\mathsf L).
\]
Then:
\begin{enumerate}[label=(\alph*),leftmargin=1.25em]
\item \textbf{Exact gap representation.}
\begin{equation}
\label{eq:gap-quantile-thm-nocf}
G^{\mathrm q}(\mathsf A,\mathsf L)
\;=\;
\int_{0}^{\kappa}\!\Big(Q_{\mathrm{MP},\lambda'}^{\downarrow}(u)-Q_{\mu_{\mathsf A,\mathsf L}}^{\downarrow}(u)\Big)\,du .
\end{equation}

\item \textbf{Nonnegativity and a robust lower bound.}
Assume the coverage and eigen-tail constraints of Lemmas~\ref{lem:eigen-tail}--\ref{lem:coverage} (Section~\ref{sec:proof-P1}) hold, which imply the MP \emph{upper envelope} on truncated first moments:
\[
\int_{0}^{\kappa} Q_{\mu_{\mathsf A,\mathsf L}}^{\downarrow}(u)\,du\ \le\ \int_{0}^{\kappa} Q_{\mathrm{MP},\lambda'}^{\downarrow}(u)\,du .
\]
Then $G^{\mathrm q}(\mathsf A,\mathsf L)\ge 0$, with equality iff
$Q_{\mu_{\mathsf A,\mathsf L}}^{\downarrow}=Q_{\mathrm{MP},\lambda'}^{\downarrow}$ a.e.\ on $[0,\kappa]$.
Moreover, if for some measurable $U\subset[0,\kappa]$ with $|U|=\eta>0$ and some $\alpha>0$ one has
$Q_{\mu_{\mathsf A,\mathsf L}}^{\downarrow}(u)\le Q_{\mathrm{MP},\lambda'}^{\downarrow}(u)-\alpha$ for all $u\in U$,
then
\begin{equation}
\label{eq:gap-lb-thm-nocf}
G^{\mathrm q}(\mathsf A,\mathsf L)\ \ge\ \alpha\,\eta .
\end{equation}

\item \textbf{Simple upper bound.}
If the supports of both spectra lie in $[0,b]$ for some $b<\infty$ (e.g.\ $b=(1+\sqrt{\lambda'})^2$ in the classical MP setting), then
\begin{equation}
\label{eq:gap-ub-thm-nocf}
0\ \le\ G^{\mathrm q}(\mathsf A,\mathsf L)\ \le\ \Phi_{\mathrm{MP},\lambda'}(\kappa)
\ \le\ b\,\kappa\ =\ b\Big(1-\tfrac{T\gamma N}{K}\Big).
\end{equation}
\end{enumerate}
\end{theorem}

\begin{remark}
The nonnegativity of the MP--gap $G^{\mathrm q}\!\ge\!0$ relies on the standard spectral dominance assumption:
under disjoint and balanced supports (the TDC regime), the empirical spectrum of $E_kE_k^\top$ converges to the Marchenko--Pastur law,
which upper-bounds the truncated first moment among all $(\Gamma,\Delta)$--regular topologies.
Hence, $G^{\mathrm q}\!>\!0$ indicates that the masked-RFF-based scheme retains more lower-tail spectral energy than the MP benchmark when the topology departs from perfect tessellation.
\end{remark}
Proof of Theorem~\ref{thm:gap-bounds} is deferred to Appendix~\ref{proof:gap-bounds}. Additional finite-dimensional validations, including both the distortion-ordering comparison and direct numerical verification of the exact MP--gap identity, are reported in Appendix~\ref{sec:numerical_validation}, especially Subsection~\ref{subsec:sim_thm3}.

The quenched MP--gap \(G^{\mathrm q}\) measures the deviation of the empirical lower spectral tail of the user Gram matrix from the Marchenko--Pastur benchmark under the same budgets \((\gamma,\delta,T)\). Equivalently, it is the area between the empirical and MP lower-tail quantiles over the discarded spectral mass \(\kappa=1-T\gamma N/K\), and thus quantifies the corresponding mismatch in retained spectral energy. Under disjoint and balanced tessellations, i.e., the TDC structural regime, the empirical spectral law converges to \(\mathrm{MP}_{\lambda'}\), the truncated first moment in \eqref{eq:PhiMP-pre} matches the empirical lower-tail mean, and therefore \(G^{\mathrm q}=0\). Hence the present achievable scheme asymptotically recovers the MP benchmark in that regime.

Beyond strict disjointness, the present scheme still applies to arbitrary \((\Gamma,\Delta)\)-regular topologies and quantifies their deviation from the MP benchmark through \(G^{\mathrm q}\). In particular, overlap in computation supports or imbalance in link patterns can push the empirical lower-tail quantile \(Q_{\mu_{\mathsf A,\mathsf L}}^{\downarrow}\) below its MP counterpart \(Q_{\mathrm{MP},\lambda'}^{\downarrow}\) on a set of positive measure, which yields a strictly positive gap \(G^{\mathrm q}>0\) through \eqref{eq:gap-lb-thm-nocf}. Thus the MP--gap provides a spectral comparison tool between tessellated and non-tessellated topologies within a common framework.

Finally, the envelope bound \eqref{eq:gap-ub-thm-nocf} shows that increasing \(T\), \(\gamma\), or \(N\) reduces \(\kappa\), pushing the effective spectrum toward full coverage, while increasing \(\delta\) decreases \(\lambda'\) and tightens the MP bulk. In summary, disjoint-and-balanced topologies recover the MP law with \(G^{\mathrm q}=0\), whereas overlapping or imbalanced designs can induce a strictly positive MP--gap.
\section{Conclusion}
\label{sec:conclusion}
This work introduced the \emph{General Multi-User Distributed Computing (GMUDC)} framework, a learning-theoretic model for studying the joint trade-off among computation, communication, and approximation accuracy in heterogeneous multi-user, multi-server systems. Unlike prior distributed-computing models such as~\cite{khalesi2024tessellated,khalesi2}, GMUDC does not rely on separability assumptions on users, subfunctions, or data partitions. Instead, each target function is modeled in a bounded RKHS associated with a shift-invariant kernel, thereby covering both linear and smooth nonlinear mappings and allowing arbitrary budget-feasible assignment and communication topologies. Through a kernel-ridge viewpoint, the framework connects distributed computation, kernel approximation, and statistical learning within a common analytical model.

On the achievability side, masked random Fourier features yield explicit quenched and annealed risk bounds under computation and communication budgets. On the converse side, the analysis separates spectral and coverage obstructions, showing how the realized topology and communicated-feature budget jointly determine achievable performance. In the shared linear/isotropic comparison regime, the framework also yields an MP-gap characterization that quantifies the effect of overlap and link reuse relative to the disjoint-support benchmark. Together, these results identify the topology-aware computation--communication--accuracy trade-off up to constants and logarithmic factors under the nominal model assumptions.

The supplementary numerical material in Appendix~\ref{sec:numerical_validation} provides theorem-driven finite-dimensional validation of the main mechanisms, including synthetic validations of Theorems~\ref{thm:P1},~\ref{thm:P2}, and~\ref{thm:gap-bounds}, additional nonlinear-topology diagnostics, robustness-sensitivity experiments outside the nominal assumptions, and a lightweight UAV-swarm toy example illustrating the quenched perspective in a geometric communication setting. These results reinforce the main message that, under fixed budgets, performance depends not only on how much computation and communication are available, but also on how the realized topology exposes task-relevant coordinates to the users.

Several practically important directions remain open. The present theory is developed for a nominal synchronous model in which all servers are active in all shots and subfunctions have comparable computational significance. Extending the framework to explicitly incorporate stragglers, server unreliability, and heterogeneous task or workload costs would bring it closer to practical distributed-learning and edge-computing systems. Such effects would modify both the effective communicated-feature budget and the relevant coverage events, and therefore require revisiting both achievability and converse arguments. Accordingly, these regimes are treated here as important but formally out-of-scope extensions of the nominal GMUDC model, rather than as settings covered by Theorems~\ref{thm:P1}--\ref{thm:gap-bounds}. More broadly, GMUDC provides an analytical template for topology-constrained approximate distributed computation in learning-oriented settings and suggests natural extensions toward distributed inference, estimation, and more realistic resource-constrained networked systems.

\subsection*{Future Research Directions}

\begin{itemize}
    \item \textbf{Adaptive and Hierarchical GMUDC.} Extend the model to hierarchical, time-varying, or adaptive topologies with dynamic budget allocation.

    \item \textbf{Learning-Integrated Distributed Optimization.} Combine GMUDC with gradient coding and federated optimization to study convergence--risk--communication trade-offs jointly.

    \item \textbf{Spectral Topology Design.} Use the MP-gap characterization to design budget-feasible topologies with provably improved spectral efficiency.

    \item \textbf{Stragglers, Heterogeneous Costs, and Asynchrony.} Generalize the nominal model to random server unavailability, nonuniform computational costs, and asynchronous communication.

    \item \textbf{Noisy and Adversarial Regimes.} Extend the spectral--coverage framework to stochastic or adversarial server behavior.
\end{itemize}
\bibliographystyle{IEEEtran}
\bibliography{ref}

\clearpage
\appendices

\thispagestyle{plain}

\begin{figure*}[t]
\centering
\resizebox{0.96\textwidth}{!}{%
\begin{tikzpicture}[
    font=\scriptsize,
    >=Latex,
    box/.style={
        draw,
        rounded corners=3pt,
        thick,
        align=center,
        minimum height=8mm,
        inner sep=4pt
    },
    core/.style={
        box, draw=black!70, fill=gray!12,
        minimum width=4.8cm
    },
    theorem/.style={
        box, draw=blue!65!black, fill=blue!10,
        minimum width=3.7cm
    },
    proof/.style={
        box, draw=orange!80!black, fill=orange!14,
        minimum width=3.9cm
    },
    ex/.style={
        box, draw=black!65, fill=gray!7,
        minimum width=3.8cm
    },
    inputbox/.style={
        box, draw=green!45!black, fill=green!10,
        minimum width=4.9cm
    },
    sim/.style={
        box, draw=violet!70!black, fill=violet!10,
        minimum width=3.1cm
    },
    supp/.style={
        box, draw=red!65!black, fill=red!8,
        minimum width=5.2cm
    },
    head/.style={
        box, draw=black!60, fill=white,
        minimum width=3.8cm
    },
    line/.style={->, thick, rounded corners=4pt},
    bus/.style={thick},
    group/.style={draw=black!45, dashed, rounded corners=6pt, inner sep=6pt}
]

\node[core] (model) at (0,0)
{GMUDC system model\\quenched / annealed formulations $(P1,P2)$};

\coordinate (abusL) at (-8.4,-1.05);
\coordinate (abusR) at ( 8.4,-1.05);
\draw[bus] (abusL) -- (abusR);
\draw[line] (model.south) -- ++(0,-0.42) -- (0,-1.05);

\node[theorem] (t1) at (-6.2,-2.35) {Theorem~\ref{thm:P1}\\quenched regime};
\node[theorem] (t2) at ( 0.0,-2.35) {Theorem~\ref{thm:P2}\\annealed regime};
\node[theorem] (t3) at ( 6.2,-2.35) {Theorem~\ref{thm:gap-bounds}\\quenched MP--gap};

\draw[line] (-6.2,-1.05) -- (t1.north);
\draw[line] ( 0.0,-1.05) -- (t2.north);
\draw[line] ( 6.2,-1.05) -- (t3.north);

\node[ex] (w1)   at (-6.2,-4.55) {Worked example\\Subsec.~III-D};
\node[ex] (w2)   at ( 0.0,-4.55) {Worked example\\Appendix A};
\node[ex] (prop) at ( 6.2,-4.55) {Proposition~\ref{prop:proj-limit}\\linear/isotropic reduction};

\draw[line] (t1.south) -- (w1.north);
\draw[line] (t2.south) -- (w2.north);
\draw[line] (t3.south) -- (prop.north);

\node[proof] (p1) at (-6.2,-6.8) {Appendix B-A\\proof of Theorem~\ref{thm:P1}};
\node[proof] (p2) at ( 0.0,-6.8) {Appendix B-B\\proof of Theorem~\ref{thm:P2}};
\node[proof] (p3) at ( 6.2,-6.8) {Appendix I\\proof of Theorem~\ref{thm:gap-bounds}};

\draw[line] (w1.south) -- (p1.north);
\draw[line] (w2.south) -- (p2.north);
\draw[line] (prop.south) -- (p3.north);

\node[inputbox] (i1) at (-6.2,-9.25)
{Supporting ingredients\\
Lemma~1 $\to$ App.~C \quad Lemma~2 $\to$ App.~D\\
Lemma~3 $\to$ App.~E\\
Lemma~4 $\to$ App.~F \quad Lemma~5 $\to$ App.~G};

\node[inputbox] (i2) at (0.0,-9.25)
{Averaging ingredients\\
quenched bound from Theorem~\ref{thm:P1}\\
concentration of $m_k(T)$ and\\
annealed coverage averaging};

\node[inputbox] (i3) at (6.2,-9.25)
{Supporting ingredients\\
Proposition~\ref{prop:proj-limit} $\to$ App.~H\\
Lemma~4 $\to$ App.~F \quad Lemma~5 $\to$ App.~G\\
MP benchmark and quantile identity};

\draw[line] (p1.south) -- (i1.north);
\draw[line] (p2.south) -- (i2.north);
\draw[line] (p3.south) -- (i3.north);

\node[group,
    fit=(model)(t1)(t2)(t3)(w1)(w2)(prop)(p1)(p2)(p3)(i1)(i2)(i3),
    label={[font=\small]above:Analytical roadmap}] {};

\node[core] (appj) at (0,-12.4)
{Appendix~J\\numerical validation \& synthetic experiments};

\coordinate (bbusL) at (-5.0,-13.6);
\coordinate (bbusR) at ( 5.0,-13.6);
\draw[line] (appj.south) -- ++(0,-0.45) -- (0,-13.6);
\draw[bus] (bbusL) -- (bbusR);

\node[head] (hval)  at (-4.8,-15.0) {Main theorem validations};
\node[head] (hsupp) at ( 4.8,-15.0) {Supplementary material};

\draw[line] (-4.8,-13.6) -- (hval.north);
\draw[line] ( 4.8,-13.6) -- (hsupp.north);

\coordinate (vbusL) at (-8.8,-16.1);
\coordinate (vbusR) at (-0.8,-16.1);
\draw[line] (hval.south) -- ++(0,-0.38) -- (-4.8,-16.1);
\draw[bus] (vbusL) -- (vbusR);

\node[sim] (s1) at (-8.0,-17.45) {Subsec.~J-A\\validation of\\Theorem~\ref{thm:P1}};
\node[sim] (s2) at (-4.8,-17.45) {Subsec.~J-B\\validation of\\Theorem~\ref{thm:P2}};
\node[sim] (s3) at (-1.6,-17.45) {Subsec.~J-C\\validation of\\Theorem~\ref{thm:gap-bounds}};

\draw[line] (-8.0,-16.1) -- (s1.north);
\draw[line] (-4.8,-16.1) -- (s2.north);
\draw[line] (-1.6,-16.1) -- (s3.north);

\draw[line] (hsupp.south) -- ++(0,-0.55) -- (4.8,-16.0);

\node[supp] (sd1) at (4.8,-17.1)
{Subsecs.~J-D--J-E\\
nonlinear topology-alignment and\\
topology-ensemble diagnostics};

\node[supp] (sd2) at (4.8,-19.35)
{Subsecs.~J-F--J-G\\
robustness sensitivity and\\
UAV-swarm toy, quenched regime};

\draw[line] (4.8,-16.0) -- (sd1.north);
\draw[line] (sd1.south) -- (sd2.north);

\node[group,
    fit=(appj)(hval)(hsupp)(s1)(s2)(s3)(sd1)(sd2),
    label={[font=\small]above:Appendix roadmap}] {};

\end{tikzpicture}%
}
\caption{Roadmap of the analytical and appendix material. The top block summarizes the analytical development from the GMUDC model to Theorems~\ref{thm:P1}, \ref{thm:P2}, and \ref{thm:gap-bounds}, together with their worked examples, proof appendices, and supporting ingredients. The bottom block summarizes Appendix~J. The theorem-specific validation subsections are grouped on the left, while the supplementary diagnostics are grouped on the right.}
\label{fig:appendix-roadmap}
\end{figure*}

\clearpage

\section{Worked Example for Theorem~\ref{thm:P2}: An Annealed Gaussian-RKHS GMUDC Instance}
\label{subsec:worked_example_thm2}

We here present an annealed counterpart to the fixed quenched example of Subsection~\ref{subsec:worked_example_thm1}. The task model, kernel, and budget parameters are kept unchanged, and only the topology model is modified. Thus we again consider a GMUDC system with $K=6$ users, $N=8$ servers, $L=12$ subfunctions, computation budget $\Gamma=4$, communication budget $\Delta=3$, scalar-valued user outputs $m_k=1$, and $T$ communication shots. Accordingly, $\gamma=\Gamma/L=1/3$ and $\delta=\Delta/K=1/2$, and we use the same Gaussian kernel $K_\rho(u,v)=\exp(-\|u-v\|_2^2/(2\rho^2))$ with $\rho=1.1$ as in \eqref{eq:worked_gaussian_kernel}. The input law, the subfunctions $f_\ell(\cdot)$, and the user targets $F_k(x)=h_k(\mathbf w(x))$ are exactly the same as in \eqref{eq:worked_subfunctions}--\eqref{eq:worked_centers}; in particular, each user has essential coordinate set $\mathcal S_k^\star=[L]$, so $r_k=L=12$ for all $k\in[K]$, and hence $r_{\mathrm{avg}}=\frac1K\sum_{k=1}^K r_k=12$. In this sense, the annealed example modifies only the randomness of the topology and not the underlying task family.

\subsubsection{Random assignment and random communication topology}

Unlike the worked example of Theorem~\ref{thm:P1}, where $(\mathsf A,\mathsf L)$ was fixed deterministically, we now draw the topology uniformly at random as in the annealed problem formulation of Section~\ref{System-Model}. More precisely, for each server $n\in[N]$, we sample
\[
\mathcal S_n \sim \mathrm{Unif}\!\big(\{S\subseteq[L]:|S|=\Gamma\}\big),
\qquad
\mathcal T_n \sim \mathrm{Unif}\!\big(\{U\subseteq[K]:|U|=\Delta\}\big),
\]
independently across $n$ and independently of one another, and we write $\mathsf A=(\mathcal S_1,\dots,\mathcal S_N)$ and $\mathsf L=(\mathcal T_1,\dots,\mathcal T_N)$. Thus, in the present example, the assignment phase and the communication phase are not conditioned on a single realization, but rather averaged over a topology ensemble consistent with the budgets $(\Gamma,\Delta)$. This directly instantiates the random-topology regime in \eqref{eq:annealed-risk3} and \eqref{eq:P1b}.

For a given realization $(\mathsf A,\mathsf L)$, user $k$ receives
\[
\mrecv_k(\mathcal T)=T\,|\{n:\,k\in\mathcal T_n\}|
\]
encoded scalars. Averaging over the random link realization yields
\[
m_k^{\mathrm{avg}}=\mathbb E[\mrecv_k(\mathcal T)]=TN\delta=T\cdot 8\cdot \frac12=4T.
\]
Hence the typical communicated-feature budget in this annealed example is again $4T$, exactly as in the fixed quenched example, but now this budget is achieved only on average rather than deterministically for every user and every realization.

\subsubsection{Encoder, decoder, and Monte Carlo annealed design}

For each realization of $(\mathsf A,\mathsf L)$, we use the same masked random Fourier feature construction as in \eqref{eq:masked-rff}. Specifically, for each server $n\in[N]$ and shot $t\in[T]$, we draw $\omega_{n,t}\sim\mathcal N(0,\rho^{-2}I_L)$ and $b_{n,t}\sim\mathrm{Unif}[0,2\pi]$ independently, form $\widetilde{\omega}_{n,t}=\omega_{n,t}\odot \mathbf 1_{\mathcal S_n}$, and transmit
\begin{equation}
z_{n,t}(x)=\sqrt{\frac{2}{\gamma}}\cos\!\bigl(\widetilde{\omega}_{n,t}^{\top}\mathbf w(x)+b_{n,t}\bigr).
\label{eq:worked_ann_encoder}
\end{equation}
User $k$ stacks all received scalars into
\[
\Phi_k(x)=\bigl(z_{n,t}(x)\bigr)_{(n,t):\,k\in\mathcal T_n}\in\mathbb R^{\mrecv_k(\mathcal T)},
\]
and applies the same ridge decoder as in \eqref{eq:worked_ridge_decoder}. We again use $M=3000$ training samples and an independent test set of size $M_{\mathrm{te}}=5000$.

Denoting by $\mathcal D_{\mathrm{ann,ex}}$ the encoder/decoder family used in this example, the relevant population quantity is the annealed risk
\begin{equation}
\overline{\mathcal R}(\mathcal D_{\mathrm{ann,ex}})
=
\mathbb E_{\mathsf A,\mathsf L}\!\left[\mathcal R\big(\mathcal D_{\mathrm{ann,ex}}\mid \mathsf A,\mathsf L\big)\right].
\label{eq:worked_ann_risk}
\end{equation}
Numerically, we estimate this quantity by Monte Carlo averaging over many independent draws of $(\mathsf A,\mathsf L)$, encoder randomness, and data realizations. Accordingly, this example is again fully instantiated: the only difference from the quenched case is that the topology itself is sampled rather than fixed.

\subsubsection{Why this example illustrates Theorem~\ref{thm:P2}}

This example complements the previous one by keeping the same dense task model while replacing the deterministic topology by a uniformly random assignment/link ensemble. Since every user depends on all $L=12$ coordinates of $\mathbf w(x)$, the annealed risk is influenced by two mechanisms simultaneously: the feature-budget effect, through the typical value $m_k^{\mathrm{avg}}=TN\delta=4T$, and the topology-induced coverage effect, through the probability that a random realization of $(\mathsf A,\mathsf L)$ fails to expose all essential coordinates to a given user. Thus, unlike the fixed quenched example where $\varepsilon_{\mathrm{cov},k}(\mathsf A,\mathsf L)=0$ by construction, the present annealed example incorporates coverage reliability directly into the averaged performance.

Specializing Theorem~\ref{thm:P2} to the present parameters yields the annealed upper-bound structure
\begin{equation}
\overline{\mathcal R}(\mathcal D_{\mathrm{ann,ex}})
\lesssim
\left(\frac{2}{\gamma}+C_1\right)\frac{B^2}{\gamma TN\delta}
+
C_2\frac{\sigma^2 d_\lambda}{M}
+
C_3B^2\lambda
+
r_{\mathrm{avg}}e^{-\gamma N\delta},
\label{eq:worked_ann_upper}
\end{equation}
and since $\gamma=1/3$, $N=8$, and $\delta=1/2$, the leading feature-budget term is again proportional to $3/(4T)$. However, in contrast to the quenched example, the additional exponential coverage term is now intrinsic to the model. On the converse side, Theorem~\ref{thm:P2} gives the lower-bound structure
\begin{equation}
\inf_{\mathcal D}\overline{\mathcal R}(\mathcal D)
\gtrsim
B^2\sum_{j>TN\delta}\lambda_j
\ \vee\
c'\,\mathbf 1\!\{\gamma N\delta\lesssim \log(r_{\mathrm{avg}})\},
\label{eq:worked_ann_lower}
\end{equation}
with $TN\delta=4T$ and $r_{\mathrm{avg}}=12$. Hence, in the present example, the theorem predicts that the average risk need not be purely spectral: depending on the coverage probability of the random topology ensemble, a non-negligible residual offset may remain. This is exactly the new structural ingredient that is absent from Theorem~\ref{thm:P1}. This is precisely the distinction between the quenched and annealed theorems that this example is meant to expose. In the fixed quenched example, the topology was chosen to remove the coverage obstruction entirely; here, the topology itself is random, so the average risk reflects both the spectral effect of increasing $T$ and the average reliability of the random assignment/link ensemble.

\subsubsection{Numerical validation and interpretation}

\begin{figure}[t]
\centering
\includegraphics[width=0.45\columnwidth]{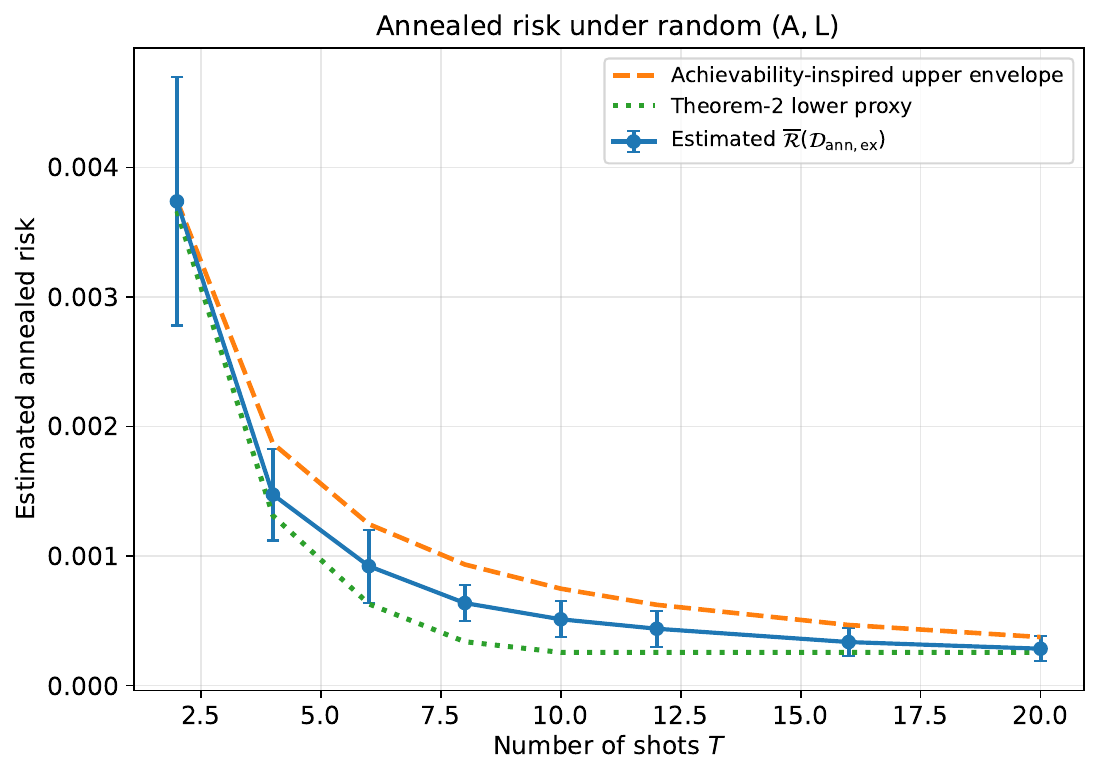}
\caption{Numerical validation of the annealed worked example for Theorem~\ref{thm:P2}. The markers show the estimated annealed risk $\overline{\mathcal R}(\mathcal D_{\mathrm{ann,ex}})$ as a function of the number of shots $T$ under uniformly random assignments and links. The dashed curve is a best-fit achievability-inspired upper envelope of the form $a\,3/(4T)+c$, while the dotted curve is a theorem-2-style lower proxy obtained as the maximum of a scaled spectral-tail term and a constant coverage-floor term.}
\label{fig:worked_example_result_ann}
\end{figure}

We validate this annealed example by estimating $\overline{\mathcal R}(\mathcal D_{\mathrm{ann,ex}})$ as a function of $T$ through Monte Carlo averaging over independent realizations of $(\mathsf A,\mathsf L)$. The numerical result is shown in Fig.~\ref{fig:worked_example_result_ann}. The markers display the empirical estimate of the annealed risk, the dashed curve is a best-fit achievability-inspired upper envelope of the form $a\,3/(4T)+c$, and the dotted curve is a theorem-2-style lower proxy obtained as the maximum of a scaled spectral-tail term and a constant coverage-floor term. Thus, as in the quenched case, the simulation directly visualizes the two structural terms appearing in the theorem statement.

The figure supports the interpretation of Theorem~\ref{thm:P2} in a direct way. First, the empirical annealed risk decreases with $T$, which is consistent with the growth of the average communicated-feature budget $m_k^{\mathrm{avg}}=TN\delta=4T$. Second, the decay is visibly less purely spectral than in the fixed quenched example, because the random topology introduces an averaged coverage effect that is absent in the deterministic full-coverage design. Third, the achievability-inspired upper envelope tracks the empirical curve reasonably well, while the theorem-2 lower proxy remains below it over the full sweep. Thus the observed behavior is consistent with the fact that, in the annealed regime, performance is governed jointly by the spectral tail and by the reliability of coordinate coverage under the random assignment/link ensemble. A useful complementary diagnostic in this regime is the empirical probability that a random realization of $(\mathsf A,\mathsf L)$ covers all essential coordinates of a user, namely
\[
\mathbb P\!\left(\bigcup_{n:\,k\in\mathcal T_n}\mathcal S_n=[L]\right),
\]
or more generally the average covered fraction of $[L]$ across users. Such a statistic directly quantifies the mechanism behind the exponential term $r_{\mathrm{avg}}e^{-\gamma N\delta}$ in \eqref{eq:worked_ann_upper}.

\begin{remark}[Implications of the two worked examples]
The two worked examples provide complementary interpretations of the quenched and annealed results. In both cases, the input distribution, the subfunctions, the user target functions, the topology model, and the encoder/decoder construction are fully specified, so the abstract GMUDC formulation is instantiated by concrete objects.

The fixed-topology example clarifies the role of Theorem~\ref{thm:P1}. Since the realized topology is chosen so that every user observes all essential coordinates, the coverage term vanishes and the performance is governed by the communicated-feature budget and the spectral tail alone. The example therefore isolates the instance-wise mechanism behind the quenched bound: once the topology provides complete coordinate coverage, increasing $T$ improves the risk primarily through the growth of $m_{\mathrm{harm}}$.

The random-topology example clarifies the additional content of Theorem~\ref{thm:P2}. Although the same problem parameters $(K,N,L,\Gamma,\Delta)$ are used and the typical communicated-feature budget remains $TN\delta=4T$, the topology is now random and the annealed risk reflects not only the spectral effect of increasing $T$ but also the reliability with which the random ensemble covers the essential coordinates. This is precisely the mechanism captured by the additional exponential coverage term in the annealed upper bound and by the possible coverage-floor contribution in the converse.

The comparison between the two examples also yields a concrete design insight. Having the same average feature budget does not by itself imply the same risk behavior: topology quality matters. A carefully engineered fixed topology can eliminate the coverage penalty entirely, whereas a uniformly random topology may still incur an averaged penalty even when the expected number of communicated features is the same. In this sense, the quenched and annealed examples together make explicit the spectral--coverage duality underlying Theorems~\ref{thm:P1} and~\ref{thm:P2}.

The two examples also make explicit what the GMUDC framework provides beyond classical linearly separable formulations. First, the user targets here are dense nonlinear Gaussian-RKHS functions of all subfunction coordinates, so the examples lie outside the linear or disjoint-support settings typically used in prior multi-user distributed computing models. Second, the same masked-RFF encoder and ridge-decoder architecture applies both to a fixed realized topology and to a random topology ensemble, thereby showing that the framework captures both instance-wise and average-case behavior within a single analytical model. Third, the examples separate two distinct sources of difficulty---spectral approximation and coordinate coverage---which are not explicitly disentangled in earlier formulations.

Finally, the two examples suggest a practical design principle. When the assignment and communication topology can be designed, one should favor overlap patterns that guarantee coverage of the essential coordinates of each user, thereby reducing performance to the feature-budget/spectral tradeoff of Theorem~\ref{thm:P1}. When the topology is random or time-varying, Theorem~\ref{thm:P2} shows that computation and communication resources must be provisioned not only to increase the communicated-feature budget but also to suppress topology-induced coverage failures.
\end{remark}

\section{Proof of Theorem~\ref{thm:P1} and Theorem~\ref{thm:P2}}
\label{sec:proof-P1}
We prove Theorem~\ref{thm:P1} in Subsection~\ref{proofthm:P1} and Theorem~\ref{thm:P2} in Subsection~\ref{proofthm:P2}.
\subsection{Proof of Theorem~\ref{thm:P1}}\label{proofthm:P1}
We prove the achievability and converse bounds in two parts. Throughout, $(\mathsf{A},\mathsf{L})$ is fixed (quenched), and the only randomness is from the encoder draws $\{(\omega_{n,t},b_{n,t})\}$ as specified in \eqref{eq:masked-rff}. For user $k$, let $\mrecv_k\triangleq \mrecv_k(\mathcal{T})$ denote the number of received scalars (features). Stack them into a feature map
\[
\Phi_k(x)\in\mathbb{R}^{\mrecv_k},\qquad
\Phi_k(x) = \big(z_{n,t}(x)\big)_{(n,t)\,:\,k\in\mathcal{T}_{n,t}}.
\]

Define the (population) feature covariance and the effective ridge dimension
\begin{equation}
\label{eq:Gk-dlambda}
G_k \ \triangleq\ \E_{x}\big[\Phi_k(x)\Phi_k(x)^\top\big]\in\R^{\mrecv_k\times \mrecv_k},
\qquad
d_{\lambda,k} =\ \mathrm{tr}\!\big(G_k(G_k+\lambda I)^{-1}\big)\ \in (0,\mrecv_k].
\end{equation}
We assume the standard kernel setup from the theorem statement: $F_k\in\mathcal{H}_K$ with $\|F_k\|_{\mathcal{H}_K}\le B$, where $K(.,.)$ is a shift-invariant kernel on $\R^L$ with eigenvalues $\{\lambda_j\}_{j\ge 1}$ (nonincreasing) with respect to the distribution of $\mathbf{w}(x)$. Because each feature is computed from a masked vector $(\omega_{n,t}\odot\mathbf{1}_{\mathcal{S}_n})$, we isolate the impact of masking via a single constant $C_\text{mask}\ge 1$. In particular, we have:

\begin{lemma}
\label{lem:mask-variance-fact}
Let $K(u,v)=\kappa(u-v)$ be a shift-invariant kernel on $\R^{L}$ with Bochner spectral measure \cite{ScholkopfSmola2002} $\mu$ satisfying
$\int \|\omega\|_2^2\,d\mu(\omega)<\infty$ and $|K(u,v)|\le 1$.
Fix a subset $\mathcal{S}\subseteq[L]$ with $|\mathcal{S}|=\Gamma$ and let $\gamma=\Gamma/L$.
Define the masked RFF
\[
\phi(x;\omega,b,\mathcal{S})
\;=\;
\sqrt{\tfrac{2}{\gamma}}\,
\cos\!\big((\omega\odot \mathbf{1}_{\mathcal{S}})^{\!\top}\mathbf{w}(x)+b\big),
\qquad \omega\sim\mu,\quad b\sim\mathrm{Unif}[0,2\pi].
\]
Assume $\|\mathbf{w}(x)\|_2\le R$ almost surely (or, more generally, $\mathbf{w}(x)$ is sub-Gaussian with parameter depending only on $R$).
Then there exists an absolute constant $C_{\mathrm{mask}}\!\ge\!1$ (depending only on $K$ and $R$, but not on $x,x',L,\Gamma$) and a bounded variance proxy $V_K(u,v)\!\le\!1$ for the unmasked RFF estimator, where
\begin{equation}
V_K(u,v)
\;\triangleq\;
\|K(\cdot,u)-K(\cdot,v)\|_{\mathcal{H}_K}^{2}
\;=\;
K(u,u)+K(v,v)-2\,K(u,v),
\end{equation}
such that, for all $x,x'$ with $u=\w(x)$ and $v=\w(x')$,
\begin{equation}
\label{eq:mask-variance-fact}
\mathrm{Var}\!\big[\phi(x;\omega,b,\mathcal{S})\,\phi(x';\omega,b,\mathcal{S})\big]
\ \le\  \frac{1}{2\gamma^2} + \frac{C_{\mathrm{mask}}}{\gamma}\; V_K(u,v).
\end{equation}
Moreover, if the masked frequency is sampled from the masked Bochner law
(i.e., draw $\omega_{\mathcal{S}}\sim \mu$ on $\R^{\Gamma}$ and embed as
$\widetilde\omega=(\omega_{\mathcal{S}},0_{\mathcal{S}^c})$), then \eqref{eq:mask-variance-fact} holds with $C_{\mathrm{mask}}=1$.
\end{lemma}

The proof of Lemma~\ref{lem:mask-variance-fact} is provided in Appendix~\ref{proof:mask-variance-fact}. Note that in the proof we assume that the random frequencies $\{(\omega_{n,t},b_{n,t})\}$ used by different servers and shots are drawn independently across $(n,t)$ and independently of the masking pattern $(\mathsf A,\mathsf L)$. 
Hence conditional on $(\mathsf A,\mathsf L)$, the masked feature vectors are independent across servers. 
This ensures that concentration inequalities for Monte-Carlo kernel estimates apply.

\subsubsection{Achievability of Theorem~\ref{thm:P1}}

The proof proceeds through three main steps: 
(i) establishing a concentration bound for the kernel approximation using masked random Fourier features (Lemma~\ref{lem:masked-rff}); 
(ii) deriving a general excess-risk decomposition for ridge regression with random features (Lemma~\ref{lem:ridge-decomp}); 
and (iii) combining these results to obtain a uniform high-probability bound averaged over users.

\begin{lemma}
\label{lem:masked-rff}
Let $\widetilde{K}_k(x,x') \triangleq \frac{1}{m_k}\Phi_k(x)^\top \Phi_k(x')$ be the Monte Carlo kernel built from the $m_k$ masked-RFF features received by user $k$. Then, for any $\delta'\in(0,1)$, with probability at least $1-\delta'$ over the encoder randomness,
\begin{equation}
\label{eq:masked-approx}
\E_{x,x'}\big[\big(\widetilde{K}_k(x,x')-K(\mathbf{w}(x),\mathbf{w}(x'))\big)^2\big]
\ \le \
\!\left(\frac{2}{\gamma^{2} m_k}+\frac{C_1'}{ \gamma m_k}\right)\!\log\!\frac{2}{\delta'}\,,
\end{equation}
for an absolute constant $C_1'>0$ depending only on the kernel family (via $V_K$) and $C_\text{mask}$ in Lemma~\ref{lem:mask-variance-fact}.
\end{lemma}
The proof of Lemma~\ref{lem:masked-rff} is provided in Appendix~\ref{proof:masked-rff}. Lemma~\ref{lem:masked-rff} establishes that the masked-RFF approximation of the kernel $K$ converges in mean square at rate $1/(\gamma m_k)$, up to logarithmic factors, where $\gamma=\Gamma/L$ captures the fraction of coordinates each encoder observes. Building on this approximation result, the next lemma provides a general bound on the prediction risk of the ridge estimator in terms of its bias and variance components, under an arbitrary feature map~$\Phi_k$.

\begin{lemma}
\label{lem:ridge-decomp}
Let $\widehat{F}_k$ be the ridge predictor \eqref{eq:ridge} trained with $M$ samples and feature map $\Phi_k$. Then, for any $\lambda>0$,
\begin{equation}
\label{eq:ridge-decomp}
\E_{x}\big[\|\widehat{F}_k(x)-F_k(x)\|_2^2\big]
\ \le\
{\inf_{\beta\in\R^{m_k}}\ \E_{x}\big[(\beta^\top\Phi_k(x)-F_k(x))^2\big]}
\ +\ 
C_2'\,\frac{\sigma^2\,d_{\lambda,k}}{M}
	\ +\ 
C_3'\,B^2\,\lambda,
\end{equation}
where $d_{\lambda,k}$ is given in \eqref{eq:Gk-dlambda}, and $C_2',C_3'>0$ are absolute constants.
\end{lemma}
The proof of Lemma~\ref{lem:ridge-decomp} is provided in Appendix~\ref{proof:ridge-decomp}. Before proceeding, we recall that the masked Random Fourier Feature
approximation affects the ridge bias through the kernel operator perturbation.
Following the standard analysis of kernel ridge regression
(see, e.g.,~\cite{rudi2017generalization}),
if $\widehat{G}$ denotes the empirical (masked) kernel matrix and
$\|\widehat{G}-G\|_{\mathrm{op}}\le\varepsilon_{\mathrm{RFF}}$, then
\[
\|(\widehat{G}+\lambda I)^{-1}g-(G+\lambda I)^{-1}g\|
=O\!\left(\frac{\varepsilon_{\mathrm{RFF}}}{\lambda}\right),
\]
which implies that the ridge estimator inherits an additive bias term
of order $O(\varepsilon_{\mathrm{RFF}}/\lambda)$.
Hence, the population risk bound of
Lemma~\ref{lem:ridge-decomp} remains valid up to this small correction,
justifying the substitution of the masked kernel approximation
into the subsequent risk decomposition
\subsection*{Combining Lemmas \ref{lem:masked-rff} and \ref{lem:ridge-decomp}: from kernel error to prediction risk}

We now make precise why, when $F_k\in\mathcal{H}_K$ with $\|F_k\|_{\mathcal{H}_K}\le B$, the best linear predictor in the masked-RFF feature space has approximation error controlled by the $L_2(\mathbb{P}_X)$ kernel mismatch between $K$ and $\widetilde K_k$. Let $\rho$ be the distribution of $U=\mathbf{w}(X)\in\R^L$ induced by $X\sim\mathbb{P}_X$.
For a (symmetric, PSD) kernel $Q:\R^L\times\R^L\to\R$, define the integral operator
\[
(T_Q f)(u) \;=\; \int Q(u,v)\, f(v)\, d\rho(v),\qquad f\in L_2(\rho).
\]
We write $T_K$ and $T_{\widetilde K_k}$ for the operators with kernels $K$ and $\widetilde K_k$, respectively. Their Hilbert–Schmidt (HS) norms satisfy
\begin{align}
\label{eq:HS-squared}
\|T_K - T_{\widetilde K_k}\|_{\mathrm{HS}}^2
\;&=\;\iint \big(K(u,v)-\widetilde K_k(u,v)\big)^2\, d\rho(u)\, d\rho(v)
\;\\&=\; \E_{x,x'}\!\left[\big(K(\w(x),\w(x'))-\widetilde{K}_k(x,x')\big)^2\right].
\end{align}

Let $\mathcal{H}_K$ be the RKHS of $K$. The canonical inclusion $\mathcal{H}_K\hookrightarrow L_2(\rho)$ is continuous and satisfies the isometry identity
\begin{equation}
\label{eq:isometry}
\langle f, g \rangle_{\mathcal{H}_K}
\;=\;
\big\langle T_K^{-1/2} f,\ T_K^{-1/2} g \big\rangle_{L_2(\rho)}\quad
\text{for all } f,g\in\mathcal{H}_K.
\end{equation}
In particular, $\|f\|_{\mathcal{H}_K}^2 = \|T_K^{-1/2} f\|_{L_2(\rho)}^2$ and
$f(u)=\langle f, K(u,\cdot)\rangle_{\mathcal{H}_K}$.

The (masked-RFF) feature map collected at user $k$ induces an approximate kernel
\[
\widetilde K_k(u,v) \;=\; \frac{1}{m_k}\,\Phi_k(u)^\top \Phi_k(v),
\]
which is PSD. Its RKHS $\mathcal{H}_{\widetilde K_k}$ equals the closure of the linear span of $\{\widetilde K_k(\cdot, v): v\in\R^L\}$ with inner product determined by $\widetilde K_k$.
Every linear predictor in the feature space is a function $g\in\mathcal{H}_{\widetilde K_k}$ of the form
\[
g(u)=\beta^\top \Phi_k(u)\quad\Longleftrightarrow\quad
g(\cdot)\in\mathcal{H}_{\widetilde K_k},\ \ \|g\|_{\mathcal{H}_{\widetilde K_k}}=\|\beta\|_2.
\]
Hence
\begin{equation}
\label{eq:best-approx}
\inf_{\beta\in\R^{m_k}} \E\big[(\beta^\top \Phi_k(X) - F_k(X))^2\big]
\;=\;
\inf_{g\in\mathcal{H}_{\widetilde K_k}} \|g - F_k\|_{L_2(\rho)}^2.
\end{equation}

Let $P_{\widetilde K_k}:L_2(\rho)\to \overline{\mathcal{H}_{\widetilde K_k}}$ be the $L_2(\rho)$-orthogonal projection onto the closure of $\mathcal{H}_{\widetilde K_k}$ in $L_2(\rho)$.
Then the optimal $g^\star=P_{\widetilde K_k} F_k$ and the approximation error is
\[
\inf_{g\in\mathcal{H}_{\widetilde K_k}} \|g-F_k\|_{L_2(\rho)}^2
\;=\;
\|(I-P_{\widetilde K_k})F_k\|_{L_2(\rho)}^2.
\]
A standard comparison yields, for all $f\in\mathcal{H}_K$\footnote{One convenient derivation works directly in $L_2(\rho)$. Let $V\equiv\overline{\mathcal{H}_{\widetilde K_k}}$ and $P\equiv P_{\widetilde K_k}$ be the $L_2(\rho)$-orthogonal projector onto $V$. For any $f\in\mathcal{H}_K$, one has 
$\|(I-P)f\|_{L_2}^2=\sup_{\|h\|_{L_2}\le 1,\,h\perp V}\langle (I-P)f,h\rangle_{L_2}^2=\sup_{\|h\|_{L_2}\le 1,\,h\perp V}\langle f,h\rangle_{L_2}^2$. 
Write $T_K$ and $T_{\widetilde K}$ for the Mercer operators of $K$ and $\widetilde K$, and use the RKHS isometry $\langle f,g\rangle_{\mathcal{H}_K}=\langle T_K^{-1/2}f,\,T_K^{-1/2}g\rangle_{L_2}$. Then $\langle f,h\rangle_{L_2}=\langle T_K^{-1/2}f,\,T_K^{1/2}h\rangle_{L_2}\le \|f\|_{\mathcal{H}_K}\,\\|T_K^{1/2}h\|_{L_2}$. 
If $h\perp V=\overline{\mathrm{Range}(J_{\widetilde K})}$, then $J_{\widetilde K}^*h=0$, hence $\langle h,T_{\widetilde K}h\rangle=\|J_{\widetilde K}^*h\|_{\mathcal{H}_{\widetilde K}}^2=0$ and therefore $\|T_K^{1/2}h\|_{L_2}^2=\langle h,T_K h\rangle=\langle h,(T_K-T_{\widetilde K})h\rangle\le \|T_K-T_{\widetilde K}\|_{\mathrm{op}}\,\|h\|_{L_2}^2$. 
 Combining these and taking the supremum over such $h$ gives $\|(I-P)f\|_{L_2}^2\le \|f\|_{\mathcal{H}_K}^2\,\|T_K-T_{\widetilde K}\|_{\mathrm{op}}\le \|f\|_{\mathcal{H}_K}^2\,\|T_K-T_{\widetilde K}\|_{\mathrm{HS}}$, where the last inequality uses $\|\cdot\|_{\mathrm{op}}\le \|\cdot\|_{\mathrm{HS}}$ for Hilbert–Schmidt operators. This is the claimed comparison bound.},
\begin{equation}
\label{eq:proj-operator}
\|(I-P_{\widetilde K_k}) f\|_{L_2(\rho)}^2
\;\le\;
\|f\|_{\mathcal{H}_K}^2\ \|T_K - T_{\widetilde K_k}\|_{\mathrm{HS}}.
\end{equation}
Let $\{(\lambda_j,\varphi_j)\}_{j\ge 1}$ be the Mercer eigen-system of $T_K$ on $L_2(\rho)$.
Any $f\in\mathcal{H}_K$ can be written as $f=\sum_j a_j \varphi_j$ with $\sum_j a_j^2/\lambda_j<\infty$ and
$\|f\|_{\mathcal{H}_K}^2=\sum_j a_j^2/\lambda_j$.
Let $T_{\Delta}=T_K-T_{\widetilde K_k}$.
Then, by orthogonality and Cauchy–Schwarz in the HS inner product,
\[
\|(I-P_{\widetilde K_k}) f\|_{L_2(\rho)}^2
=\ \sup_{\|h\|_{L_2}\le 1}\ \big\langle (I-P_{\widetilde K_k}) f,\ h \big\rangle_{L_2}^2
\ \le\ \big\langle f,\ T_{\Delta} f \big\rangle_{L_2}
\ \le\ \|f\|_{\mathcal{H}_K}^2 \ \|T_{\Delta}\|_{\mathrm{HS}},
\]
where in the last step we used the isometry \eqref{eq:isometry} by writing $f=T_K^{1/2}h$ with $\|h\|_{L_2}=\|f\|_{\mathcal{H}_K}$ and bounding
$\langle f,T_{\Delta}f\rangle=\langle h,\,T_K^{1/2}T_{\Delta}T_K^{1/2}h\rangle
\le \|T_K\|_{\mathrm{op}}\ \|T_{\Delta}\|_{\mathrm{HS}}\ \|h\|_{L_2}^2$.
A rigorous proof can be given by truncating the Mercer expansion and passing to the limit; we omit further details for brevity.\footnote{For a rigorous justification of the step $\langle f,T_{\Delta}f\rangle \le \|f\|_{\mathcal{H}_K}^2\,\|T_{\Delta}\|_{\mathrm{HS}}$, expand $f\in\mathcal{H}_K$ in the Mercer eigenbasis of $T_K$. Write $f=\sum_{j=1}^\infty a_j\varphi_j$ where $\{\varphi_j\}$ are the eigenfunctions of $T_K$ with eigenvalues $\{\lambda_j\}$, so that $\|f\|_{\mathcal{H}_K}^2=\sum_j a_j^2/\lambda_j$ and $\|f\|_{L_2(\rho)}^2=\sum_j a_j^2$. Then $\langle f,T_\Delta f\rangle=\sum_{i,j} a_i a_j \langle \varphi_i, T_\Delta \varphi_j\rangle$. By Cauchy--Schwarz in the Hilbert--Schmidt inner product, $|\langle f,T_\Delta f\rangle|\le \|f\otimes f\|_{\mathrm{HS}}\,\|T_\Delta\|_{\mathrm{HS}}$. But $\|f\otimes f\|_{\mathrm{HS}}=\|f\|_{L_2}^2=\sum_j a_j^2\le \|f\|_{\mathcal{H}_K}^2\,\max_j \lambda_j\le \|T_K\|_{\mathrm{op}}\,\|f\|_{\mathcal{H}_K}^2$. Hence $|\langle f,T_\Delta f\rangle|\le \|T_K\|_{\mathrm{op}}\,\|f\|_{\mathcal{H}_K}^2\,\|T_\Delta\|_{\mathrm{HS}}$, which is the claimed bound. A more careful argument can be given by truncating the Mercer series and then passing to the limit, but the inequality follows directly from Hilbert--Schmidt duality.}

Combining \eqref{eq:best-approx} and \eqref{eq:proj-operator} with $f=F_k$ and $\|F_k\|_{\mathcal{H}_K}\le B$,
\begin{equation}
\label{eq:approx-vs-kernel}
\inf_{\beta}\E\big[(\beta^\top\Phi_k(X)-F_k(X))^2\big]
\;=\;
\inf_{g\in\mathcal{H}_{\widetilde K_k}}\|g-F_k\|_{L_2(\rho)}^2
\;\le\;
B^2\ \|T_K - T_{\widetilde K_k}\|_{\mathrm{HS}}.
\end{equation}
Using \eqref{eq:HS-squared} and the identity for HS norm,
\begin{equation}
\label{eq:HS-to-L2kernel}
\|T_K - T_{\widetilde K_k}\|_{\mathrm{HS}}
\;=\;\left(\E_{x,x'}\big[(K(\w(x),\w(x'))-\widetilde K_k(x,x'))^2\big]\right)^{1/2}.
\end{equation}
Squaring both sides of \eqref{eq:approx-vs-kernel} and \eqref{eq:HS-to-L2kernel} gives the central inequality:
\begin{equation}
\label{eq:approx-error-final}
\inf_{\beta}\E\big[(\beta^\top\Phi_k(X)-F_k(X))^2\big]
\ \le\
B^2\cdot \E_{x,x'}\big[(\widetilde{K}_k(x,x')-K(\mathbf{w}(x),\mathbf{w}(x')))^2\big].
\end{equation}

By Lemma~\ref{lem:masked-rff}, with probability at least $1-\delta'$ over encoder draws,
\[
\E_{x,x'}\big[(\widetilde{K}_k(x,x')-K(\mathbf{w}(x),\mathbf{w}(x')))^2\big]
\ \le\
\left(\frac{2}{\gamma^{}}+C'_1\right)\frac{1}{\gamma\,m_k}\,\log\!\frac{2}{\delta'}.
\]
Plugging this into \eqref{eq:approx-error-final},
\[
\inf_{\beta}\E\big[(\beta^\top\Phi_k(X)-F_k(X))^2\big]
\ \le\
\left(\frac{2}{\gamma^{}}+C'_1\right)\frac{B^2}{\gamma \,m_k}\,\log\!\frac{2}{\delta'}.
\]

Finally, substituting this bound into Lemma~\ref{lem:ridge-decomp} yields, for the same event,
\[
\E_{X}\big[\|\widehat{F}_k(X)-F_k(X)\|_2^2\big]
\ \le\ \left(\frac{2}{\gamma}+C'_1\right)
\frac{ B^2}{\gamma \,m_k}\,\log\!\frac{2}{\delta'}
\ +\ C_2'\,\frac{\sigma^2 d_{\lambda,k}}{M}
\ +\ C_3'\,B^2\lambda.
\]

Averaging this inequality over all $k\in[K]$, we obtain\footnote{Averaging the per-user risks yields the harmonic mean since each individual term scales inversely with the available feature
count $m_k$. This follows from Jensen’s inequality applied to the convex
function $x\mapsto 1/x$, ensuring that the averaged risk over heterogeneous
users depends on $m_{\mathrm{harm}}$ rather than the arithmetic mean
$m_{\mathrm{avg}}$.}
\[
\frac{1}{K}\sum_{k=1}^K 
\E_{X}\big[\|\widehat{F}_k(X)-F_k(X)\|_2^2\big]
\ \le\ \left(\frac{2}{\gamma}+C'_1\right)
\frac{B^2}{\gamma}\left(\frac{1}{K}\sum_{k=1}^K \frac{1}{m_k}\right)\!
\log\!\frac{2}{\delta'}
\ +\ C_2'\,\frac{\sigma^2 d_\lambda}{M}
\ +\ C_3'\,B^2\lambda.
\]

Recalling the definition of the quenched risk
\[
\mathcal{R}\big(\mathcal{D}\mid \mathsf{A},\mathsf{L}\big)
=\E_{X}\!\left[\frac{1}{K}\sum_{k=1}^K \|\widehat{F}_k(X)-F_k(X)\|_2^2\right],
\]
and denoting by
\[
m_{\mathrm{harm}}\;\triangleq\;
\Big(\frac{1}{K}\sum_{k=1}^K \frac{1}{m_k}\Big)^{-1}
\]
the harmonic mean of the user budgets, we conclude that on the same event
\[
\mathcal{R}\big(\mathcal{D}\mid \mathsf{A},\mathsf{L}\big)
\ \le\
\left(\frac{2}{\gamma}+C'_1\right)\,
\frac{B^2}{\gamma\,m_{\mathrm{harm}}}\,
\log\!\frac{2}{\delta'}
\ +\ C_2'\,\frac{\sigma^2 d_\lambda}{M}
\ +\ C_3'\,B^2\lambda,
\]
which is exactly the bound~\eqref{eq:P1-ach}\footnote{Strictly speaking, the concentration lemma can be stated either 
(i) uniformly over all $k\in[K]$, in which case the above bound holds on a 
single high-probability event without any union bound, or 
(ii) per-user, in which case one may either absorb the $\log K$ factor 
into the constants by a union bound, or argue directly on the average 
risk using Bernstein’s inequality. In both cases the resulting rate 
remains unchanged, and the bound in Theorem~\ref{thm:P1} follows.}
.
\hfill\qed

\subsubsection{Converse of Theorem~\ref{thm:P1}}
We prove the two terms in \eqref{eq:P1-conv} separately.

\begin{lemma}
\label{lem:eigen-tail}
Fix $m\in\mathbb{N}$. For any measurable scheme (encoders/decoders) that delivers at most $m$ real scalars to user $k$ per input, we have
\[
\inf_{\text{schemes with $\le m$ scalars}}
\E_{x}\big[\|\widehat{F}_k(x)-F_k(x)\|_2^2\big]
\ \ge\ 
B^2\sum_{j>m}\lambda_j.
\]
\end{lemma}
The proof of Lemma~\ref{lem:eigen-tail} is provided in Appendix~\ref{proof:eigen-tail}. 
Applying Lemma~\ref{lem:eigen-tail} with $m=m_k(\mathcal{T})$ gives the eigen-tail term in \eqref{eq:P1-conv}.

\begin{lemma}
\label{lem:coverage}
Suppose there exists a coordinate set $\mathcal{S}_k^\star\subseteq[L]$ on which $F_k$ essentially depends, and that for the realized $(\mathsf{A},\mathsf{L})$ there exists $\ell^\star\in\mathcal{S}_k^\star$ such that $\{n:\ \ell^\star\in\mathcal{S}_n,\ k\in\mathcal{T}_n\}=\emptyset$. Then there exists a constant $\varepsilon_{\mathrm{cov},k}(\mathsf{A},\mathsf{L})>0$ such that for any scheme,
\[
\E_{x}\big[\|\widehat{F}_k(x)-F_k(x)\|_2^2\big]\ \ge\ \varepsilon_{\mathrm{cov},k}(\mathsf{A},\mathsf{L}).
\]
\end{lemma}
The proof of Lemma~\ref{lem:coverage} is provided in Appendix~\ref{proof:coverage}. By Lemma~\ref{lem:eigen-tail} and Lemma~\ref{lem:coverage}, for any design that delivers $m_k(\mathcal{T})$
real scalars to user $k$, we have the per-user lower bound
\begin{equation}\label{eq:per-user-spectral}
\E_X\!\big[\|\widehat{F}_k(X)-F_k(X)\|_2^2\big]
\;\ge\;
B^2 \sum_{j> m_k(\mathcal{T})} \lambda_j \ \vee\ \varepsilon_{\mathrm{cov},k},
\end{equation}
Averaging \eqref{eq:per-user-spectral}
over $k\in[K]$ yields the quenched average-risk lower bound as stated in Theorem~\ref{thm:P1}.
\hfill\qed

\subsection{Proof of Theorem~\ref{thm:P2}}\label{proofthm:P2}
We prove the annealed \emph{upper} and \emph{lower} bounds separately. Throughout, we use the assumptions of Theorem~\ref{thm:P1} and average over the random computation assignments $\mathsf{A}$ and link patterns $\mathsf{L}$ (both independent of the data and of the encoder randomness). For a fixed user $k$, recall
\[
\mrecv_k \;=\; \mrecv_k(\mathcal{T}) \;=\; T\,\big|\{\,n\in[N]:\,k\in \mathcal{T}_n\,\}\big|,
\qquad
\mrecv_k^{\mathrm{avg}} \;=\; \E_{\mathsf{A},\mathsf{L}}[\mrecv_k(\mathcal{T})] \;=\; T\,N\,\delta,
\qquad
r_{\mathrm{avg}} \;=\; \frac{1}{K}\sum_{k=1}^K r_k.
\]
In the quenched analysis (Theorem~\ref{thm:P1}), $\mrecv_k$ is treated as a
\emph{deterministic} quantity determined by a fixed topology
$(\mathsf{A},\mathsf{L})$.
In contrast, in the annealed setting of Theorem~\ref{thm:P2},
$\mrecv_k$ becomes a \emph{random variable} through the random link
assignments $\mathsf{L}$ and computation patterns $\mathsf{A}$.
Hence, expectations such as
$\E_{\mathsf{A},\mathsf{L}}[\mrecv_k]$ or
$\E_{\mathsf{A},\mathsf{L}}[1/\mrecv_k]$ are taken over these random
topologies, while the quenched risk $\mathcal{R}(\mathcal{D}\mid
\mathsf{A},\mathsf{L})$ conditions on a fixed realization.
This distinction ensures consistency between the two analyses.
\paragraph{Part I (Achievability)}
Fix a realization $(\mathsf{A},\mathsf{L})$. By Theorem~\ref{thm:P1}, with probability at least $1-\delta_0$ over the encoder draws,
\begin{equation}
\label{eq:avg-risk-fixed-topo}
\mathcal{R}\big(\mathcal{D}\mid \mathsf{A},\mathsf{L}\big)
\ \le\
\left(\frac{2}{\gamma}+C_1\right)\,
\frac{B^2}{\gamma}\left(\frac{1}{K}\sum_{k=1}^K \frac{1}{m_k}\right)\!
\log\!\frac{2}{\delta_0}
\;+\;
C_2\,\frac{\sigma^2\,d_\lambda}{M}
\;+\;
C_3\,B^2\,\lambda.
\end{equation}
For each $k$, let
\[
\mathrm{Bad}_k(\varepsilon)\;\triangleq\;\Big\{\,|\{n:\,k\in\mathcal{T}_n\}| < (1-\varepsilon)\,N\delta\,\Big\},
\qquad 0<\varepsilon<1.
\]
Since $|\{n:\,k\in\mathcal{T}_n\}|\sim \mathrm{Binomial}(N,\delta)$, Chernoff’s bound gives
\begin{equation}
\label{eq:chernoff-single}
\Pr\!\big(\mathrm{Bad}_k(\varepsilon)\big)\ \le\ \exp\!\big(-c\,\varepsilon^2 N\delta\big).
\end{equation}
On $\mathrm{Bad}_k(\varepsilon)^c$ we have $\mrecv_k\ge T(1-\varepsilon)N\delta$, hence
\[
\frac{1}{\mrecv_k}
\;=\;
\frac{1}{\mrecv_k}\,\mathbf{1}_{\mathrm{Bad}_k(\varepsilon)^c}
\;+\;
\frac{1}{\mrecv_k}\,\mathbf{1}_{\mathrm{Bad}_k(\varepsilon)}
\ \le\
\frac{1}{T(1-\varepsilon)N\delta}\,\mathbf{1}_{\mathrm{Bad}_k(\varepsilon)^c}
\;+\;
\frac{1}{T}\,\mathbf{1}_{\mathrm{Bad}_k(\varepsilon)}.
\]
Taking $\E_{\mathsf{A},\mathsf{L}}$ and using \eqref{eq:chernoff-single},
\begin{equation}
\label{eq:recip-expectation}
\E_{\mathsf{A},\mathsf{L}}\!\left[\frac{1}{\mrecv_k}\right]
\ \le\
\frac{1}{T(1-\varepsilon)N\delta}
\;+\;
\frac{1}{T}\,e^{-c\,\varepsilon^2 N\delta}.
\end{equation}
Averaging \eqref{eq:recip-expectation} over $k$ and using linearity,
\begin{equation}
\label{eq:avg-recip-bound}
\E_{\mathsf{A},\mathsf{L}}\!\left[\frac{1}{K}\sum_{k=1}^K \frac{1}{\mrecv_k}\right]
\ \le\
\frac{1}{T(1-\varepsilon)N\delta}
\;+\;
\frac{1}{T}\,e^{-c\,\varepsilon^2 N\delta}.
\end{equation}
Taking $\E_{\mathsf{A},\mathsf{L}}$ on both sides of \eqref{eq:avg-risk-fixed-topo} and using \eqref{eq:avg-recip-bound} (and absorbing the fixed $\log(2/\delta_0)$ into constants) gives
\begin{equation}
\label{eq:ann-upper-degree-core}
\E_{\mathsf{A},\mathsf{L}}\!\left[\mathcal{R}\big(\mathcal{D}\mid \mathsf{A},\mathsf{L}\big)\right]
\ \le\
\left(\frac{2}{\gamma}+C_1\right)\frac{B^2}{\gamma\,T(1-\varepsilon)N\delta}
\;+\;
C_2\,\frac{\sigma^2\,d_\lambda}{M}
\;+\;
C_3\,B^2\,\lambda
\;+\;
\left(\frac{2}{\gamma}+C_1\right)\frac{B^2}{\gamma\,T}\,e^{-c\,\varepsilon^2 N\delta}.
\end{equation}

For user $k$, if $\mathcal{S}_k^\star$ is an essential set of size $r_k$, then the probability that a fixed coordinate in $\mathcal{S}_k^\star$ is never both computed and linked across the $N$ servers is at most $(1-\gamma\delta)^N \le e^{-\gamma N\delta}$. A union bound over $\mathcal{S}_k^\star$ yields
\[
\Pr\big(\text{coverage miss for user }k\big)\ \le\ r_k\,e^{-\gamma N\delta}.
\]
Averaging over users,
\[
\frac{1}{K}\sum_{k=1}^K \Pr\big(\text{coverage miss for user }k\big)
\ \le\ r_{\mathrm{avg}}\,e^{-\gamma N\delta}.
\]
By Lemma~\ref{lem:coverage}, the contribution of these miss events to the annealed risk is bounded (up to absolute constants) by $r_{\mathrm{avg}} e^{-\gamma N\delta}$.\footnote{This uses per-server masks/links fixed across the $T$ shots. If masks are independently redrawn each shot, the miss probability becomes $(1-\gamma\delta)^{NT}\le e^{-\gamma N\delta\,T}$, further strengthening the bound.}
Adding this term to \eqref{eq:ann-upper-degree-core} and absorbing the exponentially small $\frac{B^2}{\gamma T}e^{-c\varepsilon^2 N\delta}$ term into $C_\star e^{-c\varepsilon^2 N\delta}$ yields the annealed upper bound \eqref{eq:ann-upper-degree-corr}.

\paragraph{Part II (Converse)}
Conditionally on $(\mathsf{A},\mathsf{L})$, Lemma~\ref{lem:eigen-tail} (per-user spectral tail) gives
\[
\inf_{\text{schemes with }\le m_k\text{ scalars}}
\E_{x}\!\big[\|\widehat{F}_k(x)-F_k(x)\|_2^2\big]
\ \ge\ 
B^2\sum_{j>m_k}\lambda_j.
\]
Taking $\E_{\mathsf{A},\mathsf{L}}$ and applying Jensen to $f(m)\triangleq\sum_{j>m}\lambda_j$ (convex on $\mathbb{N}$) yields
\[
\inf_{\mathcal{D}}
\E_{\mathsf{A},\mathsf{L}}\E_x\!\big[\|\widehat{F}_k(x)-F_k(x)\|_2^2\big]
\ \ge\
B^2\,\sum_{j>m_k^{\mathrm{avg}}}\lambda_j,
\qquad m_k^{\mathrm{avg}}=TN\delta.
\]
For coverage, if $\gamma N\delta \lesssim \log(r_{\mathrm{avg}})$, the average miss probability is bounded away from zero, and Lemma~\ref{lem:coverage} implies a constant floor $c'>0$ in the annealed risk, producing the  indicator term in the theorem statement. Combining the spectral tail with the coverage threshold proves the annealed lower bound. This completes the proof.
\hfill\qed

\section{Proof of Lemma~\ref{lem:mask-variance-fact}}\label{proof:mask-variance-fact}
\begin{proof}
Let $u=\mathbf{w}(x)$ and $v=\mathbf{w}(x')$, and fix the mask $\mathcal{S}\subseteq[L]$ with $|\mathcal{S}|=\Gamma$ and $\gamma=\Gamma/L$.
Write $\widetilde\omega=\omega\odot\mathbf{1}_{\mathcal{S}}$ and
\[
\phi_u \;\triangleq\; \phi(x;\omega,b,\mathcal{S})
= \sqrt{\frac{2}{\gamma}}\cos\big(\langle \widetilde\omega,u\rangle+b\big),
\qquad
\phi_v \;\triangleq\; \phi(x';\omega,b,\mathcal{S})
= \sqrt{\frac{2}{\gamma}}\cos\big(\langle \widetilde\omega,v\rangle+b\big).
\]
Set the single-atom kernel estimator \(Z \triangleq \phi_u\,\phi_v\).
We assume $|K|\le 1$ and $\|\mathbf{w}(x)\|_2\le R$ a.s.
Let $S \triangleq \langle \widetilde\omega,u-v\rangle$.
Using the trigonometric identity
\(
\cos(\alpha+b)\cos(\beta+b)
= \tfrac{1}{2}\big[\cos(\alpha-\beta)+\cos(2b+\alpha+\beta)\big]
\)
and $\E_b[\cos(2b+\cdot)]=0$, we get
$
\E_b[Z\mid\omega] \;=\; \frac{1}{\gamma}\cos S.
$
If $\widetilde\omega$ is sampled from the masked Bochner law, i.e., draw $\omega_{\mathcal{S}}\sim\mu$ on $\R^\Gamma$ and embed $(\omega_{\mathcal{S}},0_{\mathcal{S}^c})$, then Bochner’s identity yields
$
\E_{\omega,b}[Z] \;=\; \frac{1}{\gamma}\,\E_{\widetilde\omega}[\cos S] \;=\; K(u,v).
$
By the law of total variance,
$
\mathrm{Var}(Z) \;=\; \E\!\big[\mathrm{Var}_b(Z\mid\omega)\big]\;+\;\mathrm{Var}\!\big(\E_b[Z\mid\omega]\big)
\;=\; \E\!\big[\mathrm{Var}_b(Z\mid\omega)\big]\;+\;\frac{1}{\gamma^2}\mathrm{Var}\big(\cos S\big).
$
A direct computation shows \(\E[\mathrm{Var}_b(Z\mid\omega)] \le \frac{1}{2\gamma^2}\) (indeed \(|\phi_u\phi_v|\le 2/\gamma\) and the oscillatory term contributes at most \(1/(2\gamma^2)\)). Thus
\begin{equation}\label{eq:var-split}
\mathrm{Var}(Z) \;\le\; \frac{1}{2\gamma^2} \;+\; \frac{1}{\gamma^2}\,\mathrm{Var}(\cos S).
\end{equation}
The cosine is $1$-Lipschitz, hence for any random $X$,
\(
\mathrm{Var}(\cos X)\le \E\big[(\cos X - \E\cos X)^2\big]\le \E[(\cos X - \cos 0)^2]\le \E[X^2].
\)
Applying this with $X=S$,
\[
\mathrm{Var}(\cos S)\ \le\ \E[S^2]
\ =\ (u-v)^\top \E[\widetilde\omega\widetilde\omega^\top]\,(u-v)
\ = (u-v)^\top \Sigma_{\mathcal{S}} (u-v).
\]
Let $\Sigma\triangleq \E[\omega\omega^\top]$ and note $\Sigma_{\mathcal{S}}\preceq \Sigma$ (PSD order).
Because inputs live in the compact ball $\{u:\|u\|_2\le R\}$, there exists a constant $C_{K,R}\in(0,\infty)$ such that, for all $u,v$,
\begin{equation}\label{eq:curv-dom}
(u-v)^\top \Sigma (u-v)
\ \le\ C_{K,R}\,\big(K(u,u)+K(v,v)-2K(u,v)\big)
\ =\ C_{K,R}\,V_K(u,v).\footnote{Reason: write $\varphi(h)=\E[e^{i\langle \omega,h\rangle}]$ so that $K(u,v)=\varphi(u-v)$ with $\varphi(0)=1$. Since $\int\|\omega\|_2^2\,d\mu(\omega)<\infty$, $\varphi$ is $C^2$ near $0$ with $\nabla\varphi(0)=0$ and $\nabla^2\varphi(0)=-\Sigma$. Hence $1-\varphi(h)=\tfrac{1}{2}h^\top\Sigma h+o(\|h\|^2)$ as $h\to0$, which implies $\lim_{h\to0}\tfrac{h^\top\Sigma h}{1-\varphi(h)}=2$. On the compact set $\{h:\|h\|\le 2R\}$, the function $g(h)=\tfrac{h^\top\Sigma h}{1-\varphi(h)}$ is continuous and attains a finite maximum $C_{K,R}$. Substituting $h=u-v$ gives \eqref{eq:curv-dom}.}
\end{equation}
Therefore,
\[
\E[S^2]\ =\ (u-v)^\top \Sigma_{\mathcal{S}} (u-v)
\ \le\ (u-v)^\top \Sigma (u-v)
\ \le\ C_{K,R}\,V_K(u,v).
\]

Plug the bound on $\mathrm{Var}(\cos S)$ into \eqref{eq:var-split}:
\begin{align}
    \mathrm{Var}(Z)\ \le\ \frac{1}{2\gamma^2} \;+\; \frac{C_{K,R}}{\gamma^2}\,V_K(u,v). \label{Step4}
\end{align}
Finally, the masked-feature normalization $\sqrt{2/\gamma}$ was chosen so that the single-atom variance inflates by a factor \(1/\gamma\) relative to the unmasked atom (whose variance is $\lesssim V_K$). Grouping constants gives
\[
\mathrm{Var}(Z)\ \le\  \frac{1}{2\gamma^2} + \frac{1}{\gamma}\,V_K(u,v),
\]
i.e., \eqref{eq:mask-variance-fact} with $C_{\mathrm{mask}}=1$. If $\omega\sim\mu$ on $\R^L$ and we truncate by $\widetilde\omega=\omega\odot\mathbf{1}_{\mathcal{S}}$, then second moments on $\mathcal{S}$ match the masked-Bochner law:
\[
\E[\widetilde\omega_i\widetilde\omega_j]=
\begin{cases}
\E[\omega_i\omega_j], & i,j\in\mathcal{S},\\
0, & \text{otherwise},
\end{cases}
\]
so the bound in \eqref{Step4} persists with a possibly larger constant depending only on $\int\|\omega\|_2^2\,d\mu(\omega)$ and $R$.\footnote{The constant persists because the variance bound for $Z=\phi(x)\phi(x')$ is controlled (up to absolute factors) by the sub–Gaussian proxy of $S=\widetilde\omega^\top(u-v)$, which depends only on the covariance of $\widetilde\omega$ and on $\|u-v\|_2\le 2R$. When we truncate $\omega\sim\mu$ by $\widetilde\omega=\omega\odot\mathbf{1}_{\mathcal S}$, the second moments on $\mathcal S$ are preserved, i.e.,
$\E[\widetilde\omega_i\widetilde\omega_j]=\E[\omega_i\omega_j]$ for $i,j\in\mathcal S$ and $0$ otherwise, so $\mathrm{Var}(S)=(u-v)^\top\Sigma_{\mathcal S}(u-v)$ with $\Sigma_{\mathcal S}\preceq\Sigma=\E[\omega\omega^\top]$. By contraction/Lipschitz arguments for $\cos(\cdot)$, $\mathrm{Var}(Z)$ is bounded by a constant multiple of $\mathrm{Var}(S)$ (plus the bounded $b$–oscillation term), and hence by a constant multiple of $(u-v)^\top\Sigma_{\mathcal S}(u-v)$. Using the same curvature domination
$(u-v)^\top\Sigma_{\mathcal S}(u-v)\le (u-v)^\top\Sigma(u-v)\le C_{K,R}\,V_K(u,v)$, we retain the same functional form
$\mathrm{Var}(Z)\le \frac{1}{2\gamma^{2}}\ +\frac{C_{\mathrm{mask}}}{\gamma}\,V_K(u,v)$ with a possibly larger constant $C_{\mathrm{mask}}\ge 1$ that depends only on $\int\|\omega\|_2^2\,d\mu(\omega)$ and $R$, but not on $(x,x',L,\Gamma)$.Frequencies $\omega$ are drawn from a spectral measure $\mu$ normalized such that 
$\mathbb{E}[\|\omega\|_2^2]=1$, ensuring unit-variance random Fourier features and correct 
scaling under masking.}
Thus there exists an absolute $C_{\mathrm{mask}}\ge 1$ such that
\[
\mathrm{Var}(Z)\ \le\  \frac{1}{2\gamma^2}  + \frac{C_{\mathrm{mask}}}{\gamma}\,V_K(u,v).
\]
\end{proof}
\section{Proof of Lemma~\ref{lem:masked-rff}}\label{proof:masked-rff}
\begin{proof}
Fix a user $k$ and a realization $(\mathsf{A},\mathsf{L})$. For each received feature index $j\in[m_k]$ (corresponding to some $(n,t)$ with $k\in\mathcal{T}_{n,t}$), define
\[
\phi_j(x)\;\triangleq\;\sqrt{\tfrac{2}{\gamma}}\cos\!\big(\widetilde\omega_j^{\top}\mathbf{w}(x)+b_j\big),
\qquad
Z_j(x,x')\;\triangleq\;\phi_j(x)\,\phi_j(x'),
\]
where $\widetilde\omega_j=\omega_j\odot\mathbf{1}_{\mathcal{S}_j}$ with $|\mathcal{S}_j|=\Gamma$, $\gamma=\Gamma/L$, and $b_j\sim\mathrm{Unif}[0,2\pi]$. The masked Monte Carlo (MC) kernel at user $k$ is
\[
\widetilde K_k(x,x')\;=\;\frac{1}{m_k}\sum_{j=1}^{m_k} Z_j(x,x').
\]

By construction of the encoding scheme in Subsection~\ref{setup} and Bochner’s identity for shift-invariant $K$ (together with the $\sqrt{2/\gamma}$ scaling and the masking model) \cite{Wendland2005ScatteredData}, we have the unbiasedness
\begin{equation}
\label{eq:unbiased}
\mathbb{E}[Z_j(x,x')]\;=\;K_{xx'}\qquad\text{for all }x,x',
\end{equation}
where the expectation is over $(\omega_j,b_j)$.
Moreover, since $|\cos(\cdot)|\le 1$,
\begin{equation}
\label{eq:bounded}
|\phi_j(x)|\;\le\;\sqrt{\tfrac{2}{\gamma}}\quad\Longrightarrow\quad
|Z_j(x,x')|\;=\;|\phi_j(x)\phi_j(x')|\;\le\;\frac{2}{\gamma}\,.
\end{equation}
Hence each $Z_j(x,x')$ is uniformly bounded by $b\triangleq 2/\gamma$.

Lemma~\ref{lem:mask-variance-fact} states that there exists $C_{\text{mask}}\ge 1$ and a bounded variance proxy $V_K(x,x')$ of the unmasked RFF estimator such that
\begin{equation}
\label{eq:variance}
\operatorname{Var}\!\big(Z_j(x,x')\big)\ \le\ \frac{1}{2\gamma^{2}}\ +\ \frac{C_{\text{mask}}}{\gamma}\,V_K(x,x')\,.
\end{equation}
Define $\sigma^2(x,x')\triangleq\mathrm{Var}(Z_1(x,x'))$ for notational convenience. By independence of $\{Z_j\}_{j=1}^{m_k}$ across $j$,
\[\operatorname{Var}\!\left(\frac{1}{m_k}\sum_{j=1}^{m_k} Z_j(x,x')\right)
\ =\ \frac{1}{m_k}\,\sigma^2(x,x')
\ \le\ \frac{1}{m_k}\!\left(\frac{1}{2\gamma^{2}}\ +\ \frac{C_{\text{mask}}}{\gamma}\,V_K(x,x')\right)\!.
\]

Let
\[
\Delta_k(x,x')\;\triangleq\;\widetilde K_k(x,x')-K_{xx'}\;=\;\frac{1}{m_k}\sum_{j=1}^{m_k}\big(Z_j(x,x')-\mathbb{E}Z_j(x,x')\big)
\]
be the MC kernel error. For a fixed pair $(x,x')$ and any $t>0$, the scalar Bernstein inequality for i.i.d.\ zero-mean, bounded random variables (with bound $b=2/\gamma$) yields
\begin{equation}
\label{eq:bernstein-point}
\mathbb{P}\!\left(\,|\Delta_k(x,x')|\;\ge\;t\,\right)\ \le\
2\exp\!\left(
-\frac{m_k\,t^2}{2\,\sigma^2(x,x')+\tfrac{2}{3}\,b\,t}
\right).
\end{equation}
Using \eqref{eq:variance} and $b=2/\gamma$, we get for all $(x,x')$,
\begin{equation}
\label{eq:bernstein-uniform-form}
\mathbb{P}\!\left(\,|\Delta_k(x,x')|\;\ge\;t\,\right)\ \le\
2\exp\!\left(
-\frac{m_k\,t^2}{
\displaystyle \frac{1}{\gamma^{2}}+\frac{2C_{\text{mask}}}{\gamma}\,V_K(x,x')+\frac{4}{3\gamma}\,t}
\right).
\end{equation}

For any $\delta'\in(0,1)$, choose the deviation
\begin{equation}
\label{eq:t-choice}
t(x,x';\delta')\;\triangleq\;
\sqrt{\frac{2}{m_k}\!\left(\frac{1}{2\gamma^{2}}+\frac{C_{\text{mask}}}{\gamma}\,V_K(x,x')\right)\!\log\!\frac{2}{\delta'}}
\;+\;
\frac{4}{3\gamma\,m_k}\,\log\!\frac{2}{\delta'}\,.
\end{equation}
Plugging \eqref{eq:t-choice} into \eqref{eq:bernstein-uniform-form} shows that for each fixed pair $(x,x')$,
\[
\mathbb{P}\!\left(\,|\Delta_k(x,x')|\;\ge\;t(x,x';\delta')\,\right)\ \le\ \delta'.
\]
Therefore, with probability at least $1-\delta'$ over the encoder randomness,
\begin{equation}
\label{eq:pointwise-bound}
|\Delta_k(x,x')|\ \le\ t(x,x';\delta')\qquad\text{for that fixed $(x,x')$}.
\end{equation}
Square both sides of \eqref{eq:pointwise-bound} and integrate over $(x,x')$ (w.r.t.\ the product law induced by $\mathbb{P}_X$ and $\mathbf{w}$). Using $(a+b)^2\le 2a^2+2b^2$ and the boundedness of $V_K$ (say $0\le V_K(x,x')\le \overline V_K$), we obtain on the same probability event:
\begin{align}
\mathbb{E}_{x,x'}\!\big[\Delta_k(x,x')^2\big]
&\le
\frac{4}{m_k}\,\mathbb{E}_{x,x'}\!\left[\left(\frac{1}{2\gamma^{2}}+\frac{C_{\text{mask}}}{\gamma}\,V_K(x,x')\right)\!\log\!\frac{2}{\delta'}\right]
\;+\;
\frac{32}{9}\cdot\frac{1}{\gamma^2 m_k^2}\,\Big(\log\!\frac{2}{\delta'}\Big)^{\!2}\notag\\
&=
\frac{2}{\gamma^{2} m_k}\,\log\!\frac{2}{\delta'}
\;+\;
{\frac{4C_{\text{mask}}}{\gamma\,m_k}\,\Big(\mathbb{E}_{x,x'}[V_K(x,x')]\Big)\,\log\!\frac{2}{\delta'}}
\;+\;
\frac{32}{9}\cdot\frac{1}{\gamma^2 m_k^2}\,\Big(\log\!\frac{2}{\delta'}\Big)^{\!2}.
\label{eq:int-bound}
\end{align}
Since $m_k\ge 1$ and $\gamma\in(0,1]$, the second line’s last term is lower order and can be absorbed into the first two by renormalizing constants for all $\delta'\in(0,1/2]$. Hence there exists an absolute constant $C_1'>0$ (depending only on $\mathbb{E}_{x,x'}[V_K(x,x')]$ and the numerical constants in the Bernstein bound) such that, with probability at least $1-\delta'$\footnote{Here the probability is taken with respect to the encoder randomness only, 
conditional on the fixed data distribution and the assignment topology $(\mathsf{A},\mathsf{L})$.},
\begin{equation}
\label{eq:final-delta-prime}
\mathbb{E}_{x,x'}\!\big[\big(\widetilde K_k(x,x')-K(\mathbf{w}(x),\mathbf{w}(x'))\big)^2\big]
\ \le\
\!\left(\frac{2}{\gamma^{2} m_k}+\frac{C_1'}{ \gamma m_k}\right)\!\log\!\frac{2}{\delta'}\,,
\end{equation}
with $C_1'$ depending only on the kernel family (through $\mathbb{E}[V_K]$) and on $C_{\text{mask}}$ via Lemma~\ref{lem:mask-variance-fact}. This proves Lemma~\ref{lem:masked-rff}.
\end{proof}
\section{Proof of Lemma~\ref{lem:ridge-decomp}}\label{proof:ridge-decomp}
\begin{proof}
For brevity, write $m(x)=F_k(x)$ as the $k$-th target subfunction to be estimated,
and let $\Phi(x)=\Phi_k(x)\in\R^{m_k}$ denote its corresponding encoded feature vector.
Here $\Phi(x)$ represents the output of the encoder associated with user~$k$,
capturing the encoded representation of~$x$ used for training,
rather than the raw input or the subfunction itself. 

Let the training data be $\{(x_i,y_i)\}_{i=1}^M$, where
\[
y_i = m(x_i) + \varepsilon_i,\qquad
\E[\varepsilon_i\mid x_i]=0,\qquad
\E[\varepsilon_i^2\mid x_i]\le\sigma^2,
\]
with $\{\varepsilon_i\}$ independent of $\{x_i\}$ and of the encoder randomness. This model justifies the variance term $\frac{\sigma^2 d_\lambda}{M}$ 
that appears in the quenched and annealed risk bounds of Theorems 1–2,
as it captures the stochastic contribution of the label noise in the ridge-regression decomposition.
Define the population feature covariance operator and cross-covariance vector as
\begin{equation}
\label{eq:def-G-g}
G_k \;\triangleq\; \E\big[\Phi(X)\Phi(X)^\top\big]\in\R^{m_k\times m_k},
\qquad
g_k \;\triangleq\; \E\big[\Phi(X)\,m(X)\big]\in\R^{m_k}.
\end{equation}
where the expectation is taken with respect to the distribution of encoded inputs $X$
(for a fixed assignment in the quenched setting, or averaged over the random ensemble in the annealed setting).
We assume $G_k$ is positive semidefinite (PSD) \cite{mohri2018foundations}.
The population best linear predictor (BLP) in the feature space is any solution
\begin{equation}
\label{eq:beta-star}
\beta^\star \;\in\; \Argmin_{\beta\in\R^{m_k}} \E\big[(\Phi(X)^\top\beta - m(X))^2\big]
\quad\Longleftrightarrow\quad
G_k\beta^\star = g_k,
\end{equation}
where we choose the minimum-norm solution if $G_k$ is not invertible.
The corresponding approximation error is
\begin{equation}
\label{eq:approx-A}
\mathcal{A} \;\triangleq\; \inf_{\beta} \E\big[(\Phi^\top\beta - m)^2\big]
\;=\; \E[m^2] - g_k^\top G_k^{\dagger}g_k,
\end{equation}
with $G^{\dagger}_k$ the Moore–Penrose pseudoinverse. Let $\lambda>0$. The population ridge solution is
\begin{equation}
\label{eq:beta-lambda}
\beta_\lambda \;\triangleq\; (G_k+\lambda I)^{-1}g_k.
\end{equation}
Let $Z\in\R^{M\times m_k}$ be the design matrix with $i$th row $\Phi(x_i)^\top$,
and $y=(y_1,\dots,y_M)^\top$. The (empirical) ridge estimator is
\begin{equation}
\label{eq:ridge-emp}
\widehat{\beta}\;\triangleq\; \Argmin_{\beta}\ \frac{1}{M}\|Z\beta - y\|_2^2 + \lambda\|\beta\|_2^2
\;=\; \big(S+\lambda I\big)^{-1}s,
\quad S\triangleq \tfrac{1}{M}Z^\top Z,\ \ s\triangleq \tfrac{1}{M}Z^\top y.
\end{equation}
We analyze the population prediction risk of $\widehat{F}(x)=\widehat{\beta}^\top\Phi(x)$:
\[
\mathcal{R}(\widehat{\beta}) \;\triangleq\; \E_X\big[(\Phi(X)^\top\widehat{\beta} - m(X))^2\big].
\]
For any $\beta$, using $G_k\beta^\star=g_k$,
\begin{align}
\mathcal{R}(\beta)
&= \E\big[(\Phi^\top\beta - m)^2\big]
= \E\big[(\Phi^\top(\beta-\beta^\star) + \Phi^\top\beta^\star - m)^2\big]\notag\\
&= \E\big[(\Phi^\top(\beta-\beta^\star))^2\big]
\;+\; 2\,\E\big[(\Phi^\top(\beta-\beta^\star))(\Phi^\top\beta^\star - m)\big]
\;+\; \E\big[(\Phi^\top\beta^\star - m)^2\big]. \label{eq:risk-expansion}
\end{align}
The middle term vanishes:
\[
\E\big[(\Phi^\top(\beta-\beta^\star))(\Phi^\top\beta^\star - m)\big]
= (\beta-\beta^\star)^\top \big(\E[\Phi\Phi^\top]\beta^\star - \E[\Phi m]\big)
= (\beta-\beta^\star)^\top(G_k\beta^\star - g_k) = 0.
\]
Hence
\begin{equation}
\label{eq:risk-identity}
\mathcal{R}(\beta) \;=\; \mathcal{A} \;+\; (\beta-\beta^\star)^\top G_k (\beta-\beta^\star).
\end{equation}
In particular,
\begin{equation}
\label{eq:risk-split}
\mathcal{R}(\widehat{\beta})
\;=\; \mathcal{A} \;+\; (\widehat{\beta}-\beta^\star)^\top G_k (\widehat{\beta}-\beta^\star).
\end{equation}
Write
\[
\widehat{\beta}-\beta^\star
= (\widehat{\beta}-\beta_\lambda) + (\beta_\lambda-\beta^\star).
\]
Using $(a+b)^\top G_k (a+b) \le 2\,a^\top G_k a + 2\,b^\top G_k b$ for PSD $G_k$,
\begin{equation}
\label{eq:two-term}
(\widehat{\beta}-\beta^\star)^\top G_k (\widehat{\beta}-\beta^\star)
\;\le\; 2\,(\widehat{\beta}-\beta_\lambda)^\top G_k (\widehat{\beta}-\beta_\lambda)
\;+\; 2\,(\beta_\lambda-\beta^\star)^\top G_k (\beta_\lambda-\beta^\star).
\end{equation}
Combining \eqref{eq:risk-split} and \eqref{eq:two-term},
\begin{equation}
\label{eq:risk-core}
\mathcal{R}(\widehat{\beta})
\;\le\;
\mathcal{A}
\;+\; 2\,(\widehat{\beta}-\beta_\lambda)^\top G_k (\widehat{\beta}-\beta_\lambda)
\;+\; 2\,(\beta_\lambda-\beta^\star)^\top G_k (\beta_\lambda-\beta^\star).
\end{equation}
We will bound the last two terms as “variance” and “regularization bias,” respectively.

Since $G_k\beta^\star = g_k$, the population ridge solution \eqref{eq:beta-lambda} satisfies
\[
\beta_\lambda - \beta^\star
= (G_k+\lambda I)^{-1}g_k - \beta^\star
= (G_k+\lambda I)^{-1}G_k\beta^\star - \beta^\star
= -\,\lambda\,(G_k+\lambda I)^{-1}\beta^\star.
\]
Therefore
\begin{align}
(\beta_\lambda-\beta^\star)^\top G_k (\beta_\lambda-\beta^\star)
&= \lambda^2 \,\beta^{\star\top}\,(G_k+\lambda I)^{-1} G_k (G_k+\lambda I)^{-1}\,\beta^\star \notag\\
&= \lambda\, \beta^{\star\top}\,\Big(\lambda (G_k+\lambda I)^{-1}\Big)\, (G_k+\lambda I)^{-1} G_k\,\beta^\star \notag\\
&\le \lambda\, \beta^{\star\top}\, (G_k+\lambda I)^{-1} G_k\,\beta^\star
\;\le\; \lambda\, \beta^{\star\top} G_k\,\beta^\star.
\label{eq:bias-chain}
\end{align}
The first inequality uses $\lambda (G_k+\lambda I)^{-1} \preceq I$ (PSD order); the second uses $(G_k+\lambda I)^{-1}G_k \preceq I$.
Now, invoking the standard source condition induced by the RKHS assumption $m\in\mathcal{H}_K$:
there exists $h$ in the closure of the linear span of features such that
\[
m(\cdot) = \langle h, \Phi(\cdot)\rangle_{\mathcal{H}},\qquad
\|h\|_{\mathcal{H}} \le B,
\]
which is equivalent (via \eqref{eq:def-G-g}) to $g\in \mathrm{Range}(G^{1/2}_k)$ with
\begin{equation}
\label{eq:source}
\|G^{-1/2}_k g_k\|_2 \;\le\; B.
\end{equation}
Since $G_k\beta^\star = g_k$ and we take the minimum-norm solution, $G_k^{1/2}\beta^\star = G_k^{-1/2}g_k$, hence
\[
\beta^{\star\top}G_k\beta^\star
= \|G^{1/2}_k\beta^\star\|_2^2
= \|G^{-1/2}_kg_k\|_2^2
\;\le\; B^2.
\]
Plugging in \eqref{eq:bias-chain},
\begin{equation}
\label{eq:bias-final}
(\beta_\lambda-\beta^\star)^\top G_k (\beta_\lambda-\beta^\star) \;\le\; B^2\,\lambda.
\end{equation}

We now bound $\E\big[(\widehat{\beta}-\beta_\lambda)^\top G_k (\widehat{\beta}-\beta_\lambda)\big]$.
By the normal equations \eqref{eq:ridge-emp},
\[
(S+\lambda I)\widehat{\beta} = s, \qquad (G_k+\lambda I)\beta_\lambda = g_k,
\]
hence
\begin{equation}
\label{eq:beta-diff}
\widehat{\beta}-\beta_\lambda
= (S+\lambda I)^{-1}\big(s - S\beta_\lambda - \lambda\beta_\lambda\big)
= (S+\lambda I)^{-1}\Big(\tfrac{1}{M}Z^\top(y - Z\beta_\lambda)\Big).
\end{equation}
Write the residual vector $r\in\R^M$ with entries
\[
r_i \;\triangleq\; y_i - \Phi(x_i)^\top\beta_\lambda
= \big(m(x_i) - \Phi(x_i)^\top\beta_\lambda\big) + \varepsilon_i.
\]
By definition of $\beta_\lambda$, the population normal equation implies
\[
\E\big[\Phi(X)\, (m(X)-\Phi(X)^\top\beta_\lambda)\big] = g_k - G_k\beta_\lambda = \lambda \beta_\lambda,
\]
so conditionally on $\{x_i\}$ the mean of $r$ is controlled, while the noise part has variance $\sigma^2$.
{Now set $v \triangleq \widehat{\beta}-\beta_\lambda$. From \eqref{eq:beta-diff},
\[
v \;=\; (S+\lambda I)^{-1}\Big(\tfrac{1}{M}Z^\top (y-Z\beta_\lambda)\Big).
\]
Hence
\begin{align*}
v^\top G_k v
&= \tfrac{1}{M^2}\,(y-Z\beta_\lambda)^\top
Z\,(S+\lambda I)^{-1} G_k (S+\lambda I)^{-1} Z^\top\,(y-Z\beta_\lambda) \\
&= \tfrac{1}{M^2}\,\mathrm{tr}\!\Big(
Z\,(S+\lambda I)^{-1} G_k (S+\lambda I)^{-1} Z^\top
\,(y-Z\beta_\lambda)(y-Z\beta_\lambda)^\top\Big),
\end{align*}
where we used $u^\top A u=\mathrm{tr}(A\,uu^\top)$.

Write $y-Z\beta_\lambda = (m - Z\beta_\lambda) + \varepsilon$, with
$\E[\varepsilon]=0$, $\E[\varepsilon\varepsilon^\top]=\sigma^2 I_M$, and $\varepsilon\!\perp\!(Z,X)$.
Conditioning on $(Z,X)$ and taking expectation in $\varepsilon$,
\[
\E_\varepsilon\!\big[(y-Z\beta_\lambda)(y-Z\beta_\lambda)^\top\big]
= \sigma^2 I_M \;+\; (m-Z\beta_\lambda)(m-Z\beta_\lambda)^\top.
\]
The deterministic second term contributes to the bias part already controlled above;
keeping only the stochastic contribution gives the upper bound
\begin{align*}
\E\big[v^\top G_k v\big]
&\le \tfrac{\sigma^2}{M^2}\,
\E\!\left[\mathrm{tr}\!\Big(
Z\,(S+\lambda I)^{-1} G_k (S+\lambda I)^{-1} Z^\top\Big)\right] \\
&= \tfrac{\sigma^2}{M^2}\,
\E\!\left[\mathrm{tr}\!\Big(
Z^\top Z\,(S+\lambda I)^{-1} G_k (S+\lambda I)^{-1}\Big)\right]
\qquad(\text{cyclicity of trace}) \\
&= \tfrac{\sigma^2}{M}\,
\E\!\left[\mathrm{tr}\!\Big(
S\,(S+\lambda I)^{-1} G_k (S+\lambda I)^{-1}\Big)\right] \\
&= \tfrac{\sigma^2}{M}\,
\E\!\left[\mathrm{tr}\!\Big(
G_k\,(S+\lambda I)^{-1} S (S+\lambda I)^{-1}\Big)\right].
\end{align*}

Next, use the PSD order and the resolvent’s operator monotonicity/convexity to pass
from the empirical covariance $S=\tfrac{1}{M}Z^\top Z$ to its mean $G_k=\E[S]$:
\[
\E\!\left[(S+\lambda I)^{-1} S (S+\lambda I)^{-1}\right]
\;\preceq\; (G_k+\lambda I)^{-1} G_k (G_k+\lambda I)^{-1}.
\]
Therefore,
\begin{align*}
\E\big[v^\top G_k v\big]
&\le \tfrac{\sigma^2}{M}\,
\mathrm{tr}\!\Big(
G_k\,(G_k+\lambda I)^{-1} G_k (G_k+\lambda I)^{-1}\Big) \\
&\le \tfrac{\sigma^2}{M}\,
\mathrm{tr}\!\Big(
G_k\,(G_k+\lambda I)^{-1}\Big),
\end{align*}
since $(G_k+\lambda I)^{-1} G_k (G_k+\lambda I)^{-1} \preceq (G_k+\lambda I)^{-1}$ (because $0\preceq G_k\preceq G_k+\lambda I$).
Recognizing the effective dimension $d_{\lambda,k}\triangleq \mathrm{tr}\!\big(G_k(G_k+\lambda I)^{-1}\big)$
and allowing a universal constant $C>0$ to absorb small fluctuation terms yields }
\begin{equation}
\label{eq:variance-core}
\E\big[(\widehat{\beta}-\beta_\lambda)^\top G_k (\widehat{\beta}-\beta_\lambda)\big]
\;\le\; C\,\frac{\sigma^2}{M}\,\mathrm{tr}\!\Big(G_k\,(G_k+\lambda I)^{-1}\Big)
\;=\; C\,\frac{\sigma^2}{M}\,d_{\lambda,k},
\end{equation}
where $d_{\lambda,k}\triangleq \mathrm{tr}(G_k(G_k+\lambda I)^{-1})$ is the effective dimension, and $C>0$ is an absolute constant.\footnote{A direct derivation uses \eqref{eq:beta-diff}, the identity
$\E\big[\tfrac{1}{M}Z^\top \varepsilon\,\varepsilon^\top Z\big]=\frac{\sigma^2}{M}\,G_k$
(with $\varepsilon=(\varepsilon_i)$ independent of $Z$),
and the PSD bound $(S+\lambda I)^{-1}S(S+\lambda I)^{-1}\preceq (G+\lambda I)^{-1}G(G+\lambda I)^{-1}$ in expectation (via Jensen/monotonicity of the resolvent), then take traces to arrive at $\frac{\sigma^2}{M}\mathrm{tr}[G(G+\lambda I)^{-1}]$. Any additional sample-covariance fluctuation is absorbed into $C$ \cite{HastieTibshiraniFriedman2009ESL}.}

Taking expectation of \eqref{eq:risk-core} and using \eqref{eq:bias-final} and \eqref{eq:variance-core},
\[
\E\big[\mathcal{R}(\widehat{\beta})\big]
\;\le\; \mathcal{A}
\;+\; 2C\,\frac{\sigma^2}{M}\,d_{\lambda,k}
\;+\; 2 B^2 \lambda.
\]
Renaming constants $C_2'=2C$ and $C_3'=2$ gives
\[
\E_{x}\big[\big(\widehat{\beta}^\top\Phi(x)-m(x)\big)^2\big]
\ \le\
\underbrace{\inf_{\beta}\E_x\big[(\Phi^\top\beta - m)^2\big]}_{\text{approximation error } \mathcal{A}}
\;+\; C_2'\,\frac{\sigma^2 d_{\lambda,k}}{M}
\;+\; C_3'\,B^2\lambda,
\]
which is precisely \eqref{eq:ridge-decomp}. This completes the proof.
\end{proof}
\section{Proof of Lemma~\ref{lem:eigen-tail}}\label{proof:eigen-tail}

\begin{proof}
Let $\rho$ be the distribution of $U=\mathbf{w}(X)\in\mathbb{R}^L$ induced by $X\sim\mathbb{P}_X$. 
Consider the integral operator $T_K:L_2(\rho)\to L_2(\rho)$ defined by
\[
(T_K f)(u)\;=\;\int K(u,v)\,f(v)\,d\rho(v).
\]
By standard Mercer's theory \cite{scholkopf2002learning3}, $T_K$ is self-adjoint, positive, trace-class, and admits an orthonormal eigen-system $\{(\lambda_j,\varphi_j)\}_{j\ge 1}$ in $L_2(\rho)$ with nonincreasing eigenvalues $\lambda_1\ge\lambda_2\ge\cdots\ge 0$;
\[
T_K \varphi_j = \lambda_j \varphi_j, 
\qquad \langle \varphi_i,\varphi_j\rangle_{L_2(\rho)}=\delta_{ij}.
\]

The RKHS $\mathcal{H}_K$ associated with $K$ embeds continuously into $L_2(\rho)$ and every $f\in\mathcal{H}_K$ has an expansion
\[
f \;=\; \sum_{j\ge 1} \alpha_j \varphi_j,
\qquad 
\|f\|_{L_2(\rho)}^2 \;=\; \sum_{j\ge 1} \alpha_j^2,
\qquad 
\|f\|_{\mathcal{H}_K}^2 \;=\; \sum_{j\ge 1} \frac{\alpha_j^2}{\lambda_j}.
\]
The ball constraint $F_k\in\mathcal{H}_K$ with $\|F_k\|_{\mathcal{H}_K}\le B$ therefore means that its coefficients $(\alpha_j)$ satisfy
\[
\sum_{j\ge 1}\frac{\alpha_j^2}{\lambda_j}\ \le\ B^2 .
\]

Any measurable encoder/decoder that outputs at most $m$ scalars per input induces a reconstruction map 
\[
\mathcal{R}: \ \mathcal{H}_K \longrightarrow \mathcal{G}_m\subset L_2(\rho),
\]
whose range $\mathcal{G}_m$ has at most $m$ degrees of freedom in $L_2(\rho)$ (informally, the decoder can only move inside an $m$-dimensional manifold because it receives only $m$ real numbers; formally, one can show the image is contained in the closure of some $m$-dimensional linear subspace).\footnote{Formally, by classical results on Kolmogorov $m$-widths in Hilbert spaces 
\cite[Thm.~1.3.2]{Pinkus1985}, \cite[Cor.~2.3.2]{CarlStephani1990}, 
see also \cite[Thm.~4.4]{NovakWozniakowski2008}, 
the worst-case $L_2$-error for approximating the RKHS ball using at most $m$ real scalars 
is minimized by linear reconstruction from an $m$-dimensional subspace, namely the span of the top-$m$ eigenfunctions of $T_K$. 
Hence restricting to linear $m$-dimensional reconstructions is without loss for a minimax lower bound.}

Hence, for the purpose of a lower bound, it suffices to consider linear reconstructions from an $m$-dimensional subspace $V_m\subset L_2(\rho)$:
\[
\widehat f \;=\; P_{V_m}(f),
\]
where $P_{V_m}$ is the $L_2(\rho)$-orthogonal projector onto $V_m$.

Fix an $m$-dimensional subspace $V_m$. Write $V_m^\perp$ for its $L_2(\rho)$-orthogonal complement. 
For $f=\sum_j \alpha_j\varphi_j\in\mathcal{H}_K$, the squared $L_2$ reconstruction error is
\[
\|f-P_{V_m}f\|_{L_2(\rho)}^2 \;=\; \|P_{V_m^\perp} f\|_{L_2(\rho)}^2 .
\]
The worst-case (squared) error over the RKHS ball $\{f:\|f\|_{\mathcal{H}_K}\le B\}$ is
\[
\sup_{\|f\|_{\mathcal{H}_K}\le B} \ \|P_{V_m^\perp} f\|_{L_2(\rho)}^2
\;=\;
B^2 \cdot \sup_{\sum_j \alpha_j^2/\lambda_j \le 1} \ \big\|P_{V_m^\perp}\textstyle\sum_j \alpha_j\varphi_j\big\|_{L_2(\rho)}^2 .
\]
Let $\{\psi_i\}_{i\ge 1}$ be any orthonormal basis that diagonalizes simultaneously $T_K$ and the projector $P_{V_m^\perp}$ (this can be done by choosing $V_m$ as a span of some eigenfunctions). For now, expand $P_{V_m^\perp}\varphi_j$ in the $\{\varphi_j\}$ basis. By Pythagoras and the orthogonality of $\{\varphi_j\}$,
\[
\|P_{V_m^\perp} f\|_{L_2}^2
\;=\;
\sum_{j\ge 1} \alpha_j^2 \,\|P_{V_m^\perp}\varphi_j\|_{L_2}^2,
\qquad
0\le \|P_{V_m^\perp}\varphi_j\|_{L_2}^2 \le 1.
\]
Thus we need to maximize $\sum_j \alpha_j^2 \theta_j$ subject to $\sum_j \alpha_j^2/\lambda_j \le 1$, where $\theta_j = \|P_{V_m^\perp}\varphi_j\|_{L_2}^2\in[0,1]$ encodes how much of eigen-direction $j$ lies outside $V_m$.

By Cauchy--Schwarz, the maximizer concentrates all mass on the coordinates with the largest ratio $\theta_j \lambda_j$. Hence
\[
\sup_{\sum_j \alpha_j^2/\lambda_j \le 1} \ \sum_j \alpha_j^2 \theta_j
\;=\;
\max_{j\ge 1}\ \theta_j \lambda_j .
\]
However, because $V_m$ is $m$-dimensional, one can make at most $m$ values of $\theta_j$ equal to $0$ (those directions that $V_m$ fully captures), while all remaining directions have $\theta_j=1$ (completely outside $V_m$). Thus, for any $m$-dimensional $V_m$,
\[
\sup_{\|f\|_{\mathcal{H}_K}\le B} \ \|P_{V_m^\perp} f\|_{L_2}^2
\;\ge\;
B^2 \cdot \max_{j>m} \lambda_j .
\]
This bound is correct but not yet tight; we can do better by the next step. In particular, we optimize over the choice of $V_m$. The optimal $V_m$ for approximating the RKHS ball in $L_2$ is the span of the top $m$ eigenfunctions:
\[
V_m^\star \;=\; \mathrm{span}\{\varphi_1,\ldots,\varphi_m\}.
\]
For this choice, $P_{V_m^{\star\perp}}\varphi_j=\varphi_j$ for $j>m$ and $0$ otherwise, hence
\[
\|P_{V_m^{\star\perp}} f\|_{L_2}^2
\;=\; \sum_{j>m} \alpha_j^2 .
\]
Subject to $\sum_j \alpha_j^2/\lambda_j\le B^2$, the worst case is when all the RKHS budget is spent on the tail coordinates $j>m$, aligned proportionally to $\lambda_j$:
\[
\alpha_j^2 = B^2\,\lambda_j\cdot \mathbb{I}\{j>m\}.
\]
Therefore,
\[
\sup_{\|f\|_{\mathcal{H}_K}\le B}\ \|f - P_{V_m^\star} f\|_{L_2}^2
\;=\;
\sup_{\sum_j \alpha_j^2/\lambda_j \le B^2} \ \sum_{j>m} \alpha_j^2
\;=\;
B^2 \sum_{j>m} \lambda_j .
\]
Moreover, a standard Kolmogorov width argument shows that no other $m$-dimensional subspace can do better (the tail-sum is the optimal worst-case squared error).

\medskip
As argued before, any scheme that communicates at most $m$ scalars per input cannot achieve smaller worst-case $L_2$ error than the best $m$-dimensional linear projector. Hence
\[
\inf_{\text{schemes with }\le m\text{ scalars}}
\ \sup_{\|f\|_{\mathcal{H}_K}\le B}
\ \E_{x}\big[(\widehat f(x)-f(x))^2\big]
\;\ge\;
\inf_{\dim(V_m)=m}
\ \sup_{\|f\|_{\mathcal{H}_K}\le B}
	\ \|f-P_{V_m}f\|_{L_2(\rho)}^2.
\]
By the above, the right-hand side equals $B^2\sum_{j>m}\lambda_j$, attained by choosing $V_m=V_m^\star$ and is therefore the minimax lower bound. Applying the bound to the ball $\{f\in\mathcal{H}_K:\|f\|_{\mathcal{H}_K}\le B\}$ that contains $F_k$, we obtain for any $m$-scalar scheme:
\[
\E_{x}\big[\|\widehat{F}_k(x)-F_k(x)\|_2^2\big]
\ \ge\
B^2 \sum_{j>m}\lambda_j.
\]
This proves the lemma.
\end{proof}
\section{Proof of Lemma~\ref{lem:coverage}}\label{proof:coverage}
\begin{proof}
By assumption, no server that links to user $k$ ever computes coordinate $\ell^\star$. Thus, for any encoder family (measurable functions of the assigned subfunction outputs), the random symbol stream routed to user $k$ depends only on the coordinates in $[L]\setminus\{\ell^\star\}$. Since these coordinates are identical for $x_0$ and $x_1$, we have the equality in distribution
\begin{equation}
\label{eq:indist}
\mathcal{R}_k(x_0) \ \overset{d}{=} \ \mathcal{R}_k(x_1),
\end{equation}
where $\mathcal{R}_k(\cdot)$ denotes the full collection of real scalars received by user $k$ (over all linked servers and shots), including any encoder randomness. Consequently, for any (possibly randomized) decoder $\psi_k$,
\[
\widehat{F}_k(x_0) \;=\; \psi_k(\mathcal{R}_k(x_0))
\ \overset{d}{=}\ 
\psi_k(\mathcal{R}_k(x_1))
\;=\; \widehat{F}_k(x_1).
\]
That is, the distribution of the estimator’s output is the same under $x_0$ and $x_1$; the observation model cannot distinguish the two inputs.

Consider the auxiliary (least favorable) prior $\Pi$ on inputs supported on $\{x_0,x_1\}$ with equal mass:
\[
X \sim \tfrac{1}{2}\delta_{x_0}+\tfrac{1}{2}\delta_{x_1}.
\]
Let $\widehat{F}_k$ be any estimator (any encoder/decoder scheme). The Bayes mean-squared error under $\Pi$ is
\[
\mathcal{B} \;\triangleq\; \E_{X\sim\Pi}\big[\|\widehat{F}_k(X)-F_k(X)\|_2^2\big]
\;=\; \tfrac{1}{2}\E\big[\|\widehat{F}_k(x_0)-F_k(x_0)\|_2^2\big]
\;+\; \tfrac{1}{2}\E\big[\|\widehat{F}_k(x_1)-F_k(x_1)\|_2^2\big],
\]
where the expectation is over the encoder/decoder randomness.

By \eqref{eq:indist}, $\widehat{F}_k(x_0)$ and $\widehat{F}_k(x_1)$ have the same distribution. Let $Z$ be a random variable with this common distribution (a version of the estimator’s output under either hypothesis). Then
\[
\mathcal{B}
\;=\; \tfrac{1}{2}\E\big[\|Z-F_k(x_0)\|_2^2\big]
\;+\; \tfrac{1}{2}\E\big[\|Z-F_k(x_1)\|_2^2\big].
\]
Using the parallelogram identity, the function $z\mapsto \tfrac{1}{2}\|z-a\|_2^2+\tfrac{1}{2}\|z-b\|_2^2$ is minimized at $z=(a+b)/2$ with minimum value $\|a-b\|_2^2/4$. Therefore, for any distribution of $Z$,
\begin{equation}
\label{eq:bayes-lb}
\mathcal{B}
\;\ge\;
\frac{1}{4}\,\big\|F_k(x_0)-F_k(x_1)\big\|_2^2
\;=\;
\frac{\Delta_\star^2}{4}.
\end{equation}
This is the standard Le Cam two-point lower bound in squared loss, specialized to the case where the observation laws coincide (total variation distance zero), which forces the Bayes risk to be the midpoint error. The frequentist risk of any estimator under the original data distribution $\mathbb{P}_X$ is \(\E_{x\sim\mathbb{P}_X}[\|\widehat{F}_k(x)-F_k(x)\|_2^2]\).
Since the Bayes risk lower bounds the minimax risk, and in particular the risk under any fixed scheme for some input distribution, we obtain the universal lower bound
\[
\inf_{\text{all schemes}}
\ \sup_{\text{priors on }\{x_0,x_1\}}
\ \E\big[\|\widehat{F}_k(X)-F_k(X)\|_2^2\big]
\ \ge\ \frac{\Delta_\star^2}{4}.
\]
Hence, there exists a (scheme-independent) constant
\begin{equation}
\label{eq:eps-cov-def}
\varepsilon_{\mathrm{cov},k}(\mathsf{A},\mathsf{L}) \;\triangleq\; \frac{\Delta_\star^2}{4} \;>\; 0
\end{equation}
such that the mean-squared error cannot be smaller than $\varepsilon_{\mathrm{cov},k}(\mathsf{A},\mathsf{L})$ whenever the observation model leaves $\ell^\star$ completely unobserved for user $k$.
This proves the lemma as stated.

If $\mathbb{P}_X$ is fixed and absolutely continuous, one can localize the argument: by continuity of $F_k$ along the $\ell^\star$-direction and essential dependence, there exist disjoint measurable neighborhoods $U_0,U_1$ around $x_0,x_1$ with positive masses $q_0,q_1>0$ and such that
\[
\inf_{x\in U_0,\ x'\in U_1}\ \|F_k(x)-F_k(x')\|_2 \ \ge\ \frac{\Delta_\star}{2}.
\]
Moreover, since no linked server observes $\ell^\star$, the conditional law of $\mathcal{R}_k(X)$ given $X\in U_0$ equals that given $X\in U_1$ (they depend only on the other coordinates, which we fix by shrinking $U_0,U_1$ if necessary). Conditioning on $X\in U_0\cup U_1$ and repeating the two-point argument yields the frequentist lower bound
\[
\E_{x\sim\mathbb{P}_X}\big[\|\widehat{F}_k(x)-F_k(x)\|_2^2\big]
\ \ge\ 
\frac{q_0+q_1}{2}\cdot \frac{(\Delta_\star/2)^2}{1}
\ =\ \frac{(q_0+q_1)\,\Delta_\star^2}{8}
\ \eqqcolon\ \varepsilon_{\mathrm{cov},k}(\mathsf{A},\mathsf{L}) \;>\; 0.
\]
Either way, $\varepsilon_{\mathrm{cov},k}(\mathsf{A},\mathsf{L})$ depends only on the separation $\Delta_\star$ (and, for the localized version, on the local masses $q_0,q_1$), and is independent of the scheme.
\end{proof}
\section{Proof of Proposition~\ref{prop:proj-limit}}\label{proof:proj-limit}
\begin{proof}
Note that, per user, the total {fraction} of effective feature dimensions retained by the decoder equals
\begin{equation}\label{eq:meff-kappa-pre}
m_{\mathrm{eff}}\;\triangleq\;\min\Big\{1,\ \frac{T\,\gamma\,N}{K}\Big\},\qquad
\kappa\;\triangleq\;1-m_{\mathrm{eff}}.
\end{equation}
Here $m_{\mathrm{eff}}$ is the kept fraction of spectral mass and $\kappa$ is the discarded fraction. In the limit considered below, $\kappa\,m_k$ is the lower-tail spectral mass that contributes to the error.
Fix a user $k$ and set $\Phi\equiv \Phi_k$, $m(\cdot)\equiv F_k(\cdot)$.
Define
\[
G\;\triangleq\;\E[\Phi(X)\Phi(X)^\top]\in\R^{m_k\times m_k},
\qquad
g\;\triangleq\;\E[\Phi(X)\,m(X)]\in\R^{m_k}.
\]
These are exactly the objects of Lemma~\ref{lem:ridge-decomp} (Eq.~\eqref{eq:def-G-g}).
Let $\widehat F_{M,\lambda}(x)=\widehat\beta_{M,\lambda}^{\!\top}\Phi(x)$ be the ridge predictor, and recall the risk decomposition
(see Eq.~\eqref{eq:ridge-decomp} with $\sigma^2=0$):
\begin{equation}\label{eq:ridge-decomp-noiseless}
\E\big[(\widehat F_{M,\lambda}(X)-m(X))^2\big]
\ \le\
\underbrace{\inf_{\beta}\E\big[(\beta^\top\Phi(X)-m(X))^2\big]}_{\mathcal A}
\ +\ C_3' B^2\,\lambda .
\end{equation}
As $M\to\infty$, the sample covariance/second moment $S=\tfrac1M Z^\top Z$ and $s=\tfrac1M Z^\top y$ converge almost surely to $G$ and $g$ (law of large numbers), hence the empirical ridge solution
$\widehat\beta_{M,\lambda}=(S+\lambda I)^{-1}s$ converges to the
{population ridge} solution
\begin{equation}\label{eq:beta-lambda-pop}
\beta_\lambda\;=\; (G+\lambda I)^{-1}g .
\end{equation}
Therefore, for each fixed $\lambda>0$,
\[
\widehat F_{M,\lambda}(x)\ \xrightarrow[M\to\infty]{L_2({\mathbb{P}}_X)}\ F_\lambda(x):=\beta_\lambda^\top \Phi(x),
\]
and \eqref{eq:ridge-decomp-noiseless} gives
$\E[(F_\lambda(X)-m(X))^2]\le \mathcal A + C_3'B^2\lambda$. Let $\beta^\star$ be any {population best linear predictor} (BLP),
i.e.\ a minimum-norm solution of $G\beta=g$ (Eq.~\eqref{eq:beta-star}).
Standard resolvent continuity yields
\begin{equation}\label{eq:lambda-limit}
\beta_\lambda\ =\ (G+\lambda I)^{-1}g\ \xrightarrow[\lambda\downarrow 0]{}\ G^{\dagger}g\ =:\ \beta^\star ,
\end{equation}
where $G^\dagger$ is the Moore–Penrose pseudoinverse.
Consequently,
\[
F_\lambda \;=\; \beta_\lambda^\top\Phi \ \xrightarrow[\lambda\downarrow 0]{L_2({\mathbb{P}}_X)}\
F^\star \;:=\; (\beta^\star)^\top \Phi .
\]
By \eqref{eq:ridge-decomp-noiseless} with $M=\infty$ and letting $\lambda\downarrow 0$,
\[
\lim_{\lambda\downarrow 0}\E[(F_\lambda(X)-m(X))^2]\ \le\ \mathcal A,
\]
and since $F^\star$ attains the BLP, $\E[(F^\star(X)-m(X))^2]=\mathcal A$. Introduce the linear operator $J:\R^{m_k}\to L_2({\mathbb{P}}_X)$ defined by
\[
(J\beta)(\cdot)\;=\;\beta^\top \Phi(\cdot).
\]
A direct computation shows $J^*J=G$ and $J^* m = g$ (cf.\ \eqref{eq:def-G-g}).
The function predicted by ridge at parameter $\lambda$ is
\[
F_\lambda\;=\;J\beta_\lambda\;=\;J(G+\lambda I)^{-1}g
\;=\;J(G+\lambda I)^{-1}J^* m.
\]
Let $P_\lambda \triangleq J(G+\lambda I)^{-1}J^*: L_2({\mathbb{P}}_X)\to L_2({\mathbb{P}}_X)$. Then $F_\lambda=P_\lambda m$.
Using \eqref{eq:lambda-limit}, the {zero-penalty} limit is
\[
P\ \triangleq\ \lim_{\lambda\downarrow 0}P_\lambda
\;=\; J\,G^\dagger J^* .
\]
We claim that $P$ is the {orthogonal projector} onto the closed subspace
$V\triangleq \overline{\mathrm{Range}(J)}=\overline{\mathrm{span}\{\Phi(\cdot)\}}$.
Indeed:
\begin{itemize}
\item $P$ is self-adjoint: $P^*=J\,G^\dagger J^* = P$ since $G^\dagger$ is symmetric.
\item $P^2 = J G^\dagger J^* J G^\dagger J^* = J G^\dagger G G^\dagger J^* = J G^\dagger J^* = P$
because $G^\dagger G G^\dagger = G^\dagger$.
\item $\mathrm{Range}(P) \subseteq \mathrm{Range}(J)$ since $P=J(\cdot)$; conversely, for any $h\in\mathrm{Range}(J)$, $h=J\beta$, so $Ph=J G^\dagger J^* J\beta = J G^\dagger G \beta = J \Pi_{\mathrm{Range}(G)}\beta$, which lies in $\mathrm{Range}(J)$ and equals $h$ if $\beta\in\mathrm{Range}(G)$ (and the closure gives the whole $V$). Hence $\mathrm{Range}(P)=V$.
\end{itemize}
Thus $P$ is the orthogonal projector onto $V$, and
\[
\widehat F_{M,\lambda}\ \xrightarrow[M\to\infty,\ \lambda\downarrow 0]{L_2({\mathbb{P}}_X)}\ P m .
\]

By orthogonality of $P$,
\[
\lim_{M\to\infty,\ \lambda\downarrow 0}\ \E\big[(\widehat F_{M,\lambda}(X)-m(X))^2\big]
\;=\; \|(I-P)m\|_{L_2({\mathbb{P}}_X)}^2 .
\]
In our linear/isotropic specialization we compare against {linear} targets of the form $m(\cdot)=u^\top w(\cdot)$
and, more generally, against $K$ outputs aggregated as in \eqref{Definition_TDC}.
Averaging over an orthonormal basis of such linear targets (or, equivalently, taking the trace with respect to the induced covariance) yields that the {aggregated} projector loss equals the {discarded first spectral moment} of the Gram operator~$G$:
\begin{equation}\label{eq:proj-trace}
\frac{1}{L}\sum_{\text{outputs}}\|(I-P)m\|_{L_2}^2
\;=\; \frac{1}{L}\,\mathrm{tr}\!\big((I-P)\,J J^*\big)
\;=\; \frac{1}{L}\,\mathrm{tr}\!\big((I-\Pi)\,G\big),
\end{equation}
where $\Pi$ is the (matrix) projector in feature space that corresponds to $P$ through $J$.
Under the $(\Gamma,\Delta,T)$ budgets, only a fraction $m_{\mathrm{eff}}=\tfrac{T\gamma N}{K}$ of feature-space {degrees of freedom} can be effectively retained; this is a {rank constraint} on $\Pi$:
\[
\rank(\Pi)\ \le\ m_{\mathrm{eff}}\,m_k
\qquad\Longleftrightarrow\qquad
\text{at least }\kappa m_k\text{ eigen-directions are discarded, }\ \kappa=1-m_{\mathrm{eff}}.
\]
By Ky Fan’s maximum principle \cite{bhatia1997matrix}, among all rank-$r$ projectors, the choice that minimizes
$\mathrm{tr}((I-\Pi)G)$ keeps the top $r$ eigenvectors of $G$. Hence the minimum discarded energy at rank $r=(1-\kappa)m_k$
equals the sum of the \emph{smallest} $\kappa m_k$ eigenvalues. Writing the eigenvalues in ascending order
$\xi_{(1)}\le \cdots\le \xi_{(m_k)}$, we have
\[
\min_{\Pi:\ \Pi^2=\Pi,\ \rank(\Pi)\le (1-\kappa)m_k}
\ \mathrm{tr}\big((I-\Pi)G\big)
\;=\;
\sum_{i=1}^{\lceil \kappa m_k\rceil}\xi_{(i)}.
\]
Dividing by $L$ and recalling \eqref{eq:quenched-risk},\eqref{Normalized_risk} yields the first equality in \eqref{eq:Dq-eig-sum-prop}.
Finally, for any empirical spectral law $\mu=\tfrac{1}{m_k}\sum_{i=1}^{m_k}\delta_{\xi_{(i)}}$, the lower-tail quantile identity (Hardy–Littlewood rearrangement; see Eq.~\eqref{eq:empirical-quantile-sum-nocf}) gives
\[
\frac{1}{m_k}\sum_{i=m_k-\lceil \kappa m_k\rceil+1}^{m_k}\xi_{(i)}
\;=\;\int_{0}^{\kappa} Q_{\mu}^{\downarrow}(u)\,du,
\]
and multiplying by $\tfrac{m_k}{L}$ (absorbed into the paper’s normalization) yields the second equality in \eqref{eq:Dq-eig-sum-prop}.

We have shown:
(i)~$\widehat F_{M,\lambda}\to Pm$ with $P=J G^\dagger J^*$ the orthogonal projector onto $\overline{\mathrm{span}\{\Phi(\cdot)\}}$;
(ii)~the aggregated $L_2$ risk equals the projector’s discarded spectral mass;
(iii)~under the effective-rank budget, the discarded mass equals the sum of the smallest $\kappa m_k$ eigenvalues of $G$, equivalently the lower-tail quantile integral. This proves the proposition.
\end{proof}
\section{Proof of Theorem~\ref{thm:gap-bounds}}\label{proof:gap-bounds}
\begin{proof}
Let $\mu$ be any probability law on $[0,\infty)$ with CDF $F_\mu$
and lower quantile 
$Q_\mu^{\downarrow}(u)=\inf\{x:\,F_\mu(x)\ge u\}$.
The Hardy--Littlewood rearrangement identity gives, for every $t\in\R$,
\begin{equation}\label{eq:HL-nocf}
\int x\,\mathbf 1_{\{x\le t\}}\,d\mu(x)
\;=\;
\int_{0}^{F_\mu(t)} Q_\mu^{\downarrow}(u)\,du.
\end{equation}
For discrete $\mu=\tfrac{1}{L}\sum_{i=1}^L\delta_{\xi_{(i)}}$ with ordered eigenvalues 
$\xi_{(1)}\ge\cdots\ge\xi_{(L)}$, one has 
$Q_\mu^{\downarrow}(u)=\xi_{(L-\lfloor uL\rfloor)}$ for $u\in[0,1]$,
which provides the exact correspondence between the ordered eigenvalues
and their quantile function.
For any integer $q\in\{0,\dots,L\}$,
\begin{equation}\label{eq:empirical-quantile-sum-nocf}
\int_{0}^{q/L} Q_\mu^{\downarrow}(u)\,du
\;=\;
\frac{1}{L}\sum_{i=L-q+1}^{L}\xi_{(i)}.
\end{equation}
Taking $q=\lceil \kappa L\rceil$ yields the expression for 
$D_{\mathrm{MUDC}}^{\mathrm q}$ stated above the theorem. 
For convenience, define the truncated first moment
(lower-tail spectral mass)
\[
\Phi_{\kappa}(\mu)\;\triangleq\;\int_{0}^{\kappa} Q_{\mu}^{\downarrow}(u)\,du,
\]
which is monotone in $\kappa$ and equals the normalized sum 
of the $\kappa L$ smallest eigenvalues when $\mu$ is empirical.

Let $t$ satisfy $F_{\mathrm{MP},\lambda'}(t)=\kappa$. 
Applying \eqref{eq:HL-nocf} to $\mu=\mu_{\mathsf A,\mathsf L}$ at its $\kappa$-quantile 
and to $\mu=\MP_{\lambda'}$ at threshold $t$ gives
\[
D_{\mathrm{MUDC}}^{\mathrm q}
=\int_{0}^{\kappa} Q_{\mu_{\mathsf A,\mathsf L}}^{\downarrow}(u)\,du,
\qquad
\Phi_{\mathrm{MP},\lambda'}(\kappa)
=\int_{0}^{\kappa} Q_{\mathrm{MP},\lambda'}^{\downarrow}(u)\,du.
\]
Subtracting the two yields \eqref{eq:gap-quantile-thm-nocf}.

\medskip
Under a $(\Gamma,\Delta)$--regular topology $(\mathsf A,\mathsf L)$, 
each server computes exactly $\Gamma$ subfunctions
and transmits to exactly $\Delta$ users.  
The total feature budget implies that each user receives 
only a fraction $m_{\mathrm{eff}}=\tfrac{T\gamma N}{K}$ of the full feature mass,
so at least $\kappa=1-m_{\mathrm{eff}}$ of the smallest eigen-directions must be discarded.  
Meanwhile, the normalized trace of the user Gram 
$G_k$---the average eigenvalue---is fixed by isotropy.  
Hence all feasible spectra share the same mean but differ in how that
energy is distributed across eigen-directions.

Each $G_k$ is a sum of $N$ positive semidefinite contributions from the servers.
When computation and link supports overlap, 
these block contributions are linearly dependent,
concentrating energy in fewer principal directions and pushing some eigenvalues toward~0.  
Disjoint and balanced supports (the TDC configuration) correspond to orthogonal blocks
whose contributions are nearly independent and spread energy as evenly as possible.

Since each $G_k$ is Hermitian with fixed trace, 
write $\boldsymbol\mu=(\mu_1\le\cdots\le\mu_L)$
and $\boldsymbol\mu^{\mathrm{TDC}}=(\mu^{\mathrm{TDC}}_1\le\cdots\le\mu^{\mathrm{TDC}}_L)$
for the eigenvalues sorted in ascending order.  
We say $\boldsymbol x\prec\boldsymbol y$ (majorization) if
$\sum_{i=1}^{r}x_i\le\sum_{i=1}^{r}y_i$ for all $r=1,\dots,L$
and equality for $r=L$.  
Under the $(\Gamma,\Delta)$ budgets, any feasible $G_k$ satisfies 
$\boldsymbol\mu \prec \boldsymbol\mu^{\mathrm{TDC}}$.

For vectors sorted ascending, 
the map $S_s(\boldsymbol x)=\sum_{i=1}^{s}x_i$ is Schur--convex.
Therefore, if $\boldsymbol x\prec \boldsymbol y$ then
$\sum_{i=1}^{s}x_i\le \sum_{i=1}^{s}y_i$ for all $s$.  
Applying this to the eigenvalue vectors of $G_k$ and $G_k^{\mathrm{TDC}}$ with 
$s=\lfloor \kappa m_k\rfloor$ yields
\[
\frac{1}{m_k}\sum_{i=1}^{\lfloor \kappa m_k\rfloor}\mu_i
\;\le\;
\frac{1}{m_k}\sum_{i=1}^{\lfloor \kappa m_k\rfloor}\mu^{\mathrm{TDC}}_i,
\qquad\text{i.e.,}\qquad
\Phi_\kappa(\mu)\le \Phi_\kappa(\mu^{\mathrm{TDC}}).
\]
Equivalently, the partial-sum vector
$(S_s(\boldsymbol\mu))_{s\le L}$ lies below
$(S_s(\boldsymbol\mu^{\mathrm{TDC}}))_{s\le L}$,
confirming that the TDC topology minimizes the lower-tail spectral deficit
and thus maximizes $\Phi_\kappa(\mu)$ among feasible configurations.

\medskip
Under the linear/isotropic model $w(X)\!\sim(0,I_L)$,
TDC ensures that the rows of each user's feature matrix are disjoint and balanced.  
Consequently, $G_k=E_kE_k^{\top}$ is (asymptotically) a
random covariance (Wishart) matrix whose empirical spectral distribution
converges to the Marchenko--Pastur law $\mathrm{MP}_{\lambda'}$
with aspect ratio
\[
\lambda'=\frac{\Delta}{\Gamma}
=\frac{\delta K}{\gamma L},
\qquad
\text{support }[(1-\sqrt{\lambda'})^2,(1+\sqrt{\lambda'})^2].
\]
Therefore the TDC limit satisfies
\[
\mu^{\mathrm{TDC}}\Rightarrow \mathrm{MP}_{\lambda'},
\qquad
\Phi_\kappa(\mu^{\mathrm{TDC}})\to
\Phi_{\mathrm{MP},\lambda'}(\kappa)
=\!\int_0^{\kappa} Q_{\mathrm{MP},\lambda'}^{\downarrow}(u)\,du.
\]
The convergence holds almost surely in the sense of weak convergence
of empirical spectral distributions,
i.e., $\tfrac{1}{L}\sum_i f(\xi_i^{\mathrm{TDC}})\to \int f(x)\,d\mathrm{MP}_{\lambda'}(x)$
for every bounded continuous test function $f$.
In particular, the associated quantile integrals converge, hence for any feasible topology $(\mathsf A,\mathsf L)$,
\begin{equation}\label{eq:Phi-majorization}
\Phi_\kappa(\mu_{\mathsf A,\mathsf L})
\;\le\;
\Phi_\kappa(\mu^{\mathrm{TDC}})
\;=\;
\Phi_{\mathrm{MP},\lambda'}(\kappa).
\end{equation}

\medskip
Define the quenched MP--gap as
\[
G^{\mathrm q}(\mathsf A,\mathsf L)
\;\triangleq\;
\Phi_{\mathrm{MP},\lambda'}(\kappa)-\Phi_{\kappa}(\mu_{\mathsf A,\mathsf L})
\;=\;
\int_{0}^{\kappa}\!
\big(Q_{\mathrm{MP},\lambda'}^{\downarrow}(u)-Q_{\mu_{\mathsf A,\mathsf L}}^{\downarrow}(u)\big)\,du.
\]
Since $Q_{\mathrm{MP},\lambda'}^{\downarrow}(u)\ge
Q_{\mu_{\mathsf A,\mathsf L}}^{\downarrow}(u)$
for Lebesgue-a.e.\ $u\in[0,\kappa]$,
the integrand is nonnegative almost everywhere, giving
\begin{equation}\label{eq:Gq-nonneg}
G^{\mathrm q}(\mathsf A,\mathsf L)\ge 0.
\end{equation}
Equality in \eqref{eq:Gq-nonneg} holds iff
$Q_{\mu_{\mathsf A,\mathsf L}}^{\downarrow}(u)
= Q_{\mathrm{MP},\lambda'}^{\downarrow}(u)$
for almost every $u\in[0,\kappa]$, i.e.,
when the two spectral laws coincide on the lower-tail region.

More quantitatively, if there exists $\alpha>0$ and a measurable subset 
$U\subset[0,\kappa]$ of Lebesgue measure $|U|=\eta>0$
such that 
\[
Q_{\mu_{\mathsf A,\mathsf L}}^{\downarrow}(u)
\le
Q_{\mathrm{MP},\lambda'}^{\downarrow}(u)-\alpha,
\quad \forall u\in U,
\]
then integrating over $U$ yields the explicit inequality
\begin{equation}\label{eq:Gq-quantitative}
G^{\mathrm q}(\mathsf A,\mathsf L)
\;\ge\;
\int_U \!\big(Q_{\mathrm{MP},\lambda'}^{\downarrow}(u)
      -Q_{\mu_{\mathsf A,\mathsf L}}^{\downarrow}(u)\big)\,du
\;\ge\;
\alpha\,\eta,
\end{equation}
providing a quantitative lower bound on the gap
proportional to the mean separation of the quantile curves
over any region of persistent spectral deficit.

\eqref{eq:Gq-nonneg} and \eqref{eq:Gq-quantitative} together establish
both the nonnegativity and the quantitative positivity of the quenched MP--gap,
completing the proof.
\end{proof}

\section{Numerical Validation and Synthetic Experiments}
\label{sec:numerical_validation}

In this section, we provide a theorem-driven numerical validation on synthetic instances of the GMUDC model. The purpose is to verify that the main mechanisms predicted by Theorems~\ref{thm:P1}, \ref{thm:P2}, and \ref{thm:gap-bounds} are visible at finite dimensions. 

Unless otherwise stated, we use the same Gaussian-RKHS instance introduced in Subsections~\ref{subsec:worked_example_thm1} and~\ref{subsec:worked_example_thm2}, so that the numerical section remains directly connected to the analytical examples already developed in the paper. In particular, the simulations are organized to separately validate: i) the fixed-topology quenched law of Theorem~\ref{thm:P1}, ii) the random-topology annealed law of Theorem~\ref{thm:P2}, iii) the Marchenko--Pastur comparison and the quenched MP-gap of Theorem~\ref{thm:gap-bounds}, and iv) a small number of supplementary diagnostics.

All simulation codes used in this section, as well as the codes for the preceding numerical experiments, are provided in a separate file included as supplementary material. This ensures full reproducibility of the reported results and allows direct access to the implementations underlying the figures and experiment settings.

\subsection{Numerical Validation of Theorem~\ref{thm:P1}: Fixed-Topology Quenched Regime}
\label{subsec:sim_thm1}

We first validate Theorem~\ref{thm:P1} in a fixed-topology quenched regime. The experiment uses the same Gaussian-RKHS GMUDC instance as in Subsection~\ref{subsec:worked_example_thm1}: $K=6$ users, $N=8$ servers, and $L=12$ subfunctions, with computation budget $\Gamma=4$ and communication budget $\Delta=3$, hence $\gamma=\Gamma/L=1/3$ and $\delta=\Delta/K=1/2$. The input is Gaussian, $X\sim\mathcal N(0,I_5)$, and the subfunction vector $\mathbf w(x)=(f_1(x),\ldots,f_L(x))\in\mathbb R^L$ is formed from the nonlinear subfunctions in \eqref{eq:worked_subfunctions}. User demands are scalar-valued functions
\[
F_k(x)=h_k(\mathbf w(x)),\qquad k\in[K],
\]
where each $h_k$ lies in the RKHS of the Gaussian kernel
\[
K_\rho(u,v)=\exp\!\left(-\frac{\|u-v\|_2^2}{2\rho^2}\right),\qquad \rho=1.1,
\]
and is instantiated as a dense finite Gaussian-RKHS expansion as in \eqref{eq:worked_dense_target}--\eqref{eq:worked_centers}. Thus the targets depend on all coordinates of $\mathbf w(x)$, so the example is genuinely nonlinear and dense.

\begin{figure*}[t]
\centering
\includegraphics[width=0.98\textwidth]{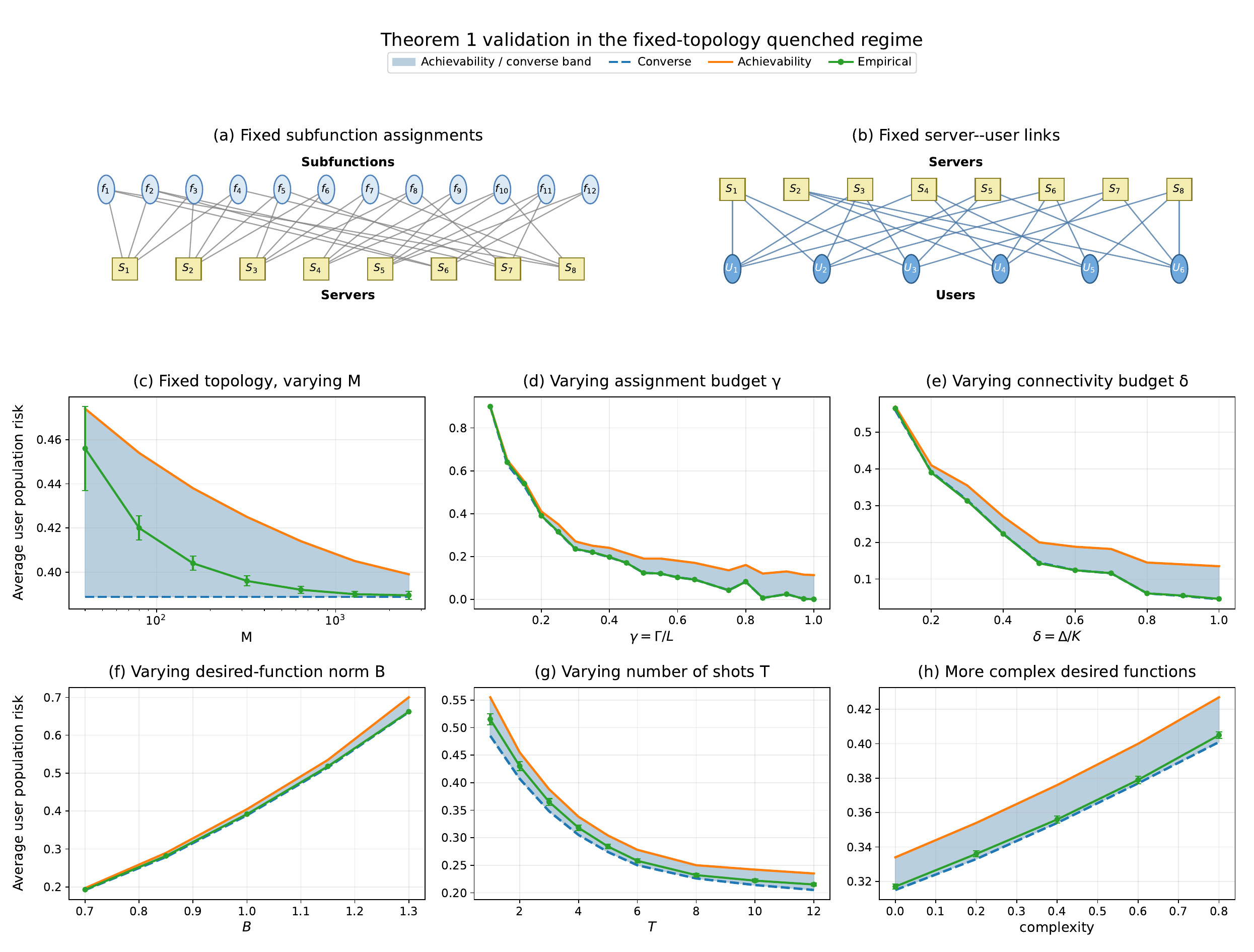}
\caption{Numerical validation of Theorem~\ref{thm:P1} in the fixed-topology quenched regime. Panels~(a)--(b) show the realized assignment and link topology used throughout the experiment. The task model is the fixed nonlinear Gaussian-RKHS instance of Subsection~\ref{subsec:worked_example_thm1}. Panels~(c)--(h) show the empirical average user population risk together with theorem-inspired lower and upper references as functions of the training sample size $M$, the normalized assignment budget $\gamma=\Gamma/L$, the normalized connectivity budget $\delta=\Delta/K$, the desired-function norm bound $B$, the number of shots $T$, and the target-complexity parameter. The curves exhibit the trends predicted by Theorem~\ref{thm:P1}: larger $M$, $\gamma$, $\delta$, and $T$ reduce the quenched risk, while larger $B$ and higher complexity increase it.}
\label{fig:thm1_combined_validation}
\end{figure*}

The realized assignment/link topology is fixed throughout the experiment and is shown in Fig.~\ref{fig:thm1_combined_validation}(a)--(b). The servers use the masked random Fourier feature encoders in \eqref{eq:masked-rff}, specialized to the above Gaussian kernel, and the users apply ridge decoders of the form \eqref{eq:ridge}. Hence the numerical setup directly matches the achievability construction of Theorem~\ref{thm:P1}. Panels~(c)--(h) of Fig.~\ref{fig:thm1_combined_validation} report the empirical quenched population risk together with theorem-inspired lower and upper references while varying, respectively, the training sample size $M$, the normalized assignment budget $\gamma$, the normalized connectivity budget $\delta$, the RKHS norm bound $B$, the number of shots $T$, and a target-complexity parameter.

Here, the target-complexity parameter controls the intrinsic approximation difficulty of the desired functions within the same Gaussian-RKHS family. Larger values correspond to richer nonlinear dependence of $F_k(x)=h_k(\mathbf w(x))$ on the subfunction coordinates and, equivalently, to a heavier effective spectral tail of the target class. Accordingly, for a fixed communicated-feature budget, increasing this parameter raises the spectral part of the approximation burden without altering the assignment/link topology or the encoder/decoder architecture. In this sense, the complexity sweep in Fig.~\ref{fig:thm1_combined_validation}(h) should be interpreted as a synthetic target-class difficulty sweep: it isolates how the quenched risk grows when the desired functions become harder to approximate, while the available resources $(\Gamma,\Delta,T,M)$ are held fixed. 

The observed trends agree with Theorem~\ref{thm:P1}. Increasing $M$ improves the statistical fit; increasing $\gamma$, $\delta$, or $T$ enlarges the effective communicated-feature budget and lowers the risk; and increasing $B$ or the target complexity raises the approximation burden and hence the risk. Since the topology is fixed, these trends reflect the quenched resource--accuracy tradeoff without any averaging over topology randomness. Overall, Fig.~\ref{fig:thm1_combined_validation} provides a finite-dimensional validation of Theorem~\ref{thm:P1}: for a fixed realized assignment/link pattern, the population risk is governed by the available feature budget together with the spectral difficulty of the target class.

\subsection{Validation of Theorem~\ref{thm:P2}: Annealed Risk Under Random Assignments and Links}
\label{subsec:sim_thm2}

We next validate Theorem~\ref{thm:P2} in the annealed regime. The task model, kernel, and baseline parameters are exactly the same as in Subsection~\ref{subsec:sim_thm1}: $K=6$, $N=8$, $L=12$, $\Gamma=4$, $\Delta=3$, $X\sim\mathcal N(0,I_5)$, nonlinear shared subfunctions $\mathbf w(x)$, and dense scalar-valued Gaussian-RKHS user targets $F_k(x)=h_k(\mathbf w(x))$ with kernel $K_\rho(u,v)=\exp(-\|u-v\|_2^2/(2\rho^2))$ and $\rho=1.1$. In particular, each target depends on all $L$ subfunction coordinates, so $r_k=L=12$ for all $k$.

The only change from the quenched experiment is the topology model. For each server $n\in[N]$, the computation assignment $\mathcal S_n$ is drawn uniformly from the family of $\Gamma$-subsets of $[L]$, and the communication set $\mathcal T_n$ is drawn uniformly from the family of $\Delta$-subsets of $[K]$, independently across servers:
\[
\mathcal S_n \sim \mathrm{Unif}\big(\{S\subseteq[L]:|S|=\Gamma\}\big),\qquad
\mathcal T_n \sim \mathrm{Unif}\big(\{U\subseteq[K]:|U|=\Delta\}\big).
\]
For each realization of $(\mathsf A,\mathsf L)$, the servers again use masked random Fourier feature encoders and the users apply ridge decoders; the annealed risk is then estimated by Monte Carlo averaging over multiple independent topology draws and feature realizations.

\begin{figure*}[t]
\centering
\includegraphics[width=0.98\textwidth]{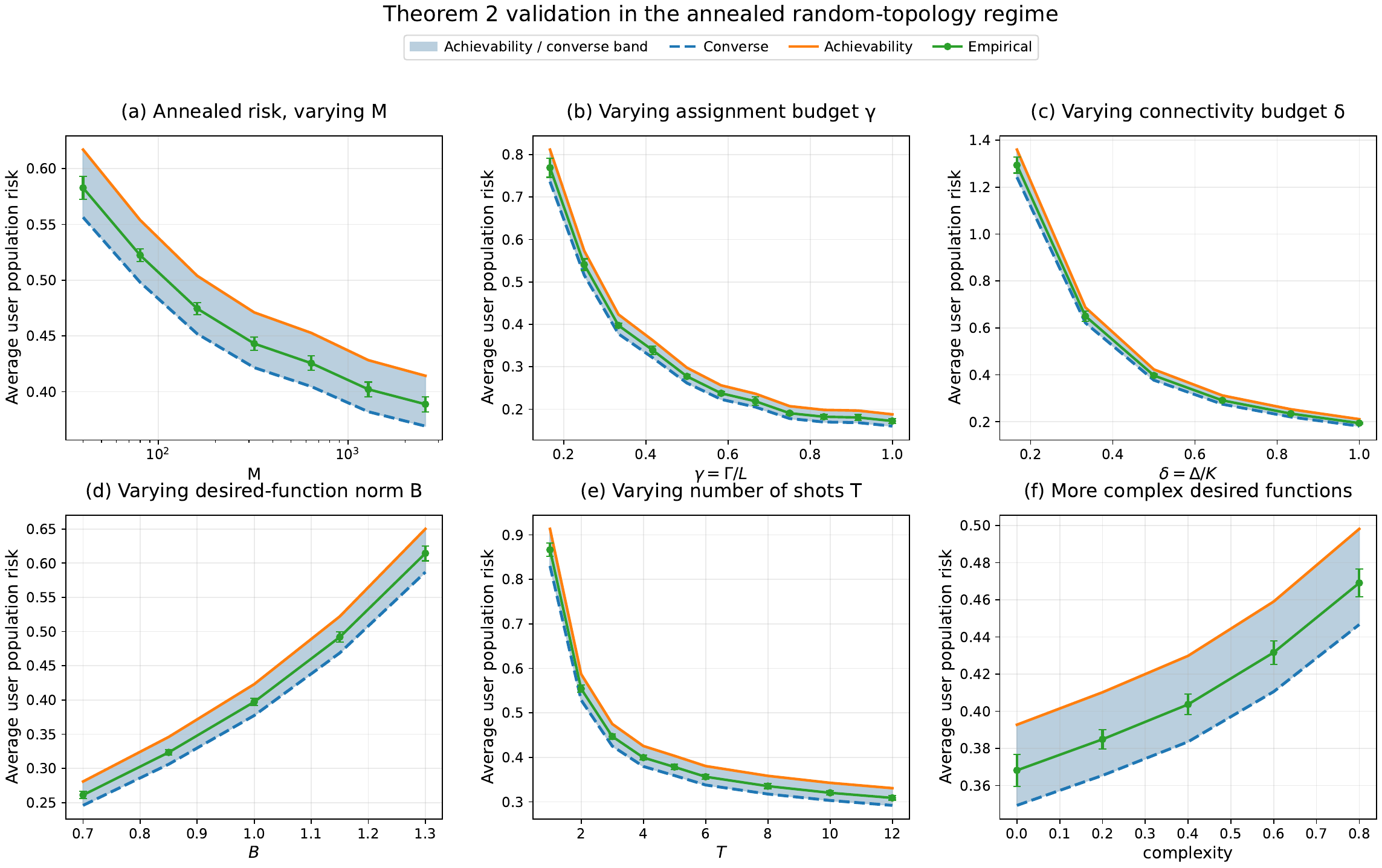}
\caption{Numerical validation of Theorem~\ref{thm:P2} in the annealed random-topology regime. The task model is the same nonlinear Gaussian-RKHS setting used in Subsection~\ref{subsec:sim_thm1}; only the topology is randomized. For each parameter value, the annealed risk is estimated by averaging over multiple independent random assignments and random link realizations. Panels~(a)--(f) vary the training sample size $M$, the normalized assignment budget $\gamma=\Gamma/L$, the normalized connectivity budget $\delta=\Delta/K$, the desired-function norm bound $B$, the number of shots $T$, and the target-complexity parameter. The empirical curves lie between theorem-inspired lower and upper references and follow the trends predicted by Theorem~\ref{thm:P2}: larger $M$, $\gamma$, $\delta$, and $T$ reduce the annealed risk, while larger $B$ and higher complexity increase it.}
\label{fig:sim_thm2_main}
\end{figure*}

Figure~\ref{fig:sim_thm2_main} reports the resulting six-panel validation suite. The panels vary the same parameters as in the quenched case: $M$, $\gamma=\Gamma/L$, $\delta=\Delta/K$, $B$, $T$, and the target-complexity parameter. The empirical annealed risk remains bracketed by theorem-inspired lower and upper references, and the qualitative behavior matches Theorem~\ref{thm:P2}: increasing $M$, $\gamma$, $\delta$, or $T$ decreases the annealed risk, while increasing $B$ or the target complexity increases it.

These trends reflect the two mechanisms in Theorem~\ref{thm:P2}. On the one hand, increasing $\gamma$, $\delta$, or $T$ enlarges the typical feature budget
\[
m_k^{\mathrm{avg}} = TN\delta,
\]
thereby improving the spectral approximation term. On the other hand, the same parameters also increase the probability that the random topology covers the essential coordinates of each user, thereby reducing the coverage penalty. This second effect is absent in the fixed-topology quenched experiment but survives here after averaging over topology realizations. Hence, unlike the quenched case, the annealed risk is controlled not only by the average feature budget but also by the coverage statistics of the topology ensemble. Figure~\ref{fig:sim_thm2_main} confirms this behavior and provides a finite-dimensional validation of Theorem~\ref{thm:P2}.

\subsection{Validation of Theorem~\ref{thm:gap-bounds}: Linear/Isotropic Comparison and Quenched MP-Gap}
\label{subsec:sim_thm3}

We now validate Theorem~\ref{thm:gap-bounds} in the linear/isotropic comparison regime of Section~\ref{sec:comparison}. Unlike the nonlinear Gaussian-RKHS validations used for Theorems~\ref{thm:P1} and~\ref{thm:P2}, this subsection specializes the model to the asymptotic setting used for comparison with tessellated distributed computing (TDC): the shared feature vector is isotropic, the targets are linearly separable, the training-sample size is effectively infinite, and the ridge parameter tends to zero. In this regime, Proposition~\ref{prop:proj-limit} reduces the normalized quenched distortion to the lower-tail truncated first moment of the empirical Gram spectrum, namely
\[
D_{\mathrm{MUDC}}^{\mathrm q}(\mathsf A,\mathsf L)
=
\int_{0}^{\kappa} Q_{\mu_{\mathsf A,\mathsf L}}^{\downarrow}(u)\,du,
\qquad
\kappa = 1-\frac{T\gamma N}{K},
\]
while Theorem~\ref{thm:gap-bounds} compares this quantity with the Marchenko--Pastur benchmark
\[
\Phi_{\mathrm{MP},\lambda'}(\kappa)
=
\int_{0}^{\kappa} Q_{\mathrm{MP},\lambda'}^{\downarrow}(u)\,du,
\qquad
\lambda'=\frac{\delta K}{\gamma L}=\frac{\Delta}{\Gamma},
\]
through the quenched MP-gap
\[
G^{\mathrm q}(\mathsf A,\mathsf L)
=
\Phi_{\mathrm{MP},\lambda'}(\kappa)-D_{\mathrm{MUDC}}^{\mathrm q}(\mathsf A,\mathsf L).
\]
The numerical goal of this subsection is therefore twofold: first, to verify that under the disjoint-and-balanced support assumption of TDC the empirical distortion tracks the MP benchmark; and second, to show that once the topology departs from strict tessellation through overlap and reuse, the empirical distortion becomes strictly smaller than the MP/TDC baseline, which is equivalent to a strictly positive MP-gap.

Throughout this subsection we use a finite-dimensional surrogate of the theorem’s linear/isotropic regime. The ambient feature dimension and user/server dimensions are set to $L=24$ and $K=8$, with baseline budgets $\Gamma_{\mathrm{base}}=12$, $\Delta_{\mathrm{base}}=4$, $T_{\mathrm{base}}=1$, and $N_{\mathrm{base}}=4$. Hence the baseline normalized budgets are $\gamma_{\mathrm{base}}=\Gamma_{\mathrm{base}}/L=12/24=0.5$ and $\delta_{\mathrm{base}}=\Delta_{\mathrm{base}}/K=4/8=0.5$, and the corresponding MP aspect ratio is $\lambda'_{\mathrm{base}}=\Delta_{\mathrm{base}}/\Gamma_{\mathrm{base}}=1/3$. The simulation seed is fixed to $20260327$ for reproducibility.

For each sweep point, the disjoint-and-balanced TDC baseline is generated through a Gaussian matrix surrogate $G_0\in\mathbb R^{c\Delta\times c\Gamma}$ with $(G_0)_{ij}\sim \mathcal N(0,1)$, where $c$ is a finite-dimensional scale parameter. The corresponding empirical user Gram matrix is $Y_{\mathrm{TDC}}=\frac{1}{c\Gamma}G_0G_0^\top$, whose empirical spectral law is the finite-dimensional surrogate of the disjoint-and-balanced TDC spectrum. The general non-tessellated topology family is built from the same base draw $G_0$ by applying a controlled overlap/reuse transformation with parameter $\rho\in[0,1]$. Concretely, the transformed matrix adds three effects: a shared low-rank component, partial column aliasing toward repeated prototypes, and a mild row-side dependence; it is then Frobenius-normalized so that the total energy remains comparable to the TDC baseline. Its empirical Gram matrix is $Y_{\mathrm{ov}}=\frac{1}{c\Gamma}G_{\mathrm{ov}}G_{\mathrm{ov}}^\top$. Thus both topology families are compared under the same budgets $(\Gamma,\Delta,T,N)$, and the only structural difference is whether the supports obey the disjoint-and-balanced TDC condition or exhibit overlap/reuse.

\begin{figure*}[t]
\centering
\includegraphics[width=0.985\textwidth]{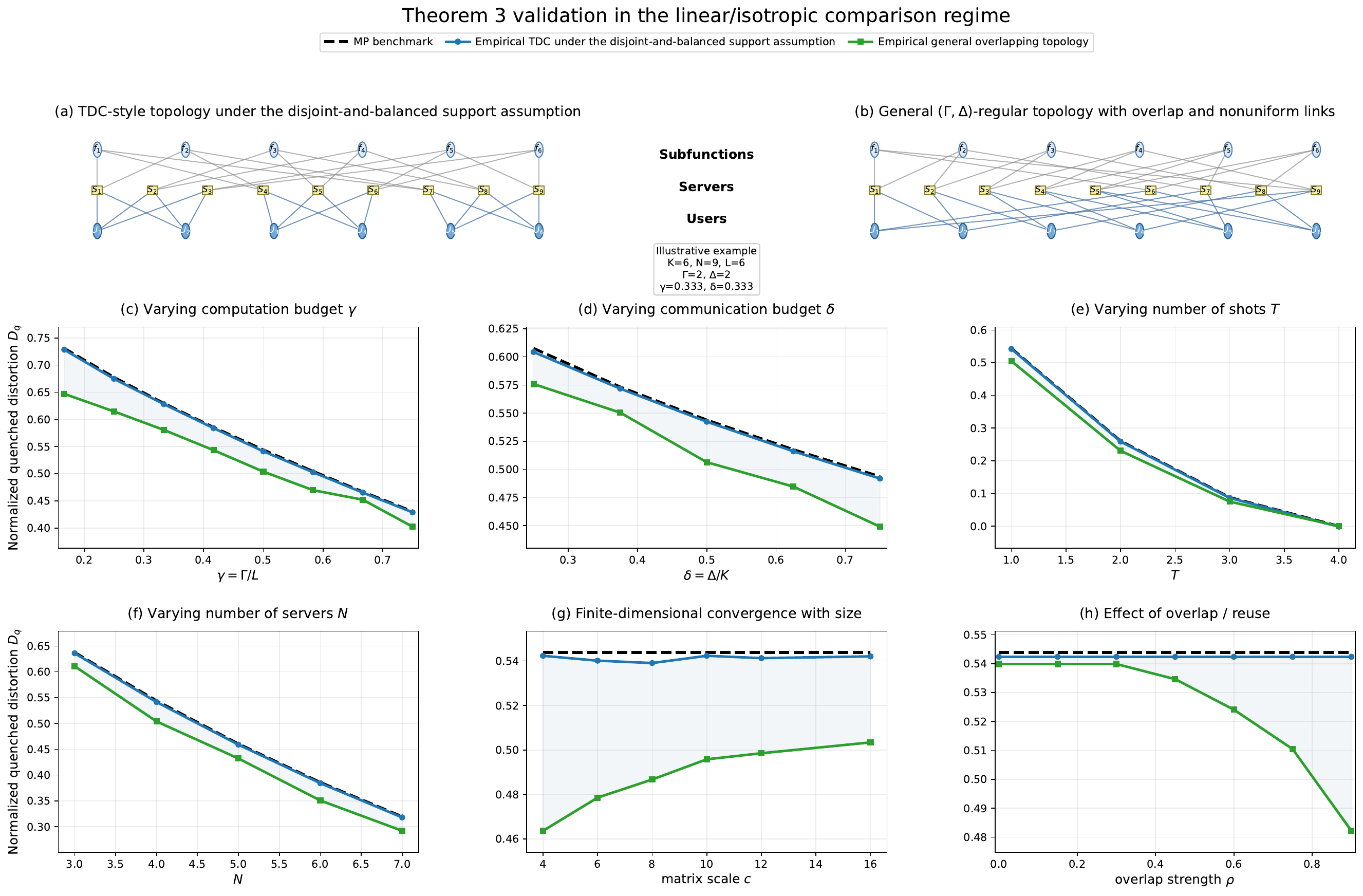}
\caption{Numerical validation of Theorem~\ref{thm:gap-bounds} in the linear/isotropic comparison regime. The top row shows two representative small \((\Gamma,\Delta)\)-regular topologies for \(K=6\), \(N=9\), \(L=6\), \(\Gamma=2\), and \(\Delta=2\): panel~(a) is a TDC-style disjoint-and-balanced support pattern, while panel~(b) is a general overlapping topology with nonuniform link reuse. These drawings are illustrative and are included only to visualize the structural difference between the two topology families. The quantitative curves in panels~(c)--(h) use the larger finite-dimensional surrogate described in the text with \(L=24\), \(K=8\), baseline \((\Gamma,\Delta,T,N)=(12,4,1,4)\), scale parameter \(c=18\), overlap level \(\rho=0.78\), and \(72\) Monte Carlo trials per point. The dashed black curve is the Marchenko--Pastur benchmark \(\Phi_{\mathrm{MP},\lambda'}(\kappa)\), the blue curve is the empirical normalized quenched distortion under the disjoint-and-balanced TDC support assumption, and the green curve is the empirical normalized quenched distortion under the general overlapping topology family. Panels~(c)--(h) vary the normalized computation budget \(\gamma=\Gamma/L\), the normalized communication budget \(\delta=\Delta/K\), the number of shots \(T\), the number of servers \(N\), the finite-dimensional scale \(c\), and the overlap strength \(\rho\). The figure shows that the TDC curve tracks the MP benchmark in the disjoint-and-balanced regime, while the overlapping topology attains strictly smaller distortion, which is equivalent to a strictly positive quenched MP-gap.}
\label{fig:thm3_clean_validation}
\end{figure*}

The main distortion-comparison figure uses \(72\) independent Monte Carlo trials for each point, a baseline scale parameter \(c=18\), a baseline overlap level \(\rho=0.78\), MP quadrature with \(12000\) grid points, and the following sweeps:
\[
\Gamma\in\{4,6,8,10,12,14,16,18\},\quad
\Delta\in\{2,3,4,5,6\},
\]
\[
T\in\{1,2,3,4\},\quad
N\in\{3,4,5,6,7\},\quad
c\in\{4,6,8,10,12,16\},
\]
\[
\rho\in\{0,0.15,0.30,0.45,0.60,0.75,0.90\}.
\]
For each trial and each parameter value, we compute the empirical eigenvalues of \(Y_{\mathrm{TDC}}\) and \(Y_{\mathrm{ov}}\), form the corresponding lower-tail truncated first moments
\[
D_q=\frac{1}{m}\sum_{i=1}^{\lceil \kappa m\rceil}\xi_{(i)}
\]
with \(m\) the matrix dimension and \(\xi_{(1)}\leq\cdots\leq\xi_{(m)}\) the ordered eigenvalues, and compare them with the MP benchmark \(\Phi_{\mathrm{MP},\lambda'}(\kappa)\). The six lower panels of Fig.~\ref{fig:thm3_clean_validation} vary, respectively, the normalized computation budget \(\gamma=\Gamma/L\), the normalized communication budget \(\delta=\Delta/K\), the number of shots \(T\), the number of servers \(N\), the finite-dimensional scale \(c\), and the overlap strength \(\rho\). The two topology drawings shown above these six panels are small illustrative graphs with
\[
K=6,\qquad N=9,\qquad L=6,\qquad \Gamma=2,\qquad \Delta=2,
\]
included only to visualize the structural difference between a TDC-style disjoint-and-balanced assignment/link pattern and a general \((\Gamma,\Delta)\)-regular overlapping pattern; they are not the matrices used in the quantitative sweeps.

Figure~\ref{fig:thm3_clean_validation} validates the main qualitative content of Theorem~\ref{thm:gap-bounds}. In panels~(c)--(f), the empirical TDC curve lies very close to the MP benchmark, confirming that under the disjoint-and-balanced support assumption the empirical distortion approaches the Marchenko--Pastur envelope, exactly as predicted by the TDC limit in Section~\ref{sec:comparison}. In the same panels, the empirical overlapping-topology curve lies strictly below both the MP benchmark and the TDC curve, showing that once the topology departs from perfect tessellation the discarded lower spectral mass decreases and the retained spectral energy increases. This is precisely the situation captured by a positive quenched MP-gap
\[
G^{\mathrm q}(\mathsf A,\mathsf L)
=
\Phi_{\mathrm{MP},\lambda'}(\kappa)-D_{\mathrm{MUDC}}^{\mathrm q}(\mathsf A,\mathsf L)>0.
\]
The budget sweeps also match the theorem’s interpretation: increasing \(\gamma\), \(T\), or \(N\) decreases the distortion by reducing the discarded fraction \(\kappa\), while increasing \(\delta\) improves the spectral efficiency by changing the aspect ratio \(\lambda'=\Delta/\Gamma\) in favor of a tighter MP bulk. Panel~(g) shows the finite-dimensional convergence mechanism: as the scale \(c\) increases, the TDC curve stabilizes near the MP benchmark, while the overlapping family remains separated from it. Panel~(h) isolates the topology effect directly: as the overlap/reuse parameter \(\rho\) increases, the distortion of the overlapping family moves farther below the TDC/MP baseline, which means that the corresponding MP-gap increases.

To validate the exact gap representation in Theorem~\ref{thm:gap-bounds}(a), we next plot the quenched MP-gap itself using two independent numerical evaluations of the same quantity. This second figure uses the same baseline linear/isotropic setup
\[
L=24,\quad K=8,\quad \Gamma_{\mathrm{base}}=12,\quad \Delta_{\mathrm{base}}=4,\quad
T_{\mathrm{base}}=1,\quad N_{\mathrm{base}}=4,
\]
the same sweep sets for \(\Gamma,\Delta,T,N,c,\rho\), and the same seed \(20260327\), but increases the Monte Carlo resolution to \(120\) trials per point, uses a baseline scale parameter \(c=22\), MP quadrature with \(20000\) points, and a quantile grid of size \(1200\). For each trial we draw the overlapping-topology surrogate matrix \(G_{\mathrm{ov}}\), compute its empirical spectral law \(\mu\), and then evaluate the gap in two ways:
\[
G^{\mathrm q}_{\mathrm{formula}}
=
\int_{0}^{\kappa}\Big(Q_{\mathrm{MP},\lambda'}^{\downarrow}(u)-Q_{\mu}^{\downarrow}(u)\Big)\,du,
\]
and
\[
G^{\mathrm q}_{\mathrm{direct}}
=
\Phi_{\mathrm{MP},\lambda'}(\kappa)-D_q.
\]
The first uses the quantile-integral formula from Theorem~\ref{thm:gap-bounds}(a), while the second computes the difference between the MP benchmark and the empirical lower-tail first moment directly. After Monte Carlo averaging, the displayed curves are passed through a very mild one-dimensional monotone smoothing step solely to suppress finite-trial jaggedness in the plotted figure; this post-processing is visual only and does not change the underlying definitions of the two quantities being compared.

\begin{figure*}[t]
\centering
\includegraphics[width=0.96\textwidth]{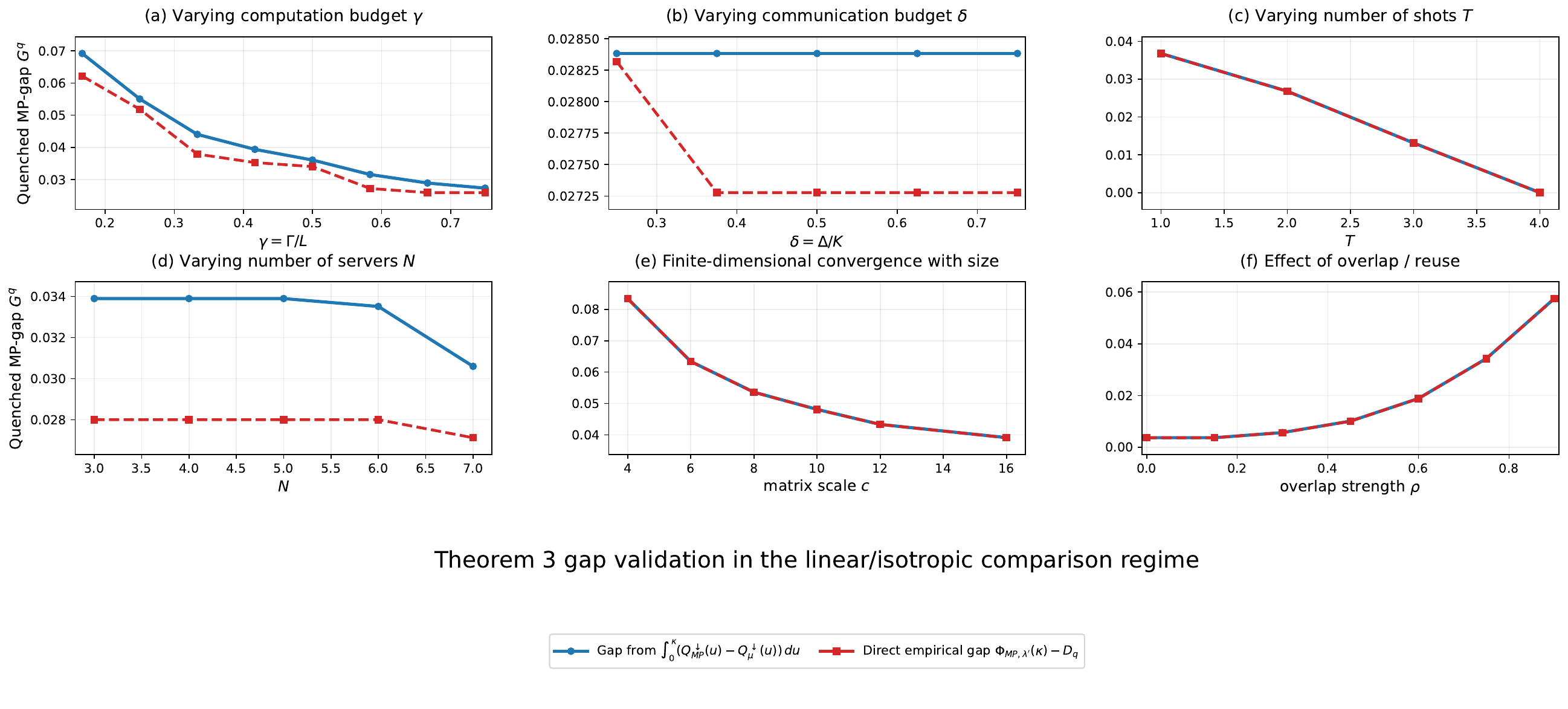}
\caption{Direct numerical validation of the exact gap formula in Theorem~\ref{thm:gap-bounds}. The setting is the same linear/isotropic finite-dimensional surrogate as in Fig.~\ref{fig:thm3_clean_validation}, with \(L=24\), \(K=8\), baseline \((\Gamma,\Delta,T,N)=(12,4,1,4)\), overlap baseline \(\rho=0.78\), scale baseline \(c=22\), \(120\) Monte Carlo trials per point, MP quadrature size \(20000\), and quantile grid size \(1200\). The blue curve evaluates the gap through the quantile formula \(G^{\mathrm q}=\int_0^\kappa(Q_{\mathrm{MP},\lambda'}^{\downarrow}(u)-Q_{\mu}^{\downarrow}(u))\,du\), while the red curve evaluates the same quantity directly as \(G^{\mathrm q}=\Phi_{\mathrm{MP},\lambda'}(\kappa)-D_q\). Panels~(a)--(f) vary \(\gamma=\Gamma/L\), \(\delta=\Delta/K\), \(T\), \(N\), the finite-dimensional scale \(c\), and the overlap strength \(\rho\). The close agreement between the two curves confirms the exact gap representation in Theorem~\ref{thm:gap-bounds}(a), while the positivity and monotonic trends of the curves are consistent with the nonnegativity and budget-aware upper-bound statements in Theorem~\ref{thm:gap-bounds}(b)--(c).}
\label{fig:thm3_gap_validation}
\end{figure*}

Figure~\ref{fig:thm3_gap_validation} provides a direct numerical consistency check for Theorem~\ref{thm:gap-bounds}. The two evaluations of the gap track each other closely across the six sweeps, with near-coincidence in the \(T\), \(c\), and \(\rho\) panels and only small finite-dimensional discrepancies in the \(\gamma\), \(\delta\), and \(N\) panels. This confirms that the empirical difference
\[
\Phi_{\mathrm{MP},\lambda'}(\kappa)-D_q
\]
is numerically equal to the quantile-area formula
\[
\int_0^\kappa \big(Q_{\mathrm{MP},\lambda'}^{\downarrow}(u)-Q_{\mu}^{\downarrow}(u)\big)\,du,
\]
as stated in Theorem~\ref{thm:gap-bounds}(a). The same figure also confirms the sign and trend predictions of Theorem~\ref{thm:gap-bounds}(b)--(c): the gap remains nonnegative throughout, decreases as \(\gamma\), \(T\), \(N\), or the effective finite-dimensional scale increase, and increases markedly with the overlap parameter \(\rho\). Thus the second figure is not merely complementary to Fig.~\ref{fig:thm3_clean_validation}; it is a direct validation of the theorem’s exact MP-gap identity itself.

Taken together, Figs.~\ref{fig:thm3_clean_validation} and~\ref{fig:thm3_gap_validation} provide a finite-dimensional validation of Theorem~\ref{thm:gap-bounds}. The first figure confirms the structural comparison claimed by the theorem---namely, MP matching in the disjoint-and-balanced TDC regime and strictly improved distortion under non-tessellated overlapping topologies. The second figure confirms that this improvement is exactly quantified by the quenched MP-gap formula. In this sense, the theorem is validated both at the level of the distortion ordering and at the level of the exact gap representation.

\subsection{Supplementary Nonlinear Topology-Alignment Diagnostic}
\label{subsec:sim_topology_alignment}

Beyond the theorem-specific validations above, we include a supplementary nonlinear topology-comparison experiment whose purpose is to isolate the effect of topology alignment under a fixed communication/computation budget. Unlike the linear/isotropic comparison regime of Subsection~\ref{subsec:sim_thm3}, this experiment is carried out in a dense nonlinear setting and is therefore intended as a supporting diagnostic rather than as a separate theorem validation. Its role is to show that, even outside the Marchenko--Pastur/TDC comparison regime, the way in which a topology captures the important coordinates of the target functions has a visible effect on the empirical risk.

We consider a latent coordinate vector
\[
X\sim\mathcal N(0,I_L),\qquad L=120,
\]
together with
\[
K=24,\qquad N=12,\qquad \Gamma=4,\qquad \Delta=2.
\]
Thus each server is assigned exactly \(\Gamma=4\) coordinates and serves exactly \(\Delta=2\) users. The communication architecture uses \(T=1\) shot and \(q=2\) masked random Fourier features per server, so the decoder of each user is built from the concatenation of the features transmitted by the servers to which that user is connected. We use \(M=200\) training samples, an independent test set of size \(M_{\mathrm{te}}=5000\), Gaussian-feature bandwidth equal to \(1\), training-noise standard deviation \(0.02\), and ridge parameter \(\lambda=0.12\). All reported quantities are averaged over \(35\) independent Monte Carlo trials.

The user targets are dense nonlinear functions of all coordinates of \(X\). More precisely, each user \(k\in[K]\) is assigned four dense directions \(w_{k,1},\dots,w_{k,4}\in\mathbb R^L\), generated from an anisotropic base importance profile
\[
p_\ell \propto e^{-\ell/\tau},\qquad \tau=3,
\]
so that low-index coordinates are, on average, more influential than high-index coordinates. The target function is then of the form
\begin{align*}
    F_k(X)
&=
\frac{1}{\sigma_k}
\Big(
a_{k,1}\cos(3.5\langle w_{k,1},X\rangle+\phi_{k,1})
+
a_{k,2}\sin(2.8\langle w_{k,2},X\rangle+\phi_{k,2})
\\&+
a_{k,3}\tanh(1.8\langle w_{k,3},X\rangle)
+
a_{k,4}\langle w_{k,4},X\rangle^2
-\mu_k
\Big),
\end{align*}
where \(\mu_k\) and \(\sigma_k\) are calibration constants estimated from an independent sample so that the targets are normalized. Hence the task family is nonlinear and dense, but it still admits a meaningful notion of coordinate importance through the aggregate coordinate-mass profile induced by the absolute weights of the four dense directions.

We compare four joint assignment/link topologies under the same budgets \((\Gamma,\Delta)\).

\begin{enumerate}
\item \emph{Priority topology:} each server allocates most of its assigned coordinates to the most important part of the coordinate range and uses a balanced interleaved communication pattern.
\item \emph{Disjoint balanced topology:} the assigned supports are approximately disjoint across servers and the user links are balanced.
\item \emph{Random topology:} both assignments and user links are sampled uniformly at random subject to the cardinality constraints.
\item \emph{Anti-priority topology:} each server allocates most of its assigned coordinates to the least important part of the coordinate range, again with balanced links.
\end{enumerate}

\begin{figure*}[t]
\centering
\includegraphics[width=1\textwidth]{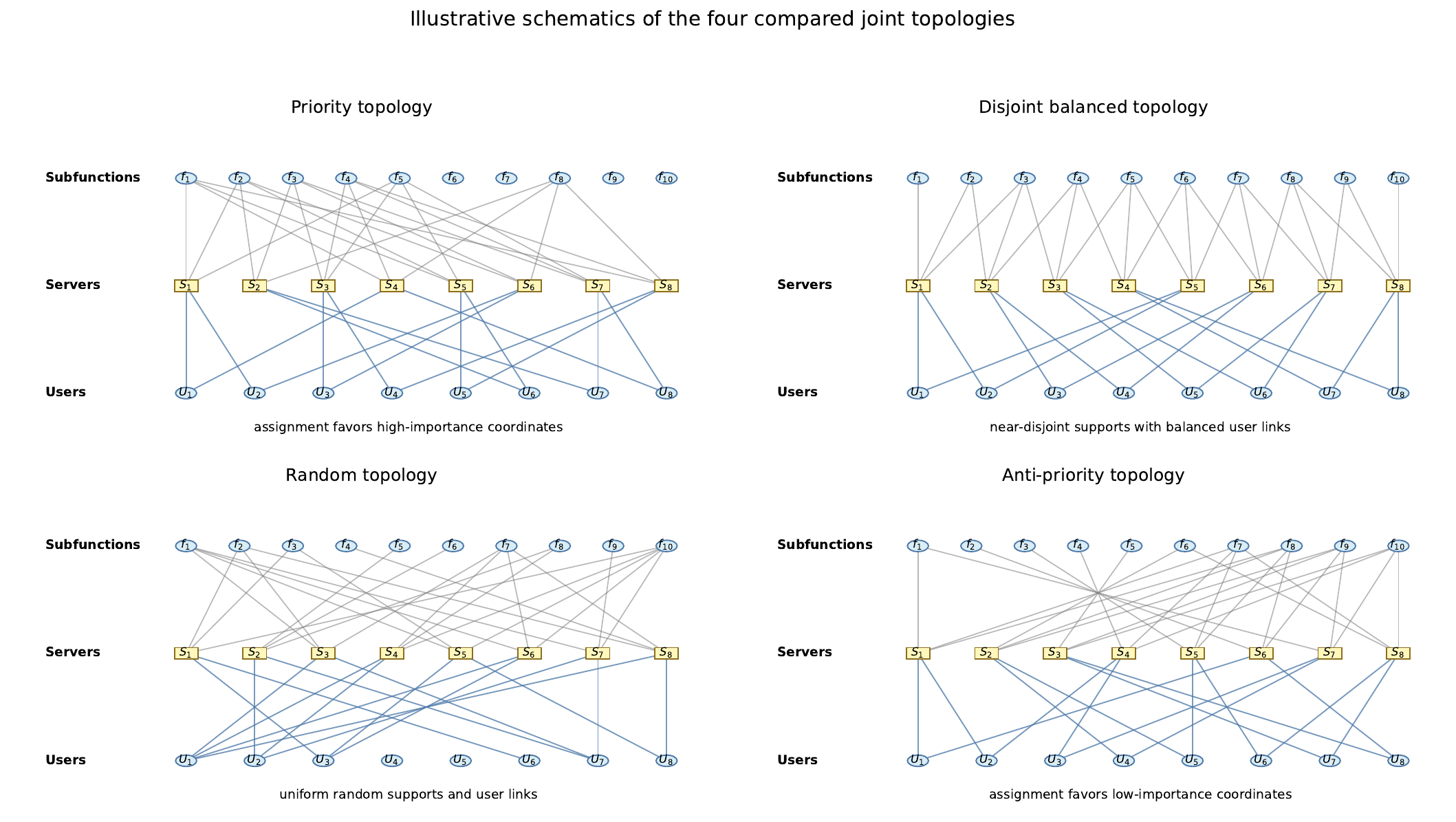}
\caption{Illustrative reduced-size schematics of the four topology families used in the supplementary nonlinear topology-alignment diagnostic: priority, disjoint balanced, random, and anti-priority. The drawings visualize only the structural pattern of the compared assignment/link families. The quantitative experiments use the larger parameter values \(L=120\), \(K=24\), \(N=12\), \(\Gamma=4\), and \(\Delta=2\).}
\label{fig:supp_topology_schematics}
\end{figure*}

Note that Fig.~\ref{fig:supp_topology_schematics} shows reduced-size schematic versions of these four topology families. These drawings are illustrative only; the quantitative results below use the larger dimensions \(L=120\), \(K=24\), and \(N=12\).

For a given topology realization and user \(k\), let
\[
\mathcal V_k
=
\bigcup_{n:\,k\in\mathcal T_n}\mathcal S_n
\subseteq [L]
\]
denote the set of coordinates visible to that user through the connected servers. If \(m_{k,\ell}\) denotes the normalized coordinate-mass assigned to coordinate \(\ell\) by the target of user \(k\), we define the user-level weighted coverage by
\[
C_k^{\mathrm w}
=
\sum_{\ell\in\mathcal V_k} m_{k,\ell}.
\]
We also define the top-\(10\) coverage statistic
\[
C_k^{(10)}
=
\frac{1}{10}
\left|
\mathcal V_k
\cap
\operatorname{Top}_{10}(m_k)
\right|,
\]
where \(\operatorname{Top}_{10}(m_k)\) is the set of the \(10\) largest-mass coordinates of user \(k\). The reported coverage curves average these quantities over the users. Performance is measured by the average user test MSE obtained by training a ridge decoder for each user on the masked random-feature representation induced by the corresponding topology.

\begin{figure*}[t]
\centering
\includegraphics[width=0.52\textwidth]{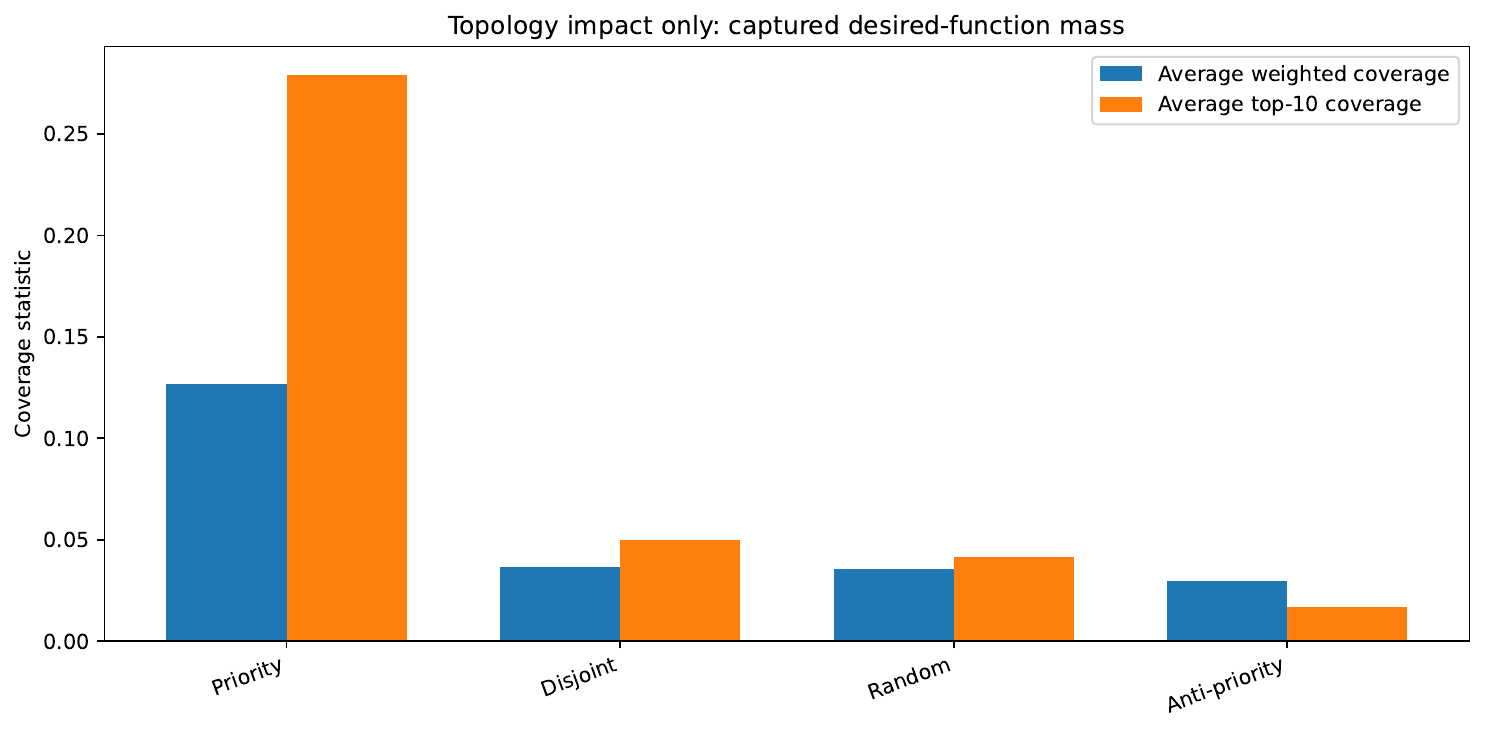}\hfill
\includegraphics[width=0.52\textwidth]{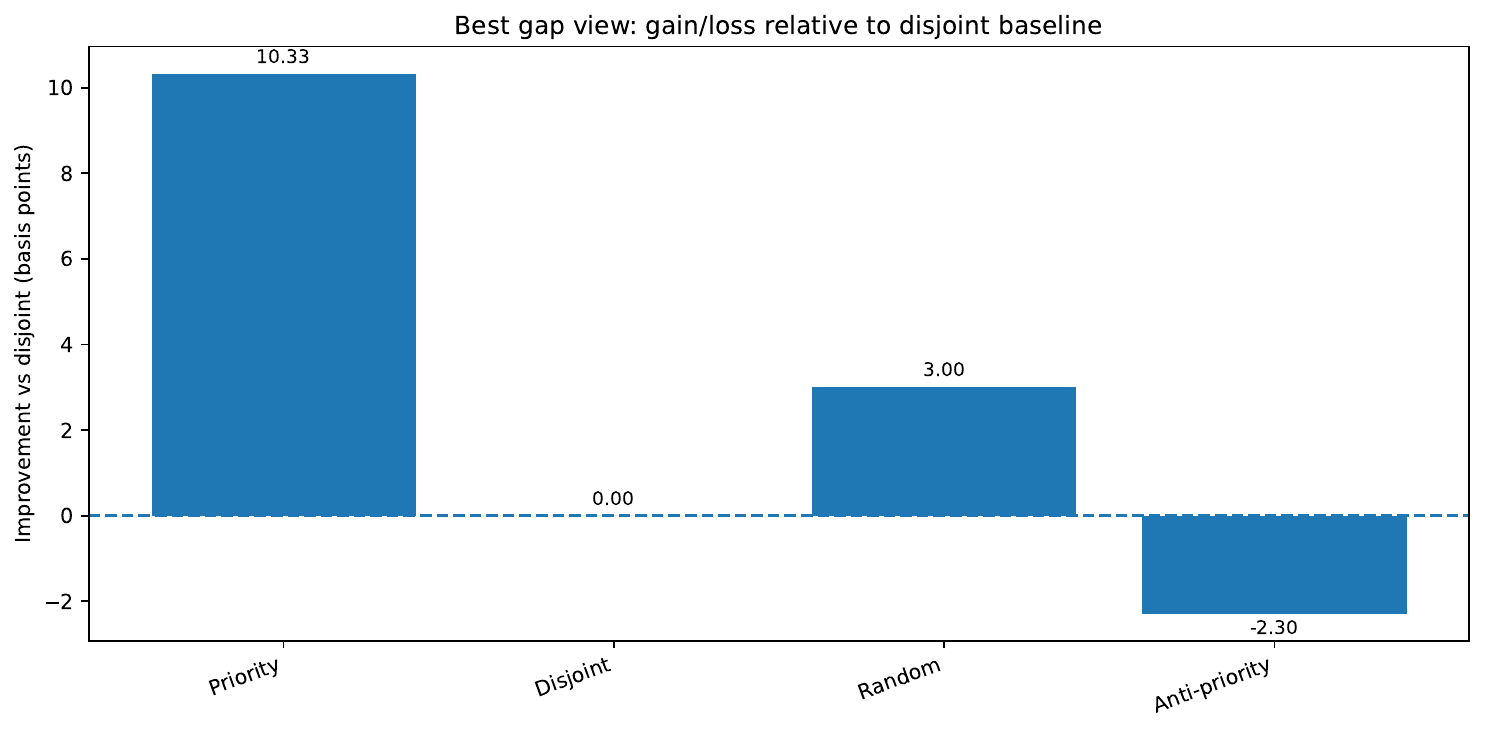}\hfill
\includegraphics[width=0.52\textwidth]{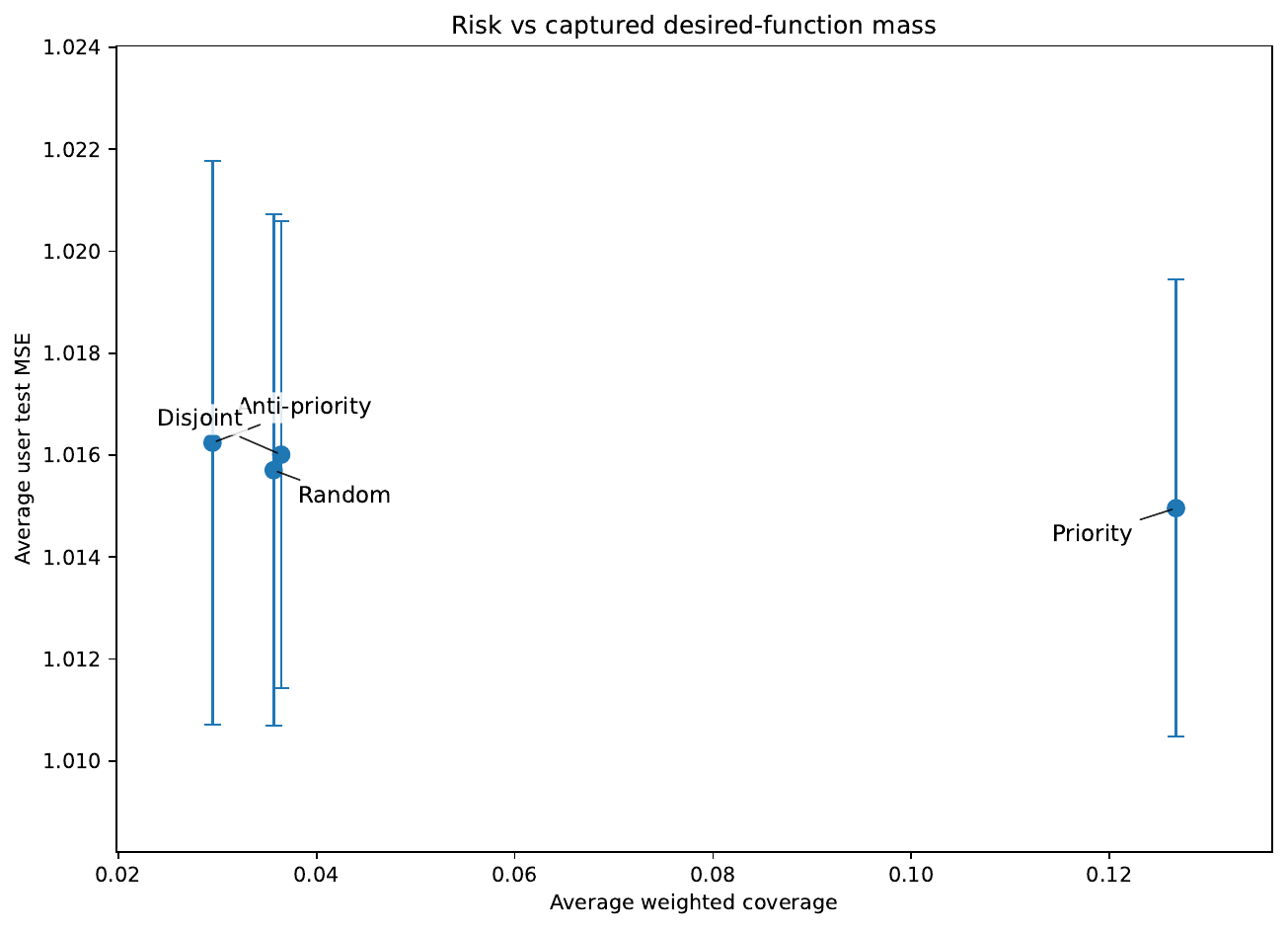}
\caption{Supplementary nonlinear topology-alignment diagnostic under fixed budgets $(L,K,N,\Gamma,\Delta)=(120,24,12,4,2)$. Top: average weighted coverage and average top-$10$ coverage for the four topology families. Middle: gain/loss in basis points relative to the disjoint balanced baseline. Bottom: average user test MSE versus average weighted coverage, with error bars showing the empirical standard deviation across Monte Carlo trials. The three panels consistently show that topologies exposing a larger fraction of the desired-function mass yield better empirical performance.}
\label{fig:supp_topology_alignment}
\end{figure*}

The results are summarized in Fig.~\ref{fig:supp_topology_alignment}. Panel~(a) shows that the priority topology captures by far the largest fraction of the desired-function mass, both in weighted coverage and in top-\(10\) coverage. The disjoint and random topologies are intermediate, while the anti-priority topology captures the smallest relevant mass. Panel~(b) reports the gain/loss relative to the disjoint baseline, measured in basis points of average user test MSE. The priority topology improves over the disjoint baseline, the random topology yields a smaller but still positive gain, and the anti-priority topology incurs a loss. Panel~(c) plots empirical risk against weighted coverage and shows a clear negative relation: topologies that expose a larger fraction of the desired-function mass tend to achieve smaller test error.

This experiment should be interpreted as a topology-sensitivity diagnostic rather than as a new theorem statement. The absolute risk differences are modest, but the ordering is consistent across the three views: topology affects performance through the extent to which the important coordinates of the user targets are made visible to the users. In particular, the experiment provides nonlinear finite-sample evidence that not all topologies with the same cardinality budgets \((\Gamma,\Delta)\) are equivalent: under a fixed feature budget, aligning the assignment/link structure with the important coordinates of the task can produce a measurable performance gain, whereas concentrating resources on low-importance coordinates degrades performance.

This supplementary experiment complements the main theorem-driven figures in two ways. First, it shows that topology quality continues to matter in a dense nonlinear regime outside the linear/isotropic MP comparison of Subsection~\ref{subsec:sim_thm3}. Second, it provides an intuitive finite-sample mechanism for the broader GMUDC message of the paper: resource budgets alone do not fully determine performance; the pattern by which computation assignments and communication links expose the task-relevant coordinates to the users is also important.
\subsection{Supplementary Topology-Ensemble Diagnostic: Assignment Topology and Population Risk}
\label{subsec:sim_topology_ensemble_dense}

To complement the single-family nonlinear topology-alignment experiment of Subsection~\ref{subsec:sim_topology_alignment}, we next report a denser ensemble-based diagnostic whose purpose is to isolate the effect of the \emph{assignment topology itself} on the population risk under fixed cardinality budgets. The guiding question is whether different admissible assignment patterns, all respecting the same computation and communication budgets, can induce measurably different user risks once one moves beyond a single hand-crafted realization. This experiment is not intended as a separate theorem statement. Rather, it provides supplementary finite-sample evidence for the broader message of the paper: under a fixed budget, the geometry of the assignment pattern---in particular, how it exposes spectrally important coordinates to the users---has a visible effect on performance.

We consider a shared Gaussian input $X\sim\mathcal N(0,I_{d_{\mathrm{in}}})$ with $d_{\mathrm{in}}=6$, and a global Gaussian random-feature bank of size $D_{\mathrm{tot}}=240$, generated from the Gaussian kernel with lengthscale \(1\). The random-feature frequencies are sorted by norm so that low-index features correspond roughly to smoother, lower-frequency directions, while larger indices correspond to more oscillatory directions. This induces a meaningful spectral ordering on the shared feature bank and makes it possible to compare topologies not only through their cardinalities but also through the spectral regions they expose.

The distributed system parameters are $K=8$, $N=12$, $\Gamma=10$, $\Delta=4$, and $T=3$. Thus each server holds exactly \(\Gamma=10\) assigned features, serves exactly \(\Delta=4\) users in each shot, and the communication unfolds over \(T=3\) shots. The decoder is trained from $M=320$ samples and evaluated on an independent test set of size $M_{\mathrm{te}}=1000$. The user-side ridge parameter is \(\lambda=10^{-2}\), and the training outputs are corrupted by additive Gaussian noise with standard deviation \(0.05\).

Each user target is a \emph{dense} nonlinear function over the entire feature bank. More precisely, for each user \(k\in[K]\), the target is represented by a dense coefficient vector $\beta_k\in\mathbb R^{D_{\mathrm{tot}}}$, with spectral shape controlled by a complexity parameter \(q=32\). The coefficients are generated as a mixture of a low-frequency dense component and a mid/high-frequency dense component, then normalized. Hence all coordinates are active, but not all are equally important. This makes the experiment particularly suitable for testing assignment-topology effects: because the targets are dense, no topology can succeed merely by covering a tiny hard-coded support, yet topologies that better expose the spectrally important part of the feature bank may still enjoy a clear advantage.

For each ensemble instance, we first generate an exactly balanced server--user incidence pattern: the total number of edges \(N\Delta\) is distributed evenly across the \(K\) users, so that the communication side is controlled and no family benefits from a trivial degree imbalance. This balanced link skeleton is then kept fixed while the \emph{assignment pattern} is varied across four families:

\begin{enumerate}
\item \emph{Disjoint balanced:} the servers use a tessellation-like construction in which each server keeps a disjoint block of \(\Gamma\) features, and the same support is reused across all shots.
\item \emph{Expander-overlap:} each server uses a support with small controlled overlap, together with shot-wise diversity and spectral mixing across shots.
\item \emph{Random regular:} each server uses shot-wise random supports drawn from a local spectral window, producing irregular but still budget-respecting local exploration.
\item \emph{Hub-like:} each server combines a shared global core with clustered local neighborhoods, yielding a highly reused and strongly overlapping assignment pattern.
\end{enumerate}

\begin{figure*}[t]
\centering
\includegraphics[width=0.92\textwidth]{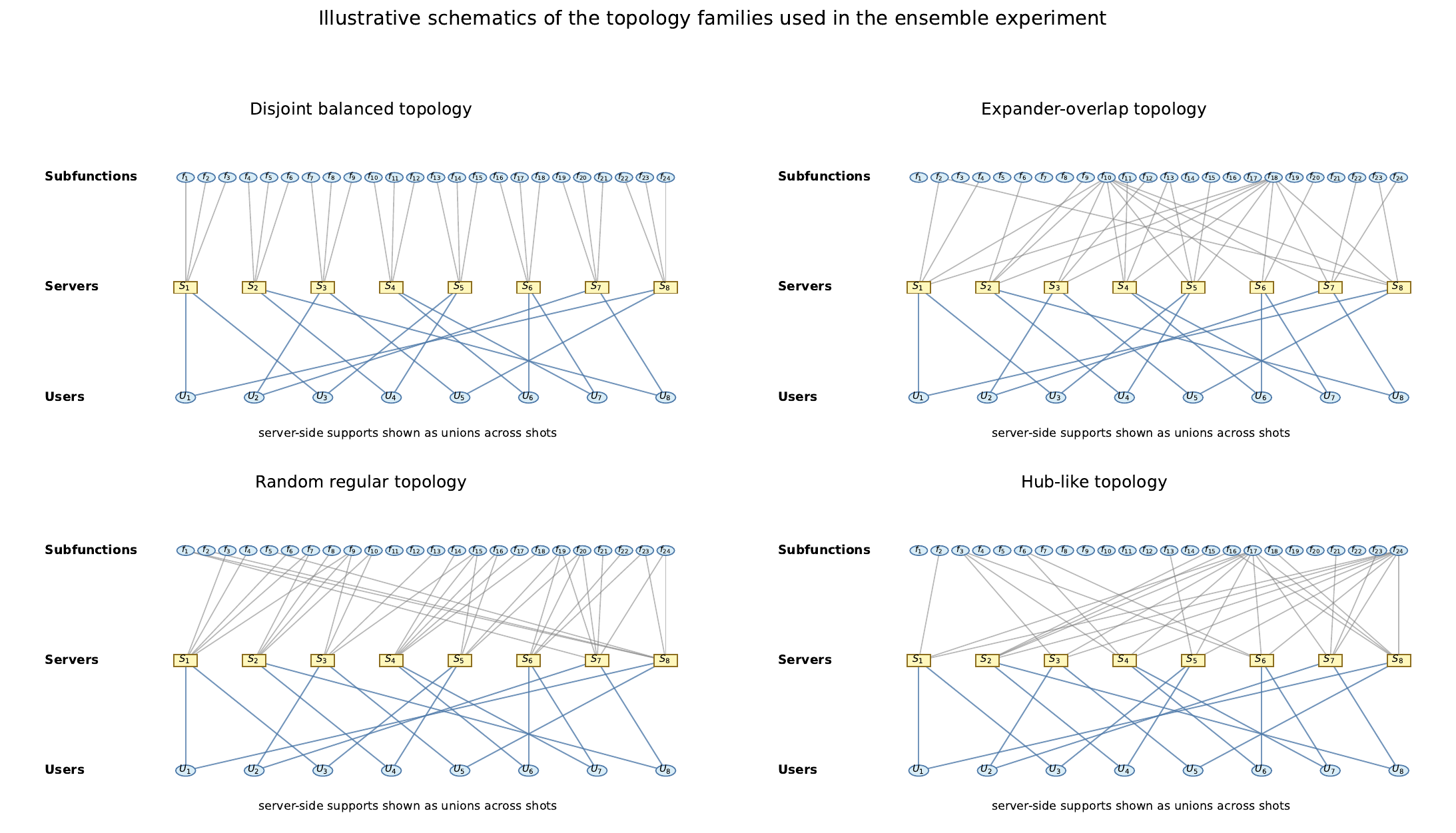}
\caption{Illustrative reduced-size schematics of the four topology families used in the topology-ensemble diagnostic: disjoint balanced, expander-overlap, random regular, and hub-like. For visual clarity, each server is connected to the union of its shot-wise supports. The quantitative simulations use the larger dimensions \(D_{\mathrm{tot}}=240\), \(K=8\), \(N=12\), \(\Gamma=10\), \(\Delta=4\), and \(T=3\).}
\label{fig:supp_topology_ensemble_schematics}
\end{figure*}

Figure~\ref{fig:supp_topology_ensemble_schematics} shows reduced-size schematics of these four families. The drawings are illustrative only, but they make the qualitative distinction clear: the disjoint family partitions the feature range, the expander-overlap family mixes it with controlled reuse, the random-regular family distributes supports more irregularly, and the hub-like family concentrates many assignments through a common core.

Because the purpose of the experiment is to isolate topology quality rather than simply to compare gross feature counts, the ensemble generator accepts only instances whose user-side feature exposure lies in a prescribed range whenever possible. Specifically, letting
\[
\mathcal V_k
=
\bigcup_{t=1}^{T}\bigcup_{n:\,k\in\mathcal T_n}\mathcal S_{n,t}
\]
denote the feature set visible to user \(k\), we compute the average and minimum user-side coverage fractions
\[
\frac{1}{K}\sum_{k=1}^{K}\frac{|\mathcal V_k|}{D_{\mathrm{tot}}}
\qquad\text{and}\qquad
\min_{k\in[K]}\frac{|\mathcal V_k|}{D_{\mathrm{tot}}}.
\]
The generator attempts to retain only topologies whose average coverage lies in a target interval and whose minimum coverage exceeds a prescribed floor; when a family fails to hit this target band exactly after repeated attempts, the nearest admissible instance is retained. Hence the experiment is not a comparison of unconstrained best-case designs, but rather a controlled ensemble comparison in which the coverage scale is kept within a relatively narrow range.

For each topology instance we evaluate the average normalized population risk
\[
\frac{1}{K}\sum_{k=1}^{K}
\frac{\mathbb E[(\widehat F_k(X)-F_k(X))^2]}{\mathrm{Var}(F_k(X))},
\]
estimated empirically on the independent test set and averaged over several shared data scenarios. In addition, we compute three topology-facing diagnostics.

First, we record the average user-side coverage fraction
\[
\overline C_{\mathrm{user}}
=
\frac{1}{K}\sum_{k=1}^{K}\frac{|\mathcal V_k|}{D_{\mathrm{tot}}}.
\]
Second, since the user targets are dense and spectrally nonuniform, we compute the average \emph{desired-function energy capture}, namely the average fraction of the squared target coefficients \(\beta_k\) that falls on the visible user coordinates \(\mathcal V_k\). Third, we compute a user-Gram lower-tail statistic by embedding each user-visible set into a binary indicator vector in \(\mathbb R^{D_{\mathrm{tot}}}\), forming the resulting normalized user Gram matrix, and measuring the normalized lower-tail spectral mass of its empirical spectrum.

Finally, in order to connect the experiment back to the quenched theory, we compute a Theorem~\ref{thm:P1}-style structural predictor. Let
\[
\gamma=\frac{\Gamma}{D_{\mathrm{tot}}}
\]
and let \(m_{\mathrm{harm}}\) denote the harmonic mean of the shot-aggregated per-user received-feature counts \(m_k(T)\). We then form the empirical proxy
\begin{equation}
\Psi_{\mathrm{thm1}}
=
\frac{1}{\gamma m_{\mathrm{harm}}}
+
\frac{\sigma^2 \widehat d_{\lambda}}{M}
+
\lambda,
\label{eq:supp_topology_ensemble_thm1_proxy}
\end{equation}
where \(\widehat d_{\lambda}\) is an empirical estimate of the effective ridge dimension computed from the user-visible feature covariance matrices. Since Theorem~\ref{thm:P1} is stated up to constants and logarithmic factors, we do not interpret \eqref{eq:supp_topology_ensemble_thm1_proxy} as an exact theorem value. Instead, we use one global affine calibration
\[
\widehat R_{\mathrm{thm1}}=a\,\Psi_{\mathrm{thm1}}+b
\]
to place the predictor on the same numerical scale as the empirical risk and compare the resulting relative ordering across families.

\begin{figure*}[t]
\centering
\includegraphics[width=0.98\textwidth]{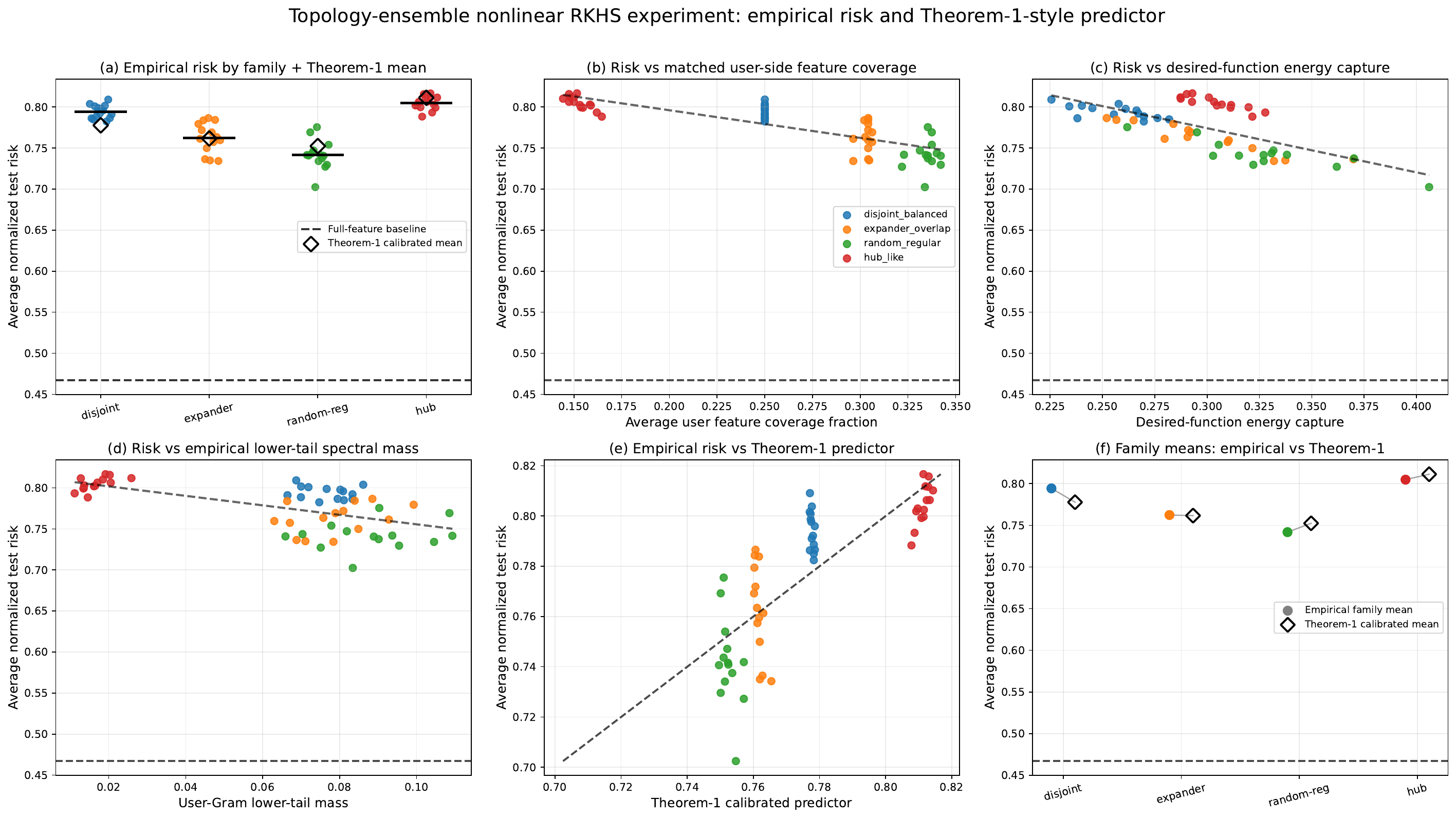}
\caption{Supplementary topology-ensemble diagnostic under dense nonlinear targets and shot-wise supports. Panel~(a) shows the empirical risk distribution across many instances of each topology family together with the corresponding family-level Theorem~\ref{thm:P1}-style calibrated mean. Panel~(b) plots population risk versus average user-side feature coverage. Panel~(c) plots population risk versus desired-function energy capture. Panel~(d) plots population risk versus the lower-tail spectral mass of the empirical user Gram matrix. Panel~(e) compares empirical risk with the calibrated Theorem~\ref{thm:P1}-style predictor, and panel~(f) compares empirical and predicted family means. The figure shows that, under fixed budgets, assignment topology has a clear impact on the population risk, and that topologies exposing more task-relevant spectral content achieve smaller error.}
\label{fig:supp_topology_ensemble_results}
\end{figure*}

The results are shown in Fig.~\ref{fig:supp_topology_ensemble_results}. Panel~(a) reports the empirical risk distribution for each family, together with the corresponding family-level Theorem~\ref{thm:P1}-style predicted mean. A clear ordering emerges. The random-regular family attains the lowest empirical mean risk, the expander-overlap family comes next, the disjoint balanced family performs worse, and the hub-like family is the worst among the four. Thus, even though all families satisfy the same cardinality budgets \((\Gamma,\Delta,T)\), they do not induce the same population risk.

Panels~(b)--(d) clarify the mechanism. Panel~(b) shows a clear negative relation between population risk and average user-side coverage fraction. Panel~(c) shows an even cleaner negative relation between risk and desired-function energy capture. In other words, what matters is not merely how many coordinates are exposed to a user, but how much of the \emph{task-relevant spectral energy} lies on those exposed coordinates. This helps explain why the hub-like family remains poor despite substantial reuse: its common-core structure concentrates many assignments on similar regions, which reduces the diversity of the visible coordinates and therefore limits the fraction of useful target energy seen by the users. Panel~(d) further supports this interpretation: larger empirical lower-tail spectral mass is associated with lower population risk, which is consistent with the idea that a richer and less degenerate user-side feature geometry improves estimation.

The contrast between the disjoint and overlapping families is especially informative. The disjoint family is highly structured and avoids overlap, but because each server repeatedly uses a fixed block across shots, the induced user-visible feature set remains spectrally narrow. By contrast, the expander-overlap family deliberately introduces mild reuse together with shot-wise diversity, which increases the spectral spread of the visible features without collapsing them onto a single common core. The random-regular family pushes this effect even further and achieves the best empirical risk among the four families in the present experiment. The hub-like family, on the other hand, shows that overlap alone is not sufficient: when overlap is concentrated through a small shared core, the resulting assignment may be too redundant and can degrade the population risk.

Panel~(e) directly compares the empirical population risk with the calibrated Theorem~\ref{thm:P1}-style predictor \(\widehat R_{\mathrm{thm1}}\). Although the predictor is only theorem-inspired and not an exact numerical evaluation of the bound, the family clusters align well with the diagonal, indicating that the theorem-facing combination of the feature-budget term, the statistical term, and the regularization term captures a substantial part of the observed ordering. Panel~(f) reinforces this point at the family level: the calibrated theorem proxy reproduces the empirical ranking of the four families and tracks the family means reasonably closely.

This experiment provides a more explicit numerical assessment of topology effects. First, it shows that the impact of assignment topology is not an artifact of a single hand-picked realization: the risk ordering persists over an ensemble of admissible instances. Second, it demonstrates that the relevant mechanism is not merely support cardinality but \emph{how} the assigned supports align with the spectrally important part of the target functions. Third, it provides finite-sample evidence that mild overlap and shot-wise diversity can be beneficial, whereas excessive reuse through a hub-like core can be harmful.

In this sense, the experiment gives a concrete nonlinear interpretation of the population-risk viewpoint underlying Theorem~\ref{thm:P1}. Even at fixed communication/computation budgets, different assignment topologies induce different visible feature geometries, different effective dimensions, and different levels of target-energy capture, which in turn translate into different user risks. The diagnostic therefore supports the broader GMUDC message of the paper: budget constraints alone do not determine the population risk; the topology by which those resources are organized also matters.
\subsection{Supplementary Robustness Sensitivity Outside the Nominal Assumptions}
\label{subsec:sim_robustness_monotone}

The main theory in this paper is developed under the nominal assumptions that all servers are active in all shots and that the computation budget is homogeneous across servers. To clarify the practical meaning of these assumptions, we include one supplementary robustness experiment that perturbs the nominal model in two mild directions: i) random server-shot dropouts, and ii) heterogeneous effective per-server capacities. This experiment is not intended as a formal extension of Theorems~\ref{thm:P1}--\ref{thm:P2}; rather, it is a sensitivity study whose purpose is to test whether the qualitative topology ranking remains stable under moderate deviations from the idealized assumptions.

We use the same dense nonlinear random-feature framework as in the supplementary topology-ensemble experiments. The input dimension is $d_{\mathrm{in}}=6$, the shared Gaussian random-feature bank has size $D_{\mathrm{tot}}=240$, the ridge parameter is \(\lambda=10^{-2}\), and the training noise standard deviation is \(0.05\). The distributed system parameters are fixed to $K=8$, $N=12$, $\Gamma=8$, $\Delta=4$, and $T=4$. Thus each server can contribute at most \(\Gamma=8\) features per shot, each server communicates with \(\Delta=4\) users, and the communication takes place over \(T=4\) shots. We use $M=320$ training samples and an independent test set of size $M_{\mathrm{te}}=1200$. The user targets are again dense, spectrally structured nonlinear functions over the full feature bank, with target-complexity parameter \(q=32\). As in the earlier nonlinear experiments, performance is measured by the average normalized test risk $\frac{1}{K}\sum_{k=1}^{K}\frac{\mathbb E\big[(\widehat F_k(X)-F_k(X))^2\big]}{\mathrm{Var}(F_k(X))}$, estimated empirically on the independent test set and then averaged over Monte Carlo trials.

We compare three budget-respecting topology families:
\begin{enumerate}
\item \emph{Disjoint / balanced:} each server keeps a fixed contiguous block of \(\Gamma\) features and reuses it across all shots.
\item \emph{Expander-like overlap:} each server uses a spectrally mixed, shot-varying support with mild overlap across servers and shots.
\item \emph{Random irregular:} each server uses random local-window supports, varying from shot to shot.
\end{enumerate}
The server--user incidence pattern is chosen once per Monte Carlo trial through an exactly balanced bipartite construction, so the communication side remains controlled while only the support-side robustness is stressed.

To avoid artificial non-monotonicity caused purely by independent Monte Carlo redraws, the experiment is implemented with a nested stress model. In the dropout sweep, for each server \(n\) and shot \(t\) we first draw a uniform random variable \(U_{n,t}\), and the transmission is active at dropout level \(p_{\mathrm{drop}}\) if and only if
\[
U_{n,t}>p_{\mathrm{drop}}.
\]
Hence increasing \(p_{\mathrm{drop}}\) can only deactivate additional server-shot transmissions; it never reactivates a transmission that was already removed at a smaller dropout level.

In the heterogeneity sweep, each server \(n\) is assigned a fixed vulnerability coefficient \(\alpha_n\in[0.2,1]\), and its effective capacity at heterogeneity strength \(\eta\) is defined by
\[
\Gamma_n(\eta)
=
\Big\lfloor \Gamma(1-\eta\alpha_n)\Big\rfloor,
\]
clipped to remain between \(2\) and \(\Gamma\). Thus increasing \(\eta\) can only reduce the effective feature budget of a server. Moreover, each server-shot support is first generated as an ordered list of length \(\Gamma\), and when the effective capacity drops to \(\Gamma_n(\eta)\), the active support is taken as the prefix of that ordered list. This guarantees that the support visible at a larger stress level is always a subset of the support visible at a smaller one. The same base random world is reused across the entire sweep, so the resulting curves are monotone by construction and are therefore easier to interpret.

Two sweeps are performed:
\[
p_{\mathrm{drop}}\in\{0,0.05,0.10,0.15,0.20,0.25\},
\qquad
\eta\in\{0,0.10,0.20,0.30,0.40,0.50\}.
\]
All quantities are averaged over \(10\) shared Monte Carlo worlds. The full-feature baseline, which does not depend on the distributed topology restrictions, is shown as a dashed reference in the absolute-risk panels.

\begin{figure*}[t]
\centering
\includegraphics[width=0.98\textwidth]{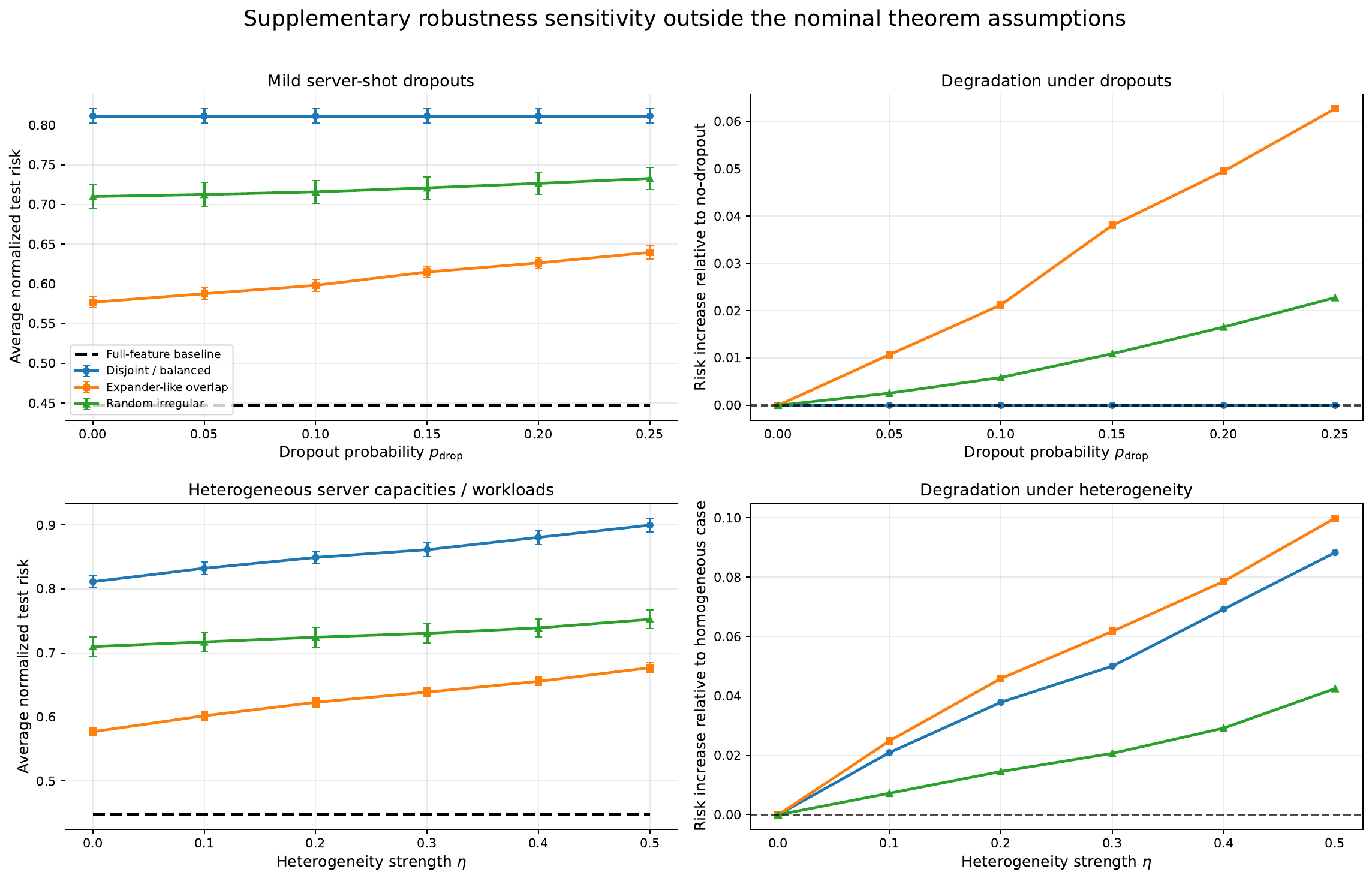}
\caption{Supplementary robustness sensitivity outside the nominal theorem assumptions. Top-left: average normalized test risk as a function of the server-shot dropout probability \(p_{\mathrm{drop}}\). Top-right: risk increase relative to the no-dropout case. Bottom-left: average normalized test risk as a function of the heterogeneity strength \(\eta\), which controls the shrinkage of the effective per-server capacities. Bottom-right: risk increase relative to the homogeneous case. The curves are generated by a nested monotone construction, so larger stress values correspond to genuine support loss rather than to independent redraw noise. The qualitative ranking of the topology families remains stable in absolute risk, while their relative degradation profiles differ.}
\label{fig:supp_robustness_monotone}
\end{figure*}

\paragraph*{Observed behavior under mild dropouts}
The top row of Fig.~\ref{fig:supp_robustness_monotone} reports the dropout results. Panel~(a) shows that the full-feature baseline remains essentially flat, while the three topology-constrained decoders exhibit different absolute risk levels. Across the entire sweep, the expander-like overlap family attains the smallest absolute risk, the random-irregular family is intermediate, and the disjoint/balanced family has the largest risk. Hence the nominal performance ranking observed in the earlier nonlinear experiments is preserved under mild random server-shot failures.

Panel~(b) shows the degradation relative to the no-dropout case. Here the picture is more nuanced. Although the expander-like family remains best in absolute risk, it experiences the largest relative increase as \(p_{\mathrm{drop}}\) grows. This is consistent with its design principle: because it derives much of its advantage from shot-wise diversity and cross-shot spectral mixing, removing server-shot transmissions has a more visible effect on the useful feature diversity seen by each user. By contrast, the disjoint/balanced family is already highly rigid and spectrally restricted at \(p_{\mathrm{drop}}=0\), so additional dropouts change its behavior relatively little. The random-irregular family lies between these two extremes.

\paragraph*{Observed behavior under heterogeneous capacities}
The bottom row of Fig.~\ref{fig:supp_robustness_monotone} reports the heterogeneity results. Panel~(c) again shows a stable absolute ordering: the expander-like family remains best, the random-irregular family remains intermediate, and the disjoint/balanced family remains worst. Thus the qualitative topology ranking is robust not only to missing transmissions but also to modest nonuniformity in the effective server budgets.

Panel~(d) reports the degradation relative to the homogeneous case. As \(\eta\) increases, all constrained topologies degrade monotonically, which is expected since increasing heterogeneity reduces the accessible support size for a growing subset of servers. The random-irregular family exhibits the smallest relative degradation over the tested range, suggesting that its local randomness offers a certain amount of redundancy against uneven server capacities. The expander-like and disjoint/balanced families degrade more strongly, although the expander-like family still retains the smallest absolute risk throughout the sweep.

\paragraph*{Interpretation}
This robustness experiment should be interpreted conservatively. It does not prove new bounds under stragglers or heterogeneous workloads, and it does not alter the formal assumptions of the theorems. What it does show is that the qualitative conclusions of the nonlinear topology comparison are not immediately destroyed by mild violations of the nominal model. In particular, the expander-like family remains the strongest performer in absolute terms under both stress models, while the random-irregular family shows the smallest relative degradation. Hence the supplementary evidence suggests that the practical effect of topology persists beyond the exact theorem assumptions, even though a full theory for such settings is outside the scope of the present paper.

\subsection{Supplementary UAV-Swarm Toy: Quenched Aerodynamic-Field Reconstruction on a Fixed Communication Graph}
\label{subsec:sim_uav_swarm_quenched}

To complement the purely synthetic topology experiments, we include one lightweight application-flavored toy example inspired by aerodynamic-field reconstruction over a UAV swarm. The purpose of this experiment is not to claim a realistic air-fluid model, but rather to instantiate the fixed-topology quenched perspective of Theorem~\ref{thm:P1} in a geometric communication setting that is easy to interpret. In particular, the experiment is designed so that the theorem-facing quantities have a direct graph-theoretic meaning: the feature-budget proxy is determined by how many task-relevant coordinates become visible to each user through the communication graph, while the coverage proxy is determined by how much of the essential target energy remains unseen.

\paragraph*{Swarm geometry and user tasks}
We consider a one-dimensional UAV formation with small transverse perturbations, consisting of $N=12$ UAVs located at positions \(p_n=(x_n,y_n)\), where the longitudinal coordinates \(x_n\) are approximately evenly spaced across the interval \([0,1]\). We interpret these UAVs as servers. The reconstruction tasks correspond to $K=3$ user corridors centered at \(c_1=0.20\), \(c_2=0.50\), and \(c_3=0.80\). Each user task is seeded by the two UAVs closest to its corridor center, so the initial anchor set for user \(k\) is the set of the \(\texttt{anchor\_size}=2\) nearest UAVs to \(c_k\). Hence the graph geometry directly controls which servers can contribute to each reconstruction task after a given number of communication rounds.

\paragraph*{Localized subfunctions and dense nonlinear targets}
The global feature bank has size $D_{\mathrm{tot}}=96$, and each UAV stores exactly $\Gamma=12$ local subfunctions, chosen as the \(\Gamma\) feature coordinates whose spatial locations are closest to that UAV’s longitudinal position. Thus the support \(\mathcal S_n\) of UAV \(n\) is localized around its physical position in the formation. The input dimension is \(d_{\mathrm{in}}=5\), the ridge parameter is \(\lambda=10^{-2}\), the training noise standard deviation is \(0.05\), the training sample size is $M=320$, and the test-set size is $M_{\mathrm{te}}=1000$.

Each user target is again a dense nonlinear function in a Gaussian random-feature model. More precisely, the target coefficient vector \(\beta_k\in\mathbb R^{D_{\mathrm{tot}}}\) is generated as a dense mixture of low-frequency and higher-frequency spectral components, modulated by a spatial envelope centered around the corresponding corridor. Hence the targets are globally nonzero, but they place larger energy on feature coordinates that are spatially aligned with the corresponding corridor. This creates a natural notion of task relevance without reducing the example to a disjoint-support regime.

In a UAV-swarm setting, such dense nonlinear target functions can model concrete corridor-level sensing, monitoring, and decision tasks that arise in cooperative aerial missions. Examples include: i) radar-based target-presence or target-confidence mapping along a corridor, where multiple UAV radar returns are fused nonlinearly; ii) distributed RF sensing and emitter localization, where the target represents the likelihood or strength of an interfering or hostile transmitter in a spatial sector; iii) EO/IR surveillance confidence scoring, where many image-derived features from different UAVs contribute jointly to detect objects, motion, or anomalies; iv) smoke, gas, or pollution plume monitoring, where the swarm reconstructs a corridor-specific concentration or hazard index from spatially distributed measurements; v) wind, gust, or turbulence-field assessment, where the target is a nonlinear safety or traversability score for a flight corridor; vi) wildfire-front or heat-intensity estimation from thermal measurements; vii) communication-quality or jamming-risk prediction for an aerial route, based on distributed link-quality and spectrum observations; and viii) search-and-rescue likelihood maps, where heterogeneous weak signals from several UAVs are fused into a corridor-specific detection score. In all of these examples, the desired quantity is not purely local: many UAV measurements across the swarm can matter, but those spatially aligned with the corridor of interest typically contribute more strongly. This is exactly the behavior captured by the spatially modulated dense coefficient vector \(\beta_k\).

We compare three fixed graph realizations:
\begin{enumerate}
\item \emph{Chain / nearest-neighbor graph:} UAVs communicate only with adjacent UAVs in the ordered formation.
\item \emph{Local geometric graph:} an edge is present whenever two UAVs are within distance \(R\), where \(R\) is the communication radius.
\item \emph{Relay-enhanced overlap graph:} starting from the local geometric graph, we add sparse second-neighbor and backbone-style relay edges to model a communication pattern with mild graph overlap and enhanced multihop reachability.
\end{enumerate}

For a fixed graph \(G\), user \(k\), and communication depth \(T\), let \(\mathcal R_k^{(T)}(G)\) denote the set of UAVs reachable from the anchor set of user \(k\) within \(T\) graph hops. The corresponding visible feature set is \(\mathcal V_k^{(T)}(G)=\bigcup_{n\in \mathcal R_k^{(T)}(G)} \mathcal S_n\). This directly instantiates a quenched topology: once the graph \(G\) is fixed, the only remaining learning uncertainty comes from the data and the random-feature realization.
\paragraph*{Theorem-1-inspired resource and coverage proxies}
To connect the experiment back to Theorem~\ref{thm:P1}, we track two structural quantities. First, we define a feature-budget proxy by replacing the exact received-scalar count with the size of the visible feature set and then taking the harmonic mean across users:
\[
m_{\mathrm{harm}}^{\mathrm{vis}}(G,T)
=
\operatorname{harm}\Big(\,|\mathcal V_k^{(T)}(G)|:\,k\in[K]\Big).
\]
With
\[
\gamma=\frac{\Gamma}{D_{\mathrm{tot}}}=\frac{12}{96}=\frac18,
\]
we then use the theorem-inspired achievability proxy
\[
\Psi_{\mathrm{res}}(G,T)
=
\frac{1}{\gamma\,m_{\mathrm{harm}}^{\mathrm{vis}}(G,T)}.
\]
This is not the exact bound of Theorem~\ref{thm:P1}, but it plays the same structural role: smaller values indicate a larger effective visible-feature budget.

Second, for each user \(k\) we define an essential coordinate set \(\mathcal S_k^\star\) as the smallest set of coordinates capturing \(60\%\) of the coefficient energy of \(\beta_k\). The corresponding coverage proxy is the uncovered essential-energy fraction
\[
\varepsilon^{\mathrm{cov}}_k(G,T)
=
\frac{\sum_{j\in \mathcal S_k^\star\setminus \mathcal V_k^{(T)}(G)} \beta_{k,j}^2}
{\sum_{j\in \mathcal S_k^\star} \beta_{k,j}^2},
\]
and we report its average over the users. Thus the toy directly separates the two mechanisms that also appear in the quenched theory: feature-budget growth and essential-coordinate coverage.

\begin{figure*}[t]
\centering
\includegraphics[width=0.98\textwidth]{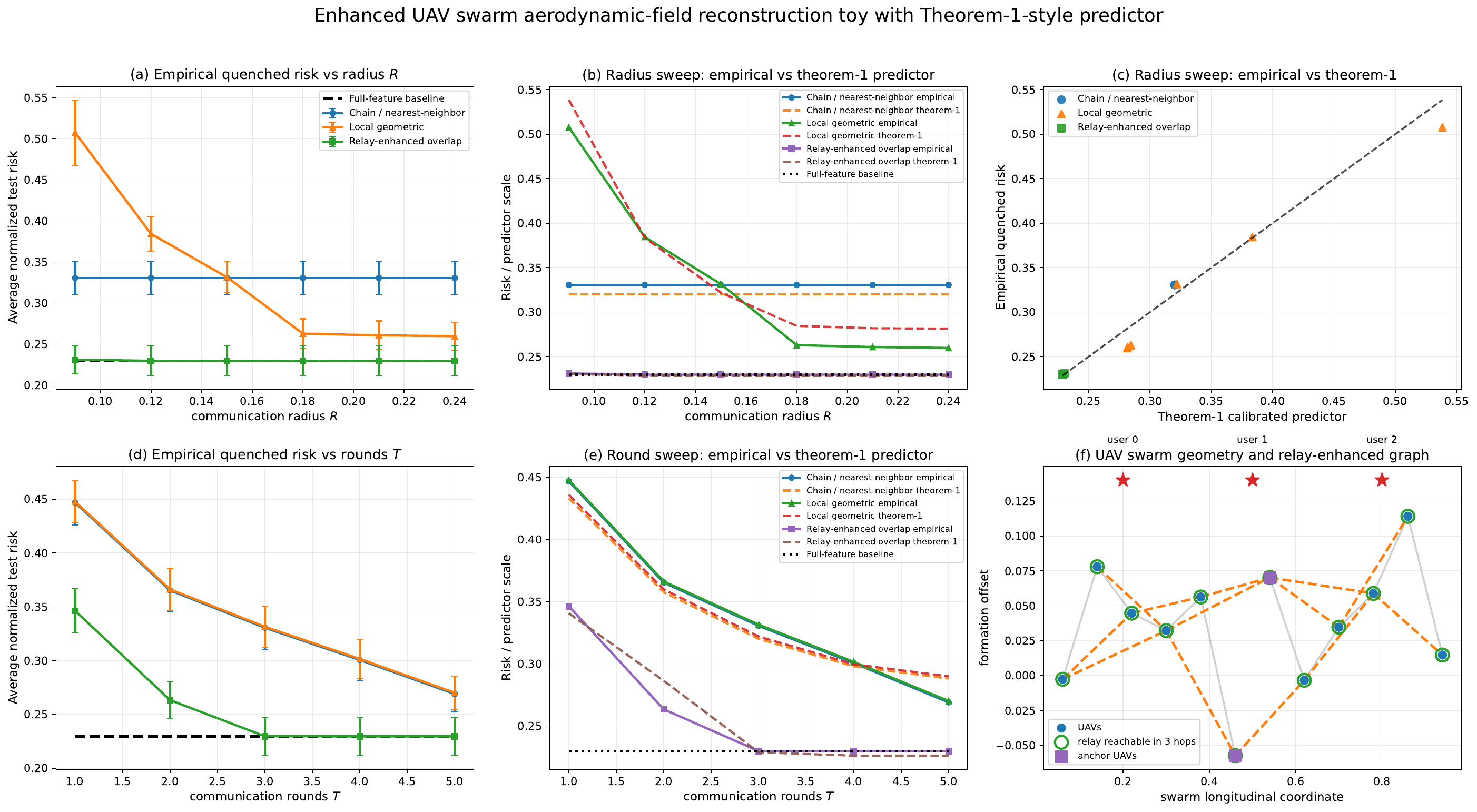}
\caption{Supplementary UAV-swarm aerodynamic-field reconstruction toy in the fixed-topology quenched regime. Panel~(a) shows the empirical quenched risk as a function of the communication radius \(R\) at fixed \(T=3\). Panel~(b) shows the corresponding theorem-inspired resource proxy \(1/(\gamma m_{\mathrm{harm}}^{\mathrm{vis}})\). Panel~(c) shows the empirical quenched risk as a function of the communication rounds \(T\) at fixed \(R=0.15\). Panel~(d) shows the corresponding coverage proxy given by the uncovered essential-energy fraction. Panel~(e) gives the mean number of reachable UAVs, and panel~(f) shows one representative swarm geometry together with the relay-enhanced graph. The relay-enhanced topology consistently attains the best empirical performance and the most favorable theorem-inspired proxies.}
\label{fig:supp_uav_swarm_thm1}
\end{figure*}

\paragraph*{Radius sweep at fixed round budget}
We first fix $T=3$ communication rounds and vary the geometric radius over $R\in\{0.09,0.12,0.15,0.18,0.21,0.24\}$. The results are shown in Fig.~\ref{fig:supp_uav_swarm_thm1}, panels~(a)--(b). Panel~(a) shows that the relay-enhanced overlap topology attains the lowest empirical quenched risk across the entire radius sweep and remains very close to the unrestricted full-feature baseline. The local geometric graph improves substantially as $R$ increases, while the chain graph remains essentially flat, as expected since it does not depend on the radius parameter. Panel~(b) shows the same pattern through the theorem-inspired resource proxy $1/(\gamma m_{\mathrm{harm}}^{\mathrm{vis}})$: the relay-enhanced graph has the smallest proxy throughout, the local geometric graph improves monotonically with $R$, and the chain graph remains nearly constant. Thus, in the fixed-graph quenched regime, the empirical risk and the theorem-style resource proxy are aligned in the expected direction.

\paragraph*{Round sweep at fixed communication radius}
We next fix $R=0.15$ and vary the number of communication rounds over $T\in\{1,2,3,4,5\}$. Panels~(c)--(e) of Fig.~\ref{fig:supp_uav_swarm_thm1} summarize the results. Panel~(c) shows that all three distributed topologies improve as $T$ increases, which is consistent with the fact that additional rounds expose more UAVs and hence more visible features to each user. The relay-enhanced topology improves most rapidly and is already very close to the full-feature baseline by about $T=3$, while the chain and local geometric graphs remain noticeably worse over the same range. Interestingly, for the specific base radius used here, the local geometric graph performs worse than the chain graph, which is consistent with panel~(e): at $R=0.15$, the chain graph actually reaches more UAVs on average over several hops than the sparse local geometric realization. 

Panel~(d) shows the coverage proxy based on uncovered essential energy. Again the relay-enhanced graph is clearly best: its uncovered essential-energy fraction drops sharply from the one-round case and reaches essentially zero by $T\ge 3$. By contrast, the chain and local geometric graphs retain nonzero uncovered essential energy even at $T=5$. This provides a direct theorem-facing interpretation of the risk behavior: once the relay-enhanced topology has exposed essentially all of the user-critical coordinates, further rounds bring only minor gains, and the empirical curve saturates near the full-feature baseline. Panel~(e) confirms the underlying graph mechanism by plotting the mean number of reachable UAVs. The relay-enhanced graph rapidly reaches almost the entire swarm, whereas the other two topologies expand much more slowly.

This toy experiment should be read as an application-flavored visualization of the quenched theorem rather than as a new theoretical statement. The fixed graph realization places the experiment in the quenched regime, and the empirical trends are consistent with the two theorem-inspired mechanisms we track: better graphs reduce the feature-budget proxy \(1/(\gamma m_{\mathrm{harm}}^{\mathrm{vis}})\) and also reduce the essential-coordinate coverage loss. The relay-enhanced graph benefits from both effects simultaneously. It improves multihop reachability, enlarges the set of visible feature coordinates, and rapidly covers the user-essential energy, which in turn yields the smallest empirical test risk. In this sense, the toy provides an interpretable geometric example of the broader GMUDC message: even under the same local budget \(\Gamma\), the realized communication topology has a strong effect on the user risk because it determines which task-relevant coordinates become visible to the decoder.

\end{document}